\documentclass[prb,10pt,twocolumn,superscriptaddress,longbibliography]{revtex4-2}
\usepackage[dvipdfmx]{graphicx}
\usepackage{dcolumn}
\usepackage{bm}
\usepackage{color}
\usepackage{amsmath,amssymb}
\begin{document}
\newcommand{\la}{\langle}
\newcommand{\ra}{\rangle}
\newcommand{\ga}{\alpha}
\newcommand{\gb}{\beta}
\newcommand{\gc}{\gamma}
\newcommand{\gs}{\sigma}
\newcommand{\vk}{{\bm{k}}}
\newcommand{\vq}{{\bm{q}}}
\newcommand{\vR}{{\bm{R}}}
\newcommand{\vQ}{{\bm{Q}}}
\newcommand{\vga}{{\bm{\alpha}}}
\newcommand{\vgc}{{\bm{\gamma}}}
\newcommand{\Ns}{N_{\text{s}}}
\newcommand{\mi}[1]{\textcolor{black}{#1}}
\newcommand{\mimt}[1]{\textcolor{black}{#1}}
\newcommand{\mic}[1]{\textcolor{black}{#1}}
\newcommand{\mir}[1]{\textcolor{black}{#1}}
\newcommand{\mib}[1]{\textcolor{black}{#1}}
\newcommand{\mig}[1]{\textcolor{black}{#1}}
\newcommand{\migr}[1]{\textcolor{black}{#1}}
\newcommand{\mimg}[1]{\textcolor{black}{#1}}
\newcommand{\bfsf}[1]{\textbf{\textsf{#1}}}
\newcommand{\avrg}[1]{\left\langle #1 \right\rangle}
\newcommand{\eqsa}[1]{\begin{eqnarray} #1 \end{eqnarray}}
\newcommand{\eqwd}[1]{\begin{widetext}\begin{eqnarray} #1 \end{eqnarray}\end{widetext}}
\newcommand{\hatd}[2]{\hat{ #1 }^{\dagger}_{ #2 }}
\newcommand{\hatn}[2]{\hat{ #1 }^{\ }_{ #2 }}
\newcommand{\wdtd}[2]{\widetilde{ #1 }^{\dagger}_{ #2 }}
\newcommand{\wdtn}[2]{\widetilde{ #1 }^{\ }_{ #2 }}
\newcommand{\cond}[1]{\overline{ #1 }_{0}}
\newcommand{\conp}[2]{\overline{ #1 }_{0#2}}
\newcommand{\nn}{\nonumber\\}
\newcommand{\cdt}{$\cdot$}
\newcommand{\bra}[1]{\langle#1|}
\newcommand{\ket}[1]{|#1\rangle}
\newcommand{\braket}[2]{\langle #1 | #2 \rangle}
\newcommand{\bvec}[1]{\mbox{\boldmath$#1$}}
\newcommand{\tr}[1]{\textcolor{black}{#1}}
\newcommand{\tc}[1]{\textcolor{black}{#1}}
\newcommand{\tb}[1]{\textcolor{black}{#1}}
\newcommand{\tbl}[1]{\textcolor{black}{#1}}
\newcommand{\tm}[1]{\textcolor{black}{#1}}
\newcommand{\tg}[1]{\textcolor{black}{#1}}
\newcommand{\tgn}[1]{\textcolor{black}{#1}}
\newcommand{\tgx}[1]{\textcolor{black}{#1}}
\newcommand{\txr}[1]{\textcolor{black}{\sout{#1}}}
\newcommand{\tbs}[1]{\textcolor{black}{\sout{#1}}}
\definecolor{green}{rgb}{0,0.75,0}
\newcommand{\tgtw}[1]{\textcolor{black}{#1}}
\definecolor{orange2}{rgb}{1,0.27,0}
\definecolor{indigo}{rgb}{0.5,0,0.7}
\newcommand{\tind}[1]{\textcolor{black}{#1}}

\preprint{APS/123-QED}

\title{
Hidden self-energies as origin of cuprate superconductivity\\
revealed by machine learning
}
\author{Youhei Yamaji}
\email{YAMAJI.Youhei@nims.go.jp}
\affiliation{Center for Green Research on Energy and Environmental Materials, National Institute for Materials Science, Namiki, Tsukuba-shi, Ibaraki, 305-0044, Japan}
\affiliation{Department of Applied Physics, The University of Tokyo, Hongo, Bunkyo-ku, Tokyo, 113-8656, Japan}
\affiliation{JST, PRESTO, Hongo, Bunkyo-ku, Tokyo, 113-8656, Japan}
\author{Teppei Yoshida}
\affiliation{Graduate School of Human and Environmental Studies, Kyoto University,
Yoshida-nihonmatsu-cho, Sakyo-ku, Kyoto, 606-8501, Japan}
\author{Atsushi Fujimori}
\affiliation{Department of Physics, University of Tokyo, Hongo, Bunkyo-ku, Tokyo, 113-0033, Japan}
\affiliation{Department of Applied Physics, Waseda University, Shinjuku-ku, Tokyo, 169-8555, Japan}
\author{Masatoshi Imada}
\affiliation{Toyota Physical and Chemical Research Institute, Nagakute, Aichi, 480-1192, Japan}
\affiliation{Research Institute for Science and Technology, Waseda University, Shinjuku-ku, Tokyo, 169-8555, Japan}

\begin{abstract}
Experimental data are the source of understanding matter.
However, measurable quantities are limited and theoretically important quantities are sometimes hidden. 
Nonetheless, recent progress of machine-learning techniques opens possibilities of exposing them only from available experimental data.
In this paper,
after establishing the reliability of the method in various careful benchmark tests,
the Boltzmann-machine method is applied to the angle-resolved photoemission spectroscopy 
spectra of cuprate high temperature superconductors,
Bi$_2$Sr$_2$CuO$_{6+\delta}$ (Bi2201) and Bi$_2$Sr$_2$CaCuO$_{8+\delta}$ (Bi2212).
We find prominent peak structures both in normal and anomalous self-energies, but they cancel in the total self-energy making the structure apparently invisible, while the peaks make universally 
dominant contributions to superconducting gap,
hence evidencing the signal that generates the high-$T_{\rm c}$ superconductivity.
The relation between superfluid density and critical temperature supports involvement of universal carrier relaxation associated with dissipative strange metals, where enhanced superconductivity is promoted by entangled quantum-soup nature of the cuprates.
The present achievement opens avenues for innovative machine-learning spectroscopy method to reveal fundamental properties hidden in direct experimental accesses. 
\end{abstract}
\pacs{
}
\maketitle

\section{Introduction}
{Momentum $k$ and energy $\omega$
dependent electron single-particle spectral function $A(k,\omega)$} can
be measured with recent revolutionarily refined resolution of angle resolved photoemission spectroscopy (ARPES)\cite{Shen}.
From $A(k,\omega)$,
the interaction effects \tgx{crucial for unconventional superconductors} can be identified in the self-energy~\cite{norman1999extraction,PhysRevLett.109.056401}.
Scanning \tgtw{tunneling} microscope (STM) and its spectra (STS) including the quasiparticle interference method~\cite{Hoffman2002} also give us insights into
{the self-energy~\cite{PhysRevLett.14.108,PhysRevB.3.4065}}.

In superconductors, the self-energy
consists of
the normal and anomalous (superconducting) contributions, $\Sigma^{\rm nor}$ and  $\Sigma^{\rm ano}$, respectively.
ARPES and STS provide us with only the total self-energy $\Sigma^{\rm tot}$ in a specific combination of these two~\cite{norman1999extraction} (see below for details). {However, to understand the superconducting mechanism, it is crucially important to extract these two separately, because they represent
different part of interaction effects: $\Sigma^{\rm ano}$ is proportional to the superconducting gap function, at the heart of superconducting properties, while normal-electron correlation effects, such as renormalized mass and life time, are encoded in $\Sigma^{\rm nor}$.}
Despite its importance, $\Sigma^{\rm ano}$ can be straightforwardly extracted
separately only when $\Sigma^{\rm nor}$ is non-singular 
as in the BCS 
{superconductivity of}
weakly correlated systems~\cite{PhysRevLett.14.108,PhysRevB.3.4065}.
\textcolor{black}{Indeed,
the decomposition of the self-energies of the BCS superconductors
has played the role of establishing
the phonon mechanism since the anomalous part contains the information of the phonon density of states,
which is crucial for the identification of the glue for the superconductivity.}
In case of the cuprate high-$T_{\rm c}$ superconductors, because of the strong electron correlation, the subject of extracting normal and anomalous self-energy separately belongs to an open enigmatic inverse problem, which has hampered the full understanding of the superconducting mechanism for decades.

Recently machine learning and data science are developing rapidly as tools to analyze accumulated data across various research domains. The expectation is to solve complex, important problems for human being, which are hardly solved by conventional tools, as in the attempt of forecasting future and designing new functional materials by utilizing existing big data. More specifically, machine learning has potential for innovating routes of exposing quantities, which is invisible in
direct 
measurements. Solving inverse problems to construct theories with predictive
\tgtw{power} by using existing experimental data is a typical target of the machine learning innovation.

In this paper, we develop a scheme of
machine-learning
technique to extract physical quantities hidden in experimental data.
To demonstrate the power of our method, we apply it to the electronic structure \tgtw{of}
the cuprate high-temperature superconductors
{under strong correlation effects manifested by the formation of the pseudogap in the normal state}. {Specifically, the Boltzmann machine~\cite{ackley1985learning,smolensky1986information} is
examined 
to extract
$\Sigma^{\rm nor}$ and  $\Sigma^{\rm ano}$
\tind{separately} 
from available 
ARPES spectra
{even when the normal self-energy is subject to prominent or singular correlation effects}.
\tind{Discovered} prominent peak structure in the energy dependence of $\Sigma^{\rm ano}$ hidden in the ARPES is shown to generate most of the superconducting gap, \tgtw{namely more than 90\% of the gap}, and hence \tind{to} be the driving force of the superconductivity in the cuprates.
\tgtw{From the extracted self-energies,
we elucidate the factors that determine the superconducting transition temperatures.}

{The organization of the present paper is as follows:
In Sec.~\ref{methods}, we summarize the essence of the regression scheme to
extract the self-energies from the spectral functions.
The details of the method is given in Sec.~\ref{detailed_method}.
The readers who \tgtw{are not interested in the technical details of the method but}
are interested \tind{only} in the results may skip \tind{Sec.~\ref{detailed_method} and directly jump} to Sec.~\ref{benchmark}.
\tgtw{Various  benchmark tests of the regression by using the present method are presented in 
Sec.~\ref{benchmark}
for simple metals and conventional  Bardeen-Cooper-Schrieffer-type
superconductors to show the reliability of the method.}
The \tgtw{main} results of \tind{the present paper studying} the regression for the cuprate superconductors are given in Sec.~\ref{results}.
The self-energies of cuprate superconductors \tgtw{extracted separately for the normal and anomalous contributions}
are shown, which \tgtw{have prominent peak structures each in the normal and anomalous part, but cancels in the spectral function}.
\tind{We also show that the prominent peak in the anomalous self-energy gives rise to the major part of the high temperature superconductivity.}
Sec.~\ref{discussion} is devoted to the discussion on the implications of the present results.
{Then, we summarize this paper and give our outlook in Sec.~\ref{summary_outlook}}.}
\section{Methods}
\label{methods}
\begin{figure}[htb]
\begin{center}
\includegraphics[width=0.5\textwidth]{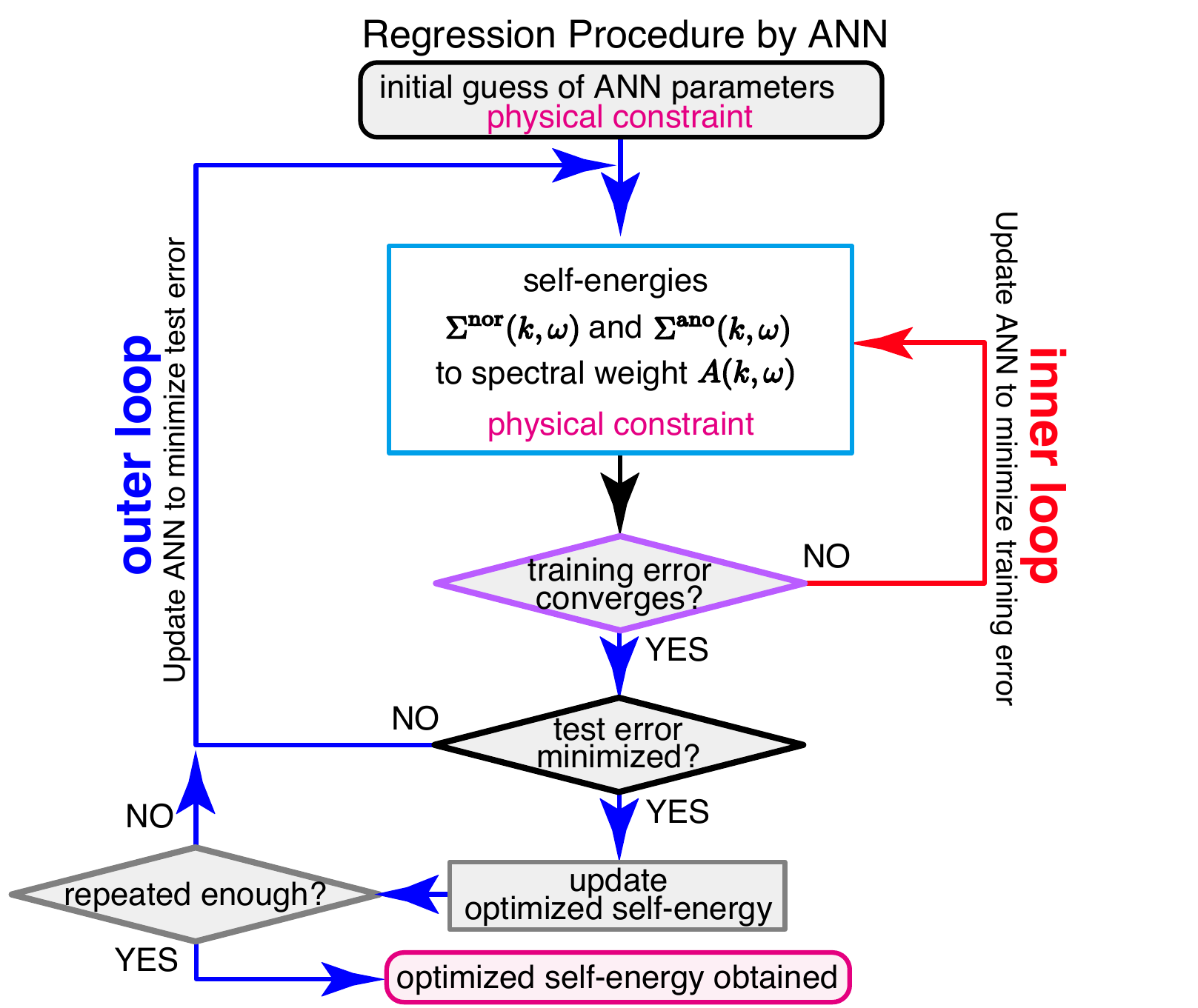}
\end{center}
\caption{
{Flow chart of machine-learning procedure.}
\noindent 
{Regression procedure of the normal self-energy $\Sigma^{\rm nor}(k,\omega)$ and anomalous self-energy $\Sigma^{\rm ano}(k,\omega)$
using the experimental spectral function.
The procedure starts from the central top (initial guess) to the training process (the inner loop) consisting of the red and black arrows to optimize all parameters in
\textcolor{black}{the artificial neural network (ANN)}.
When the error converges, the
outer loop
to decrease the test error, which
delivers the initial values for the next inner loop until the test error is minimized.
}
}
\label{Fig_Flow_Chart_Illustration_simplified}
\end{figure}
{In this section, we introduce the fundamentals and basic concept of the present method.
The detailed regression scheme is \tgtw{given} in Sec.~\ref{detailed_method}
for readers who are interested in \tgtw{technical} details.}
\subsection{Regression}
\label{method_regression}
\textcolor{black}{The present machine learning scheme is classified into the category of a general regression task, which optimizes a function $A$ to find $B$ and/or $F$, when $A$ is a nonlinear and complex functional of another function $B$ as $A=F(B)$,
where \tgtw{the} goal of
\tgtw{solving the inverse problem $B = F^{-1}(A)$ by the}
optimization of $A$ is set from a physical purpose~\cite{Theodoridis}.
Another example belonging to the regression task as an application of the machine learning is found in the use for quantum many-body problems or statistical physics problems 
(see for instance, Ref.~\onlinecite{Carleo}).}

\textcolor{black}{In the present case, $A$ is the ARPES spectral function $A(k,\omega)$, $B$ is the self-energies $\Sigma^{\rm nor}$ and $\Sigma^{\rm ano}$ and $F$ is given
\tgtw{by}
Eqs.~\tgtw{(\ref{eq:A})}
\tgtw{to}
(\ref{eq:Stot}) \tgtw{below}.
The training data is given by experimental $A(k,\omega)$ at a discrete and limited number of $\omega$.
Then the present machine learning is a typical regression task to infer
\tgtw{both}
$\Sigma^{\rm ano}(k,\omega)$ and $\Sigma^{\rm nor}(k,\omega)$ separately as continuous functions of $\omega$.}

{Here, we note that the present scheme is not a simple interpolation for
functions of $\omega$.
The regression with artificial neural network provides the self-energies $\Sigma^{\rm ano}(k,\omega)$ and $\Sigma^{\rm nor}(k,\omega)$
that simultaneously satisfy the constraints from the several rigorous and physically sound prior knowledge together with
the experimental data.}

\textcolor{black}{In the machine learning,
flexible regression models that can minimize the training error without any limitation are employed.
In contrast, when we employ phenomenological regression models,
the constraints arising from phenomenological functions substantially increase the cost function
far beyond the noise in the experimental data as demonstrated in the following discussion (see Appendix~\ref{S2.2}).
If the phenomenological form of the regression model is not a priori justified,
the machine learning scheme should have an advantage over standard phenomenological regression schemes.}

{The basic procedure
is illustrated in 
{Fig.~\ref{Fig_Flow_Chart_Illustration_simplified}}.
More technical details are found in 
the following Sec.~\ref{detailed_method} 
and the robustness, accuracy and reliability of the present machine learning are shown in detail 
in Appendices~\ref{S2.1} and \ref{S2.5}.
An important advantage of the machine learning is that the approximation converges to the correct results without overfitting and bias if one increases the data point. We do not call \tgtw{any} other approach that does not reach this systematic improvement as \tgtw{a} machine learning.}

\subsection{Green function}
\label{method_Green_function}
\textcolor{black}{We propose a theoretical method to extract
$\Sigma^{\rm nor}$ and $\Sigma^{\rm ano}$ from experimentally observed spectral functions $A(k,\omega)$
of superconductors. 
In \tgtw{the} superconducting phases, the single-particle retarded Green function at a given momentum $k$
as a function of frequency $\omega$ is given by
a diagonal component of $2\times 2$ matrix in Nambu representation,
\begin{eqnarray}
\hat{G}(k,\zeta)=
\left[
\begin{array}{cc}
\zeta - \epsilon_k - \Sigma^{\rm nor}(k,\zeta) & -\Sigma^{\rm ano}(k,\zeta) \\
-\Sigma^{\rm ano}(k,\zeta) & 
\zeta + \epsilon_k + \Sigma^{\rm nor}(k,-\zeta)^{\ast} \\ 
\end{array}
\right]^{-1},\nonumber\\
\label{eq:G}
\end{eqnarray}
with $\zeta = \omega + i \delta$ ($\delta$ is a small positive real number).
The bare dispersion is given by $\epsilon_k$.
$A(k,\omega)$ measurable by ARPES is related to $\hat{G}$ as
\begin{equation}{
A(k,\omega) = -\frac{1}{\pi}{\rm Im} \left[\{ \hat{G}(k,\zeta) \}_{11}\right]_{\delta\rightarrow +0},
\label{eq:A}
}
\end{equation}
with the normal component of the Green function
\begin{equation}{
G^{\rm nor}(k,\omega)\equiv \hat{G}(k,\omega)_{11} = [\omega-\epsilon_k-\Sigma^{\rm tot}(k,\omega)]^{-1}.
\label{eq:Gnor}
}
\end{equation}
\tgtw{Here,} the total self-energy $\Sigma^{\rm tot}$ is given by~\cite{Scalapino}
\begin{eqnarray}
\Sigma^{\rm tot}(k,\omega) 
&=& [\Sigma^{\rm nor}(k,\zeta) + W(k,\zeta)]_{\delta\rightarrow +0}, 
\label{eq:Stot}
\end{eqnarray}
with $W$ given as a specific combination,
\begin{equation}
{W(k,\omega)=\Sigma^{\rm ano}(k,\omega)^2/[\omega + \epsilon_k+\Sigma^{\rm nor}(k,-\omega)^{\ast}]. \label{eq:W}}
\end{equation}
The gap function 
\begin{equation}
{\Delta(k,\omega)=Q(k,\omega)\Sigma^{\rm ano}(k,\omega), \label{eq:Delta}}
\end{equation}
which is a measure of superconducting order, is proportional to $\Sigma^{\rm ano}(k,\omega)$ with the coefficient $Q(k,\omega)$ called the \tgtw{frequency dependent} renormalization factor
defined as 
\begin{equation}{
{Q(k,\omega) =
\left.
\frac{1}{1-[\Sigma^{\rm nor}(k,\zeta)-\Sigma^{\rm nor}(k,-\zeta)^{\ast}]/(2\zeta)}\right|_{\delta\rightarrow +0}}.
\label{eq:Q}}
\end{equation}
The $\omega \rightarrow 0$ limit of $Q$ is theoretically equivalent to the quasiparticle weight (renormalization factor) defined in Eq.~(\ref{eq:zqp}) below as we calculate in {
Appendix~\ref{S2.9} with the help of the procedure in Appendix~\ref{S1.6}}.
The real part of $\Delta(k,\omega)$, ${\rm Re} \Delta (k,\omega=0)$ is nothing but the superconducting gap \tgx{(see the definition of $\Delta$ in Eq.~(\ref{eq:Delta}))}.
{In the present report, $\delta$ is chosen to be equal to the experimental resolution as $\delta=10$ meV~\cite{kondo2011disentangling}, 
instead of taking $\delta\rightarrow 0^+$.}}  

\tr{
To estimate the density of the Cooper pairs, {mass renormalization,} and gap amplitude from the spectral function, we define
$F(k)$, \tb{$z_{\rm qp}(k)$,} and $\Delta_{0}(k)$ respectively as,
\begin{eqnarray}
F(k) &=& \tc{\int_{-\infty}^{0}d\omega \frac{1}{\pi}{\rm Im}\hat{G}(k,\omega)_{12}}, \label{eq:Fk}
\\
z_{\rm qp}^{-1}(k)&=&1 - \left.\partial {\rm Re}\Sigma^{\rm nor}(k,\omega)/\partial \omega \right|_{\omega\rightarrow 0},
\label{eq:zqp}
\\
\Delta_{0}(k) &=& \Delta (k,\omega=\Delta_{0}(k)),
\label{eq:Dqp}
\end{eqnarray}
}
\tg{where 
$z_{\rm qp}$ is called the renormalization factor (see 
Appendix~\ref{S2.9}).}
\mimt{As is well known, the gap function is interpreted by the product of the Cooper pair density and the effective attractive interaction to form the Cooper pair as the mean field acting on the Cooper pair formation.}

\subsection{Prior knowledge} 
\label{prior_knowledge}
{When we can measure entire $\omega$ dependence of $\hat{G}(k,\omega+i\delta)$ at a fixed $k$,
we can reconstruct $\Sigma^{\rm nor}(k,\omega)$ and $\Sigma^{\rm ano}(k,\omega)$ solely without any information at
{momenta other than $k$},
\textcolor{black}{because Eqs.~(\ref{eq:Gnor}) and (\ref{eq:Stot})} are all diagonal in the $k$ space.
However, we can not measure the entire $\omega$ dependence of the complete Green function matrix $\hat{G}(k,\omega+i\delta)$, which makes the information through the Kramers-Kronig relation incomplete.
{
In the literature~\cite{bok2016quantitative},
it has been assumed that the momentum dependence of the normal-state spectrum at \tgtw{a} fixed $\omega$ is a single Lorentzian curve so that $\Sigma^{\rm nor}(k,\omega)$ and $\Sigma^{\rm ano}(k,\omega)$ can be extracted separately without knowing the $\omega$ dependence at large $|\omega|$.}
In the present article, to overcome the lack of information at large $|\omega|$,
we utilize \mir{physically sound constraints} and extract $\Sigma^{\rm nor}(k,\omega)$ and $\Sigma^{\rm ano}(k,\omega)$
from experimentally observed $A(k,\omega)$ at a single fixed momentum $k$, {instead of
assuming specific momentum dependences of the spectra}.
Indeed, the present Boltzmann machine learning as detailed below successfully reproduces
$\Sigma^{\rm nor}(k,\omega)$ and $\Sigma^{\rm ano}(k,\omega)$ separately from benchmark spectral functions
without the momentum dependence of the spectra as shown in 
{Appendix~\ref{S2.5}.}} 

The physical constraints employed in the present article are classified into two categories.
The constraints in the first category are the structure of the Green function given in Eq.~(\ref{eq:G}),
the Kramers-Kronig relationship between the real and imaginary part of the self-energies, negative definiteness of ${\rm Im}\Sigma^{\rm nor}$,
and odd nature of ${\rm Im}\Sigma^{\rm ano}$ as ${\rm Im}\Sigma^{\rm ano}(-\omega)=-{\rm Im}\Sigma^{\rm ano}(\omega)$.
The constraint in the second category is the sparse and localized nature of the ${\rm Im}\Sigma^{\rm ano}$ along \tgtw{the} $\omega$ axis. In the present context,  the sparseness is defined as the property of  ${\rm Im}\Sigma^{\rm ano}$concentrated and localized in the small frequency range around the Fermi level.  It should be mentioned that the optimization procedure of the present machine learning does not explicitly impose this latter constraint because of the flexible representability of the present
(restricted) Boltzmann machine~\cite{Roux}.
The sparse nature of  ${\rm Im}\Sigma^{\rm ano}$ is, however, justified {\it a posteriori}
in the optimized solution, which turns out to satisfy physically reasonable sparseness \tgtw{although} the machine learning procedure does not explicitly impose this constraint.

{From physical grounds, the sparse and localized nature of ${\rm Im}\Sigma^{\rm ano}$ is a natural consequence as clarified in Ref.~\onlinecite{MorelAnderson}, 
where irrespective of the mechanism and symmetry of the realistic pairing, strong and long-range nature of  Coulomb repulsion in general causes severe pair breaking at large energies and suppresses $\Sigma^{\rm ano}$ at energies far away from the \tgtw{Fermi} level.}

\subsection{Overview of optimization}
{We represent the self-energies by using artificial neural networks in the present scheme,
and optimize the neural network to reproduce \tgtw{the} experimentally observed spectral functions. 
The optimization of the neural network consists of
inner and outer optimization loops
(see {Figure \ref{Fig_Flow_Chart_Illustration_simplified}}).
In the inner optimization loop, starting with given initial parameters for the neural network,
all the parameters are optimized to minimize the training error
(see the definition given in Eq.~(\ref{chi_square}) \tgtw{below}) 
by following the natural gradient method (see Appendix~\ref{S1.4} for details).
On the other hand, in the outer optimization loop,
\tgtw{test errors} (see Eq.~(\ref{overline_chi2}))
\tgtw{are} minimized by the Bayesian optimization.
The updated distribution delivers the initial values for the next inner loop.}
{The next inner loop starts again with the updated neural network parameters,
while, if the number of repetitions of the outer loop already reaches an upper limit (typically less than a hundred in the present paper),
the optimization is
\tgtw{completed}
and the current best neural network
\tgtw{gives} the optimized self-energies.
\tgtw{Here, we combine these inner loop
and outer loop optimization to avoid the overfitting,
which \tind{would be} inevitable if we \tind{would use only} the inner loop optimization with the training data.}}

\section{Technical Details}
\label{detailed_method}
In this section, we show the details of the present regression scheme
to extract the self-energy from ARPES data.
The neural network representation of the self-energy is detailed in Sec.~\ref{wavelet},
which is supplemented by the Kramers-Kronig relation in Sec.~\ref{real_part}.
The procedures to optimize the self-energy are given in Sec.~\ref{numerical_procedure}, \ref{training_error}, and \ref{test_error}. 
You may skip
\tgtw{this section}
if you are not interested in the \textcolor{black}{technical} details \textcolor{black}{of the present method}.
\subsection{\textcolor{black}{Wavelet analysis and Boltzmann machine representation of imaginary part of self-energy}} 
\label{wavelet}
\begin{figure}[htb]
\begin{center}
\includegraphics[width=0.5\textwidth]{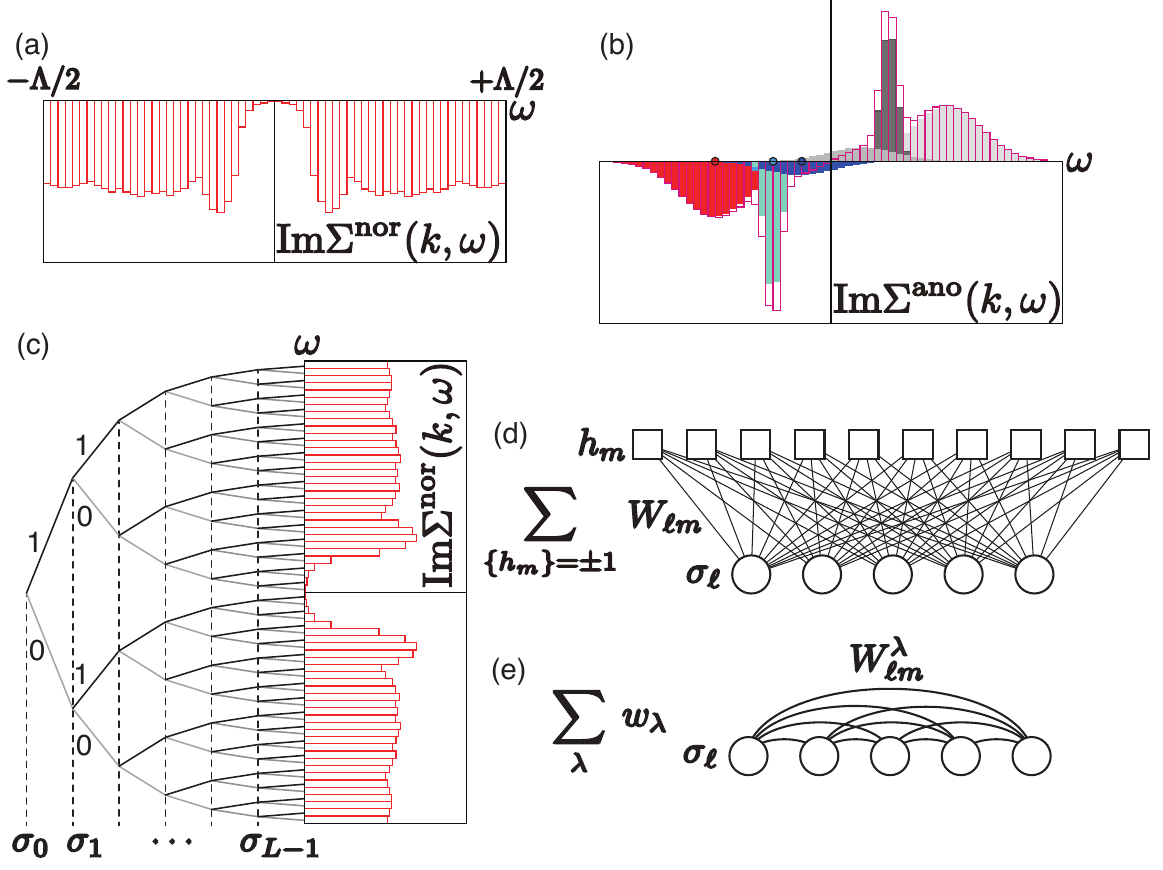}
\end{center}
\caption
{\textcolor{black}{Boltzmann machine representation of imaginary parts of self-energies.
(a) The piece-wise rectangular function representation of ${\rm Im}\Sigma^{\rm nor}$
is illustrated as a combination of red rectangles.
(b) The piece-wise rectangular function representation of ${\rm Im}\Sigma^{\rm ano}$
is shown.
While total ${\rm Im}\Sigma^{\rm ano}$ is represented by open red rectangles,
Boltzmann machine
representation generates their components as is
illustrated as filled red, cyan, and blue rectangles.
To satisfy anti-symmetry of ${\rm Im}\Sigma^{\rm ano}$, the copies of the Boltzmann machines
shown in grey rectangles are also supplemented.
(c) The wavelet-like structure of the rectangular basis set is illustrated.
From the longest wave length structure governed by $\sigma_0$ to the shortest wave length structure
controlled by $\sigma_{L-1}$, each rectangular basis (open red rectangle) is labeled by
the set of the bits $\bvec{\sigma}=(\sigma_0,\sigma_1,\dots,\sigma_{L-1})$ $(\sigma_{\ell} = \pm 1)$.
%
The structure of the
restricted Boltzmann machine for ${\rm Im}\Sigma^{\rm nor}$ and mixed distribution consisting
of Boltzmann machines for ${\rm Im}\Sigma^{\rm ano}$ are depicted in (d) and (e), respectively.}}
\label{Fig_ANN}
\end{figure}

\begin{figure*}[htb]
\begin{center}
\includegraphics[width=0.7\textwidth]{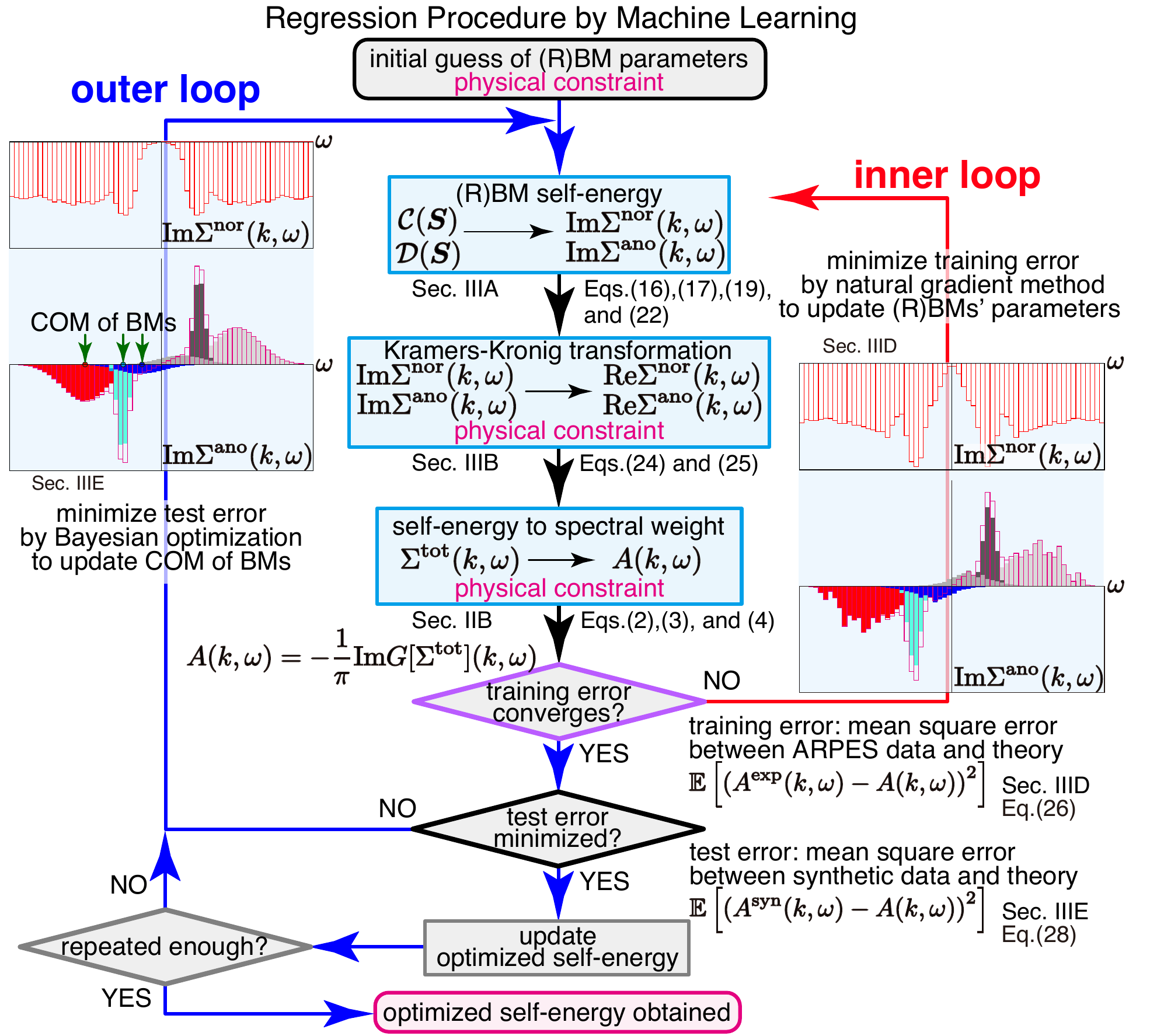}
\end{center}
\caption{
Flow chart of machine-learning procedure.
\noindent 
{Regression procedure of the normal self-energy $\Sigma^{\rm nor}(k,\omega)$ and anomalous self-energy $\Sigma^{\rm ano}(k,\omega)$
using the experimental spectral function $A^{\rm exp}(k,\omega)$. 
Here, (R)BM,  and COM (green arrows)
stand for (restricted) Boltzmann machine, and center of mass, respectively.
The procedure starts from the central top (initial guess) to the training process (the inner loop) consisting of the red and black arrows to optimize all
the BM parameters.
When the error converges, the
outer loop (blue and black arrows) updates the COM positions to decrease the test error, which
delivers the initial values for the next inner loop until the test error is minimized.
The test error is minimized by repeating the combined inner and outer loop updates.
The histograms are schematic $\omega$-dependences of ${\rm Im}\Sigma^{\rm nor/ano} (k,\omega)$ at a fixed $k$:
The stepwise representation of ${\rm Im}\Sigma^{\rm ano} (k,\omega)$ (the open purple boxes)
is obtained by antisymmetrizing the superposition of the three BM distributions (red, blue, and green filled boxes), $\mathcal{D}(\bvec{S})$,
to satisfy the odd-function property in $\omega$, where the light or dark gray histograms are added.
${\rm Im}\Sigma^{\rm nor}$ is directly given from $\mathcal{C}(\bvec{S})$ as the open red boxes.
In total, the machine learning minimizes the training (inner loop) and test (outer loop) errors given by
the average $\mathbb{E}$
over $\omega$ for Eqs.~(\ref{chi_square}) and (\ref{overline_chi2}), respectively.
}
}
\label{Fig_Flow_Chart_Illustration}
\end{figure*}

Although high-resolution ARPES data for $A(k,\omega)$ are available in experiments,
$\Sigma^{\rm nor}$ and $\Sigma^{\rm ano}$  are not directly given separately from $A(k,\omega)$,
while if $\Sigma^{\rm nor}$ and $\Sigma^{\rm ano}$ and $\epsilon_k$ are given,
$A(k,\omega)$ can be determined easily by using Eqs.~(\ref{eq:G}) and (\ref{eq:A}).
Therefore we need to solve an {underdetermined non-linear} inverse problem.
To overcome the underdetermined nature of the problem, we employ \mir{physically sound constraints justifiable even in strongly correlated electron systems} as prior knowledge as introduced above.
By incorporating these physical constraints,
we try to optimize  $\Sigma^{\rm nor}$ and $\Sigma^{\rm ano}$ so as to reproduce experimental $A(k,\omega)$.
For this purpose we employ a machine-learning method by applying a Boltzmann-machine algorithm~\cite{ackley1985learning}. 
The reliability, accuracy and robustness of the present machine learning procedure are shown
in several robustness test against noise in 
Appendix~\ref{S2.1}
and
benchmark tests in 
Appendix~\ref{S2.5}.


{
In the retarded Green function representation,
${\rm Im}\Sigma^{\rm nor}$ is negative definite and
${\rm Im}\Sigma^{\rm ano}$ 
is an odd function of $\omega$.
The negative definiteness and odd nature are guaranteed by expanding
${\rm Im}\Sigma^{\rm nor}$ and ${\rm Im}\Sigma^{\rm ano}$ with
positive definite bases $\{\theta_j\}$ as 
\eqsa{
{\rm Im}\Sigma^{\rm nor}&=&-\sum_j c_j \theta_j (\omega),\\
{\rm Im}\Sigma^{\rm ano}&=& \sum_j d_j [\theta_j (\omega) - \theta_j (-\omega)],
}
where $c_j$ and $d_j$ are real coefficients.
Due to the positive definiteness of the normal part, $c_j$ is positive.} 

{One of the simplest basis set $\{\theta_j\}$ is a set of rectangular functions,
which gives a step-wise representation of the self-energies.
To obtain a flexible and compact representation for the coefficients $c_j$ and $d_j$,
here we will combine a well-established wavelet-type representation~\cite{mallat1989theory,mallat2008wavelet,akansu2001multiresolution} 
and Boltzmann machines as follows.} The high representability of the wavelet formalism with the rectangular basis is discussed in
Appendix~\ref{S1.2}.
\textcolor{black}{The piece-wise rectangular representations of ${\rm Im} \Sigma^{\rm nor}$ and ${\rm Im} \Sigma^{\rm ano}$
are shown in Fig.~\ref{Fig_ANN}(a) and (b), respectively.}

{In this fitting, the frequency range of our interest $\omega\in[-\Lambda/2,+\Lambda/2]$ is first divided into $2^L$ grids using an integer $L$ and assign an $L$-digit binary representation as 
\begin{equation}{\bvec{\sigma} \equiv (\sigma_0,\sigma_1, \cdots, \sigma_{L-1}),
\label{eq:sigma}}
\end{equation}
where $\sigma_i={\rm mod}(I/2^{i},2)$ for the decimal representation $I(\bvec{\sigma})$ in the range
$0 \leqq I(\bvec{\sigma}) \leqq 2^L -1$ of the grid number coordinate; 
\begin{equation}{
I(\bvec{\sigma})=\sum_{\ell=0}^{L-1}\sigma_{\ell}\cdot 2^{\ell}.
}
\end{equation}
Then the unit rectangular function $\Theta^{L}_{\bvec{\sigma}}\left(x \right)$ is defined as 
\begin{eqnarray}
\Theta^{L}_{\bvec{\sigma}}\left(\omega\right)
=
\left\{
\begin{array}{cl}
 1 & {\rm for} \ x \in \left[ I(\bvec{\sigma})/2^{L},\{1+I(\bvec{\sigma})\}/2^{L}\right)\\
 0 & {\rm otherwise}\\
\end{array}
\right.
.
\nonumber\\
\label{eq:Theta}  
\end{eqnarray}
\textcolor{black}{The correspondence between the position of the rectangular functions
and the decimal representation $I(\bvec{\sigma})$ is shown in Fig.~\ref{Fig_ANN}(c).}

{Then, the basis set expansions of ${\rm Im} \Sigma^{\rm nor}$ and ${\rm Im} \Sigma^{\rm nor}$
are obtained as
\begin{eqnarray}
{\rm Im}\Sigma^{\rm nor}(\omega)&=&-
\sum_{\bvec{\sigma}}
C(\bvec{\sigma})
\Theta^{L}_{\bvec{\sigma}}\left(\frac{\omega+\Lambda/2}{\Lambda}\right),
\label{eq:Snor}
\\
{\rm Im}\Sigma^{\rm ano}(\omega)
&=&
\sum_{\bvec{{\sigma}}}
D(\bvec{{\sigma}})
\left[
\Theta^{L}_{\bvec{\sigma}}\left(\frac{\omega+\Lambda/2}{\Lambda}\right)
\right.
\nonumber\\
&&-
\left.
\Theta^{L}_{\bvec{\sigma}}\left(\frac{\Lambda/2-\omega}{\Lambda}\right)
\right],\label{mixBM}
\end{eqnarray} 
with two sets of $2^L$ fitting parameters, $C(\bvec{{\sigma}})$ and $D(\bvec{{\sigma}})$.
}

In the actual calculation, we take \textcolor{black}{$\Lambda\sim 0.8$} eV to efficiently fit the experimentally observed $A(k,\omega)$ confined to the range $\omega>-0.4$ eV.
Note that we assume ${\rm Im} \Sigma^{\rm nor}(\omega)$ is thus restricted and nonzero in the comparable range (\textcolor{black}{$\omega>-\Lambda/2$} eV, \textcolor{black}{$\Lambda\sim 0.8$} eV ) to the experimentally measurable one of $A(k,\omega)$ to fit them, while we do not impose any constraint on ${\rm Im} \Sigma^{\rm ano}(\omega)$, because $\Sigma^{\rm ano}$ is restricted within this energy range as we noted above.  We have confirmed in
Appendix~\ref{S2.1}
that the result is basically unchanged even when the energy cut-off $\Lambda$ imposed on ${\rm Im} \Sigma^{\rm nor}(\omega)$ is removed by adding possible high-energy tail or peak in the range \textcolor{black}{$\omega <-\Lambda/2$}. This means that our main finding is not altered by the experimentally unknown shape of the spectral function at $\omega<-0.4$ eV. 

Irrespective of the cut-off  $\Lambda$, the Kramers-Kronig relations given below (Eqs.~(\ref{KK-Sigmanor}) and (\ref{KK-Sigmanor}))  are applied for all the self-energies to the whole energy range ($-\infty < \omega < \infty$) in the machine learning procedure so that the optimized solution satisfies the strict causality.  In fact, for the converged $A(\omega)$, the sum rule $\int_{-\infty}^{\infty} d\omega A(\omega)=1$ is satisfied.

{As a compressed representation for $C(\bvec{{\sigma}})$ and $D(\bvec{{\sigma}})$,}
{in this paper, we {employ two different types of Boltzmann machines to enhance their representational power {and} determine each $\Sigma^{\rm nor}$ and
$\Sigma^{\rm ano}$, separately}.
${\rm Im} \Sigma^{\rm nor}$ is negative definite and a widely distributed function
within energy scale set by the Coulomb repulsion.
Thus, we use the flexible and nonnegative restricted Boltzmann machine, which will be introduced below. 
However, ${\rm Im} \Sigma^{\rm ano}$ has different properties:
It is sparse, which is justified {\it a posteriori} as we addressed already.
Therefore, 
we employ a mixture {distribution of the Boltzmann machine
without the hidden {variables} to accelerate the optimization.
There may still remain ambiguities in determining $\Sigma^{\rm nor}$ and $\Sigma^{\rm ano}$ from an observed $A(k,\omega)$.
As {proposed} below, the ambiguities are removed by {imposing} physical constraints of ${\rm Im}\Sigma^{\rm ano}$.}
}

Now we introduce the Boltzmann machine to represent $C(\bvec{\sigma})$ and
$D(\bvec{\sigma})$ in Eqs.~(\ref{eq:Snor}) and (\ref{mixBM}), respectively.
{
We first change the binary variable $\bvec{\sigma}$ introduced in Eq.~(\ref{eq:sigma}) to the Ising variable $\bvec{S}=(S_0,S_1,...S_{L-1})$ using the relation $S_{\ell}=2\sigma_{\ell}-1$ for later convenience and rewrite as ${\mathcal{C}}(\bvec{S}) (=C(\bvec{\sigma}))$ and ${\mathcal{D}}(\bvec{S}) (=D(\bvec{\sigma})) $.
Then by adding hidden Ising variables $\bvec{h}=(h_1,h_2,... )$, the Boltzmann machine is generally defined as
a Boltzmann weight for Ising {variables} $\bvec{\nu}_{\ell} = \pm1$ consisting of 
$\bvec{S}$ and $\bvec{h}$ in the notation $\bvec{\nu}=(\bvec{S},\bvec{h})$ as
\begin{equation}{
\mathcal{B}(\bvec{\nu}|\bvec{W},\bvec{B})
=\exp \left[ \sum_{\ell,m} W_{\ell m}\nu_{\ell} \nu_m + \sum_{\ell}B_{\ell}\nu_{\ell}\right],
}
\end{equation}
where $(\bvec{W})_{\ell m} = W_{\ell m}$ represents interaction among $\bvec{\nu}$,
and $(\bvec{B})_{\ell} = B_{\ell}$ represents bias fields applied to $\bvec{\nu}$.
$W_{lm}$, and $B_{\ell}$ are variational parameters to minimize the difference between the resultant $A(k,\omega)$ and the measured spectral functions.
The role of the hidden variables $\bvec{h}$ is to enhance the representability of $\mathcal{B}$ to approximate ${\mathcal{C}}(\bvec{S})$ and ${\mathcal{D}}(\bvec{S})$.}

{Thanks to the non-negativity of $\mathcal{C}(\bvec{S})$, 
it is efficiently represented by
the restricted Boltzmann machine (RBM)~\cite{smolensky1986information,Hornik,Amari1998}}, 
one of the most widely used one, as represented by
\begin{equation}{
\mathcal{C}({\bvec{S}}) = 
\sum_{\bvec{h}} \mathcal{B}_C({\bvec{S},\bvec{h}}|\bvec{W},\bvec{B}),
}
\end{equation}
where $\mathcal{B}_C$ restricts the interactions in $\mathcal{B}$ only between {\it visible} and {\it hidden} {variables} in the form $W_{\ell m}S_{\ell}h_m$.

The advantage of the RBM is that one can analytically trace out the hidden variables $h_m$,  leading to
\begin{equation}{
\displaystyle \mathcal{C}
({\bvec{S}})=
e^{b}
\prod_{m=0}^{L_{\rm h}-1} 2\cosh \left[{S_{\ell}}W_{\ell m}\right],
}
\end{equation}
where $L_{\rm h}$ is the number of the hidden variables.
Any $\omega$-dependent line shape in the energy range $[-\Lambda/2, \Lambda/2] $ can be flexibly represented by optimized Boltzmann-machine parameters, if they are nonnegative. {$L_{\rm h}$ is typically set $L_{\rm h}=2L$ to achieve a convergence with reasonable computational costs.
\textcolor{black}{The restricted Boltzmann machine representation is schematically illustrated in Fig.~\ref{Fig_ANN}(d).}

}

{For $\Sigma^{\rm ano}$,} to remove the ambiguities, we {impose} the physically required symmetries, 
\begin{eqnarray}
{\rm Re} \Sigma^{\rm ano}(\omega)&=&{\rm Re}\Sigma^{\rm ano}(-\omega) \nonumber \\
{\rm Im} \Sigma^{\rm ano}(\omega)&=&-{\rm Im}\Sigma^{\rm ano}(-\omega),
\end{eqnarray} 
which can be constrained 
{by employing the odd function Eq.~(\ref{mixBM})}.
If ${\rm Im}\Sigma^{\rm ano}$ is sparse, namely confined in a certain range of $\omega$,
${\rm Im}\Sigma^{\rm ano}$ can be better represented
by a mixture distribution, namely by a linear combination of the full Boltzmann machine
in the form 
\begin{equation}{
\textcolor{black}{\mathcal{D}({\bvec{S}}) = 
\sum_{\lambda=1}^{M}
w_{\lambda} \mathcal{B}_D(\bvec{S}|\bvec{W}^{\lambda},\bvec{B}^{\lambda})},
\label{eq:mixture}
}
\end{equation}
\textcolor{black}{where $M$ is the number of the Boltzmann machines in the linear combination.} 
\textcolor{black}{Here,} $\mathcal{B}_D$ allows only the physical variable $\bvec{S}$ without the hidden one $\bvec{h}$,
but allows the interaction between $\bvec{S}$ as 
\begin{equation}{
\mathcal{B}_D({\bvec{S}}|\bvec{W}^{\lambda},\bvec{B}^{\lambda})
=
e^{\sum_{\ell,m}S_{\ell}W_{\ell m}^{\lambda}S_m + \sum_{\ell}S_{\ell}b_{\ell}^{\lambda}}.
\label{eq:B}
}
\end{equation}
\textcolor{black}{The Boltzmann machine representation is schematically illustrated in Fig.~\ref{Fig_ANN}(e).}

Note that the linear combination of the Gaussian distributions is one of the standard procedure to approximate
a smooth function\cite{bishop2006pattern} and can be used as an initial guess of $\mathcal{D}$ (for the detailed procedure, see in
Appendix~\ref{S2.6}).
\mimg{Of course, the Boltzmann machine has representability far beyond the Gaussian distribution after the optimization.}

Since it is sufficient to take the number of variables {$S_{\ell}$ at most ${L=9}$} to fit the experiment data containing the resolution limitation,
we can explicitly take the trace summation over {$S_{\ell}$} for all $\ell$ (with $2^{L}$ terms) at each iteration step.
Therefore the drawback of
the mixture distribution of BM
beyond the RBM (namely, the complexity arising from containing the interaction term between two physical variables proportional to $W^{\lambda}_{\ell m}S_{\ell}S_m$) is not a serious problem.
We set the number of the \textcolor{black}{Boltzmann} machines in Eq.~(\ref{eq:mixture})
up to 3 \textcolor{black}{($M\leq 3$)} to promote the faster optimization of the imaginary part of the anomalous self-energy.

When the experimental spectral data contain small noise,
the step-wise representation for the imaginary part of the self-energies
may introduce a systematic increase in the test errors.
To reduce the possible increased error, we introduce a piecewise-linear representation
instead of the step-wise representation ${\rm Im}\Sigma^{\rm nor/ano}$
{(see 
Appendix~\ref{S1.2}).}
Since the estimated noise is very small for the experimental ARPES spectra of the optimally doped Bi2212 ($T_{\rm c}=90$K) at 12K,
the piecewise linear representation is helpful to achieve the comparable size of test errors with the noise in the experimental data
(see 
Appendix~\ref{S2.1} for quantitative discussions).

\subsection{Real part of self-energy}
\label{real_part}
The real part of the retarded self-energy is obtained through the Kramers-Kronig relation as,
\begin{eqnarray}
{\rm Re} \Sigma^{\rm nor}(\vec{k},\omega) &=&
\frac{1}{\pi}\mathcal{P}\int d\omega' \frac{{\rm Im} \Sigma^{\rm nor}(\vec{k},\omega')}{\omega' -\omega},
\label{KK-Sigmanor}
\\
{\rm Re} \Sigma^{\rm ano}(\vec{k},\omega) &=&
\frac{1}{\pi}\mathcal{P}\int d\omega' \frac{{\rm Im} \Sigma^{\rm ano}(\vec{k},\omega')}{\omega' -\omega},
\label{KK-Sigmaano}
\end{eqnarray}
where {a broadening factor $\delta=10$meV}
is introduced
to represent a principal value,
\eqsa{
\mathcal{P}\int d\omega' \frac{f(\omega')}{\omega'-\omega}\nonumber}
by
\eqsa{{\rm Re}\int d\omega' \frac{f(\omega')}{\omega'+i\delta-\omega}.\nonumber}
{See 
Appendix~\ref{S1.3}
for details.}
\subsection{Numerical procedure for optimization} 
\label{numerical_procedure}
We optimize the Boltzmann machines to reproduce experimentally observed spectral functions. 
The optimization of the Boltzmann machines consists of an inner and outer optimization loops
(see {Figure \ref{Fig_Flow_Chart_Illustration}}).
In the inner optimization loop, starting with given initial parameters for the Boltzmann machines,
all the parameters of the Boltzmann machine are optimized to minimize the training error defined in Eq.~(\ref{chi_square}) by following the natural gradient method detailed in Appendix~\ref{S1.4}.
On the other hand, in the outer optimization loop,
a test error defined below (Eq.~(\ref{overline_chi2}))
is minimized by the Bayesian optimization
that only 
updates the centers of mass of the distributions
in $\mathcal{D}(\bvec{S})$ by fixing other BM parameters to the values obtained in the inner loop.
The updated distribution delivers the initial values for the next inner loop.

{
To find the optimized self-energy, these inner and outer loop optimization processes
are combined as follows.
First, Boltzmann machines are initialized to follow the prior knowledge explained above.
Then, these Boltzmann machines are optimized to minimize the training error in the inner loop.
Once the inner-loop optimization converges,
the test error is evaluated by the optimized Boltzmann machines,
and the self-energies given by the present Boltzmann machines are stored as the current best ones.
When the outer loop is repeated, only if the test error becomes minimum in the whole optimization history, 
the present self-energies are stored as the current best.
Then,
the centers of mass of the $M$ Boltzmann-machine distributions are updated in the outer loop by using the Bayesian optimization scheme. 
The next inner loop starts again with the updated Boltzmann machines parameters,
while, if the number of repetitions of the outer loop already reaches an upper limit (typically less than a hundred in the present paper),
the optimization is finalized and the current best Boltzmann machines give the optimized self-energies.} For an efficient optimization, the initial condition at the largest $L$ ($L=9$) is prepared from the optimization at $L=8$ (see 
Appendix~\ref{S1.4}.)

\subsection{Minimization of training error}
\label{training_error}
{For given initial parameters of the Boltzmann machines}, the least square error of the training defined by 
\begin{equation}{
\chi^2=\frac{1}{2N_{\rm d}}\sum_j
\left\{A^{\rm exp}(\omega_j)-
f(\omega_j)A(\omega_j)
\right\}^2,
}
\label{chi_square}
\end{equation}
is minimized,
where {$N_{\rm d}$ is the number of the experimental data points,} $A^{\rm exp}(\omega)$ is an experimentally observed $A(k,\omega)$, $\{\omega_j\}$ ($j=1,2,\dots,N_{\rm d}$) is the set of frequency where $A(k,\omega)$ is observed in the experiment, {and $f(\omega_j)$ is a convolution of the Fermi-Dirac distribution and a Gaussian distribution}.
{The experimental data $A^{\rm exp}(\omega_j)$ involves the Fermi-Dirac distribution {broadened} by the resolution of the experiments.
Therefore, we introduce the convolution $f(\omega)$ of the Fermi-Dirac distribution at 12 K for Bi2212 and 11 K for Bi2201 and the Gaussian distribution with standard deviation $\sqrt{\sigma^2}=5$ meV.
{Here, we normalize the experimental data $A^{\rm exp}(\omega)$ by assuming $(1/N_{\rm d})\sum_j A^{\rm exp}(\omega_j)=n_0$, where $0 < n_0 < 1$.
In this paper, we infer $n_0=0.3$ per spin, which means that 60\% of an electron is assumed to be distributed in the experimentally  observed range ($\omega \gtrsim -0.4$ eV).} {We show that $n_0=0.3$ is indeed the optimized value of the least square fit in
Appendix~\ref{S2.7}
while the result of the self-energies does not sensitively depend on the choice of $n_0$ around 0.3. }}

\subsection{ Minimization of test error} 
\label{test_error}
{To further explore the multi-dimensional parameter space of the Boltzmann machine and find an optimized solution,
we employ the Bayesian optimization scheme 
{in the outer loop}.}
{Before performing the Bayesian optimization, as explained above, we perform sufficiently large number of optimization steps,
which is typically {$4\times 10^3$},
in the inner optimization loop 
\mimg{to minimize the training error}.
}

Then, we update the center of mass of \mimg{each component of mixture distribution represented by} Boltzmann machine in ${\rm Im}\Sigma^{\rm ano}$ defined in Eq.~(\ref{mixBM}),
{in the following procedure.
First, we extract the weight, center of mass, and variance of each Boltzmann-machine distribution by zeroth, first, and second moments as the function of $\omega$.}
To update the center of mass, we use the Bayesian optimization scheme~\cite{bishop2006pattern} depending on the history of the optimization process for the center of mass,
where
the test error $\overline{\chi^2}$ defined below is the cost function to be minimized instead of $\chi^2$, to avoid overfitting.
Then, we construct the initial values for $\mathcal{D}(\bvec{S})$ that defines ${\rm Im}\Sigma^{\rm ano}$ for the next inner loop by
using the updated center of mass with the weight and variance obtained above.
Each Boltzmann machine in the mixture distribution $\mathcal{D}(\bvec{S})$ is
initialized as the Gaussian distribution with the weight, the updated center of mass, and variance.
One may wonder why we initialize the Boltzmann machine again with the Gaussian distribution:
Since the next inner loop optimizes the Boltzmann machine again, the initialization by the Gaussian does not alter the final results, where the final convergence to the optimized self-energies is reached after the inner loop.
 The reason to reduce the distribution temporarily to the Gaussian is 
that the outer loop optimization can be handled easily since only the three lowest moments are needed.

{To define the test error, first,} we generate 
{synthetic} experimental data from the original data.
{Because the overfitting originates from reproducing detailed noisy behaviors in the experimental data finer than the experimental resolution~\cite{akaike1998information}, 
to eliminate the noise, the experimental data $A^{\rm exp}$ is fitted by a smooth function $A^{\rm fit}$ defined as
a linear combination of the Gaussian distributions~\cite{bishop2006pattern} with standard deviation $\sqrt{\sigma^2}=10$ meV equal to the experimental resolution.
Then, we can estimate amplitude of noise in the experimental data as 
\begin{equation}
\sigma_n^2 = N_{\rm d}^{-1}\sum_{j}(A^{\rm fit}(\omega_j)-A^{\rm exp}(\omega_j))^2.
\label{eq:experror}
\end{equation}

\mi{This error estimate is a standard procedure in the linear regression problem~\cite{guo2004learning}.
Inferring error or noise of the experimental data from the smoothed curve represented by an interpolation of the experimental data based on the physical assumption of smooth and continuous behavior in nature is a general established procedure in the linear regression problem.
(See for instance, Ref.~\onlinecite{bishop2006pattern}). 
This is a natural regularization to infer the generalization error reliably. There, it is important to assume the smoothness within the scale of the experimental resolution and with a frequency scale sufficiently longer than the interval of the experimental discrete points to exclude overfittings. This is achieved by the superposition of the Gaussian with the 10meV width to meet the experimental resolution, 10meV (see Ref.~\onlinecite{Theodoridis} as well). 
}

By using the probability distribution
$p(A^{\rm syn}|A^{\rm fit},\omega) \propto
\exp \left[-\left(A^{\rm syn}(\omega)-A^{\rm fit}(\omega)\right)^2/2\sigma_n^2\right]$,
we can generate synthetic experimental data {$A^{\rm syn}$}.
By assuming that $p(A^{\rm syn}|A^{\rm fit},\omega)$ well reproduces real experimental data, the cost function to avoid
the overfitting is defined by,
\begin{eqnarray}
{\overline{\chi^2} = \frac{1}{N_{\rm d}N_{\rm r}}\sum_{s=1}^{N_{\rm d}}\sum_{r=1}^{N_{\rm r}}
\{
{A^{\rm syn}_{r}}(\omega_s^{(r)})-
 f(\omega_s^{(r)})
A(\omega_s^{(r)})
\}^2},\nonumber\\
\label{overline_chi2}
\end{eqnarray}
where 
{$A^{\rm syn}_r$} is the $r$th 
{synthetic} experimental data {independently} generated by the probability distribution
{$p(A^{\rm syn}|A^{\rm fit},\omega)$}
and $\{\omega_s^{(r)}\}$ is a set of randomly chosen frequency points for each
{synthetic} data 
{$A^{\rm syn}_r$}.}
See also Appendix~\ref{S2.1}.

The optimization of the internal parameters of the Boltzmann machines \mimg{in the inner loop and the optimization of the center of mass of each mixture distribution in the outer loop is repeated together several tens.
The self-energies that give the minimum of
{ $\overline{\chi^2}$ for the test data
are called the
optimized self-energies.}}

\section{Benchmark Tests}
\label{benchmark}
In this section, \textcolor{black}{we benchmark} the performance of the present self-energy inference
by utilizing the artificial neural network representation.
First, we reproduce the normal self-energy of metals.
Since there is no anomalous component,
the regression becomes easier and less underdetermined.
As typical examples, the surface state of Be and normal state of
(La,Sr)$_2$CuO$_4$ are analyzed.

\tind{Next}, we perform the regression to reproduce
the known normal and anomalous
\tind{self-energies of superconductors}.
The present method indeed reproduces these self-energies
for a model of strong coupling Bardeen-Cooper-Schrieffer (BCS) superconductors.
The readers who are interested in the main results on the superconducting cuprates
may skip this section and go to Sec.~\ref{results}.

\begin{figure}[htb]
\begin{center}
\includegraphics[width=0.5\textwidth]{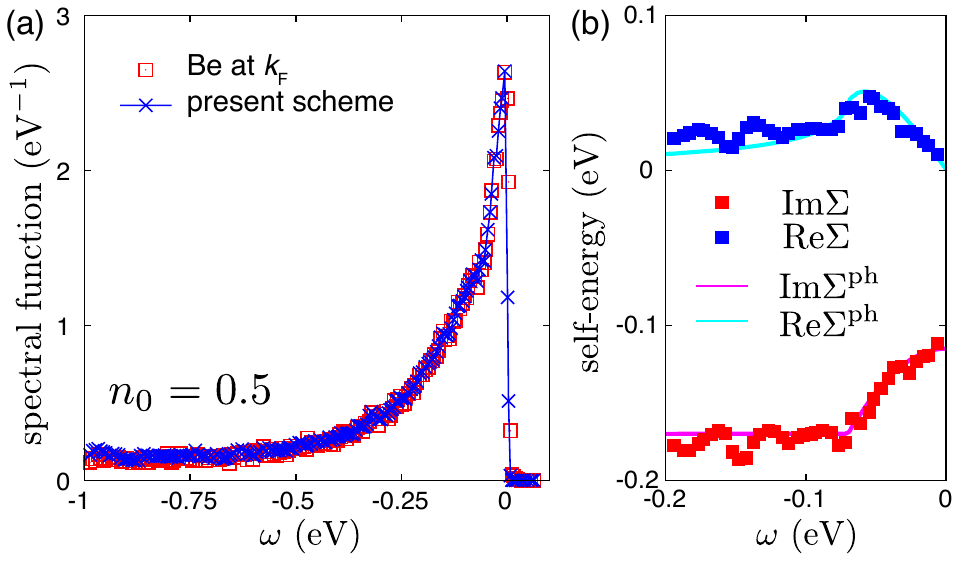}
\end{center}
\caption
{\textcolor{black}{ARPES spectrum of Be (0001) surface state at the Fermi momentum $k_{\rm F}$ and self-energy.
\textcolor{black}{In the left panel, the red open squares represent the ARPES data and blue crosses show the theoretical spectral function given by the machine-learning self-energy, which is shown in the right panel. The red (blue) closed squares represent the imaginary (real) part of the self-energy obtained by the present Boltzmann-machine learning with $n_0=0.5$
\tind{($n_0$ is the ratio of the measured weight to the total weight). For the definition of $n_0$, see Sec.~\ref{training_error}}.
The magenta (cyan) curves represent the imaginary (real) part of the self-energy obtained in Hengsberger {\it et al}.~\cite{hengsberger1999electron}
by using the Migdal-Eliashberg theory with a constant shift of the imaginary part
due to elastic scatterings. Here we note that wavy structures in $\Sigma$ shown in the right panel simply originates from the wavy structures in the experimental ARPES data superimposed on top of the sharp peak shown in the left panel (red open squares) likely to be ascribed to the experimental noise.}}
}
\label{Fig_Be_kF}
\end{figure}
\subsection{\tgtw{Normal metals}}
To benchmark the capability of the present scheme,
we analyze non-superconducting metals as trivial examples.
\subsubsection{Surface state of Be}
{The ARPES spectrum of Be (0001) surface state~\cite{hengsberger1999electron}
is analyzed as a typical metal.
In Fig.~\ref{Fig_Be_kF}(a), the spectral function at the Fermi momentum $k_{\rm F}$
\tgtw{obtained by our regression task reproduces the experimental data}.
The optimized normal self-energy is consistent with the self-energy given by
the \tgtw{combination of} Migdal-Eliashberg \tgtw{(ME)} theory
\tgtw{and} data from experiments \tgtw{by considering}
a constant shift of the imaginary part due to elastic scatterings,
as shown in Fig.~\ref{Fig_Be_kF}(b).
\tgtw{Note that our machine learning procedure does not assume the ME theory
and shows the ability of reliable regression \tind{later} even when the ME theory is not {\it a priori}
justified as in the case of strongly correlated electron systems.}}

\subsubsection{Normal state of (La,Sr)$_2$CuO$_4$}
\begin{figure*}[htb]
\begin{center}
\includegraphics[width=0.9\textwidth]{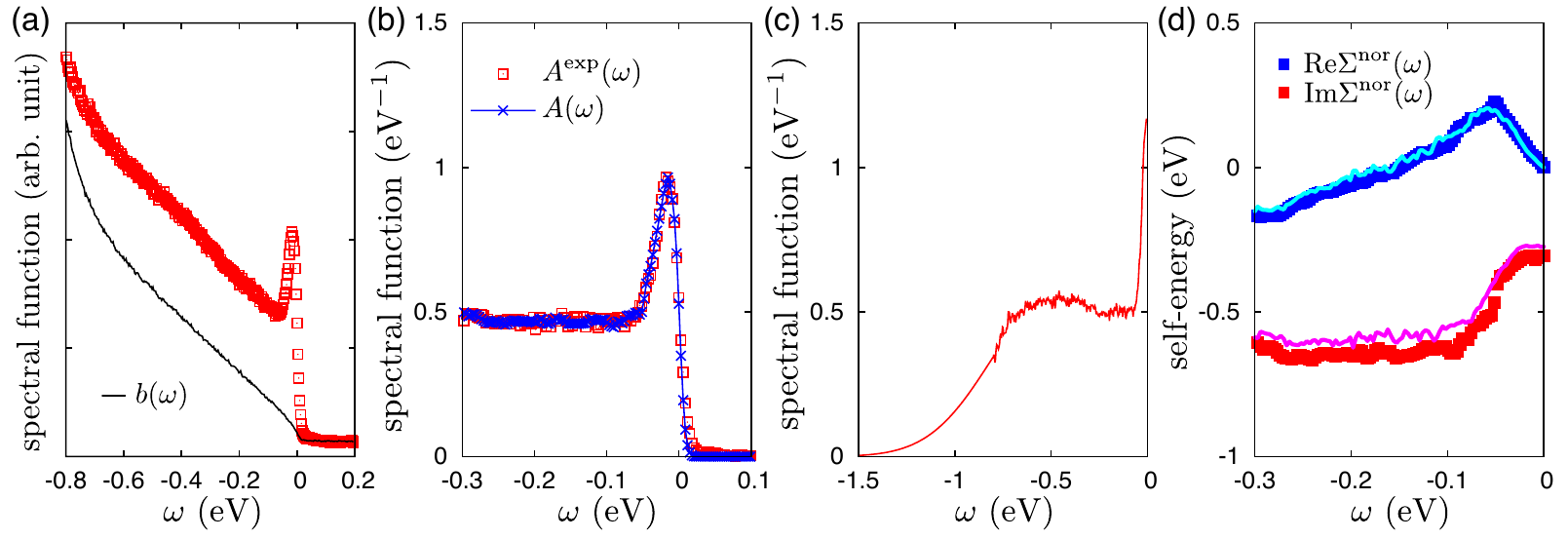}
\end{center}
\caption{
Self-energy analysis of EDC curve at the Fermi momentum $k_{\rm F}$ reported by Zhou, {\it et al}..
The raw EDC data (red squares) and background function $b(\omega)$ determined by a standard way (black solid curves) are shown in (a).
The EDC curve after subtracting $b(\omega)$ from the raw data is shown as $A^{\rm exp}(\omega)$ by red squares in (b),
in comparison with the spectral function reproduced by the Boltzmann machine self-energies (see (d)) illustrated by blue crosses.
To perform the KK transformation, a high energy Gaussian tail is added to $A^{\rm exp}(\omega)$ for $\omega < -0.8$ eV as shown in (c).
Real and imaginary parts of the normal self-energy (blue and red squares, respectively) obtained from the machine learning using the EDC curve shown in (b) are compared
with the real and imaginary parts obtained by the standard KK transformation (cyan and magenta \tgtw{solid} curves, respectively).
The standard KK scheme is detailed in Appendix~\ref{S2.5}.}
\label{Fig_LSCO}
\end{figure*}
{We have also analyzed the ARPES data of Zhou, {\it et al}.~\cite{zhou2005multiple},
to see the normal-state self-energy of (La,Sr)$_2$CuO$_4$.
After subtracting the background proposed in this paper,
we have analyzed intrinsic spectral function by the machine learning and succeeded in reproducing the spectral function from the optimized self-energy as shown in \tgtw{Fig.}~\ref{Fig_LSCO}.}

{We also compare thus obtained self-energy with the results of the standard \tgtw{Kramers-Kronig (KK)} transformation scheme and found that they are essentially consistent each other.
Here, the standard KK transformation scheme \tgtw{employed in Ref.~\onlinecite{zhou2005multiple}} consists of three steps: First, a high-energy tail is attached to the spectral function to obtain the full $\omega$ dependence of $-(1/\pi){\rm Im}G(\omega)$.
Second, by assuming the particle-hole symmetry, the KK transformation is used to obtain the real part of $G$.
Then, from the Dyson equation, the self-energy $\Sigma = \omega - G^{-1}$ is obtained.
\tgtw{\tind{Without}
the assumption on the high-energy tail
and
the particle-hole symmetry,
our regression scheme indeed reproduces the results of the standard KK transformation.}}

\subsection{BCS superconductors}
\begin{figure*}[htb]
\begin{center}
\includegraphics[width=1.0\textwidth]{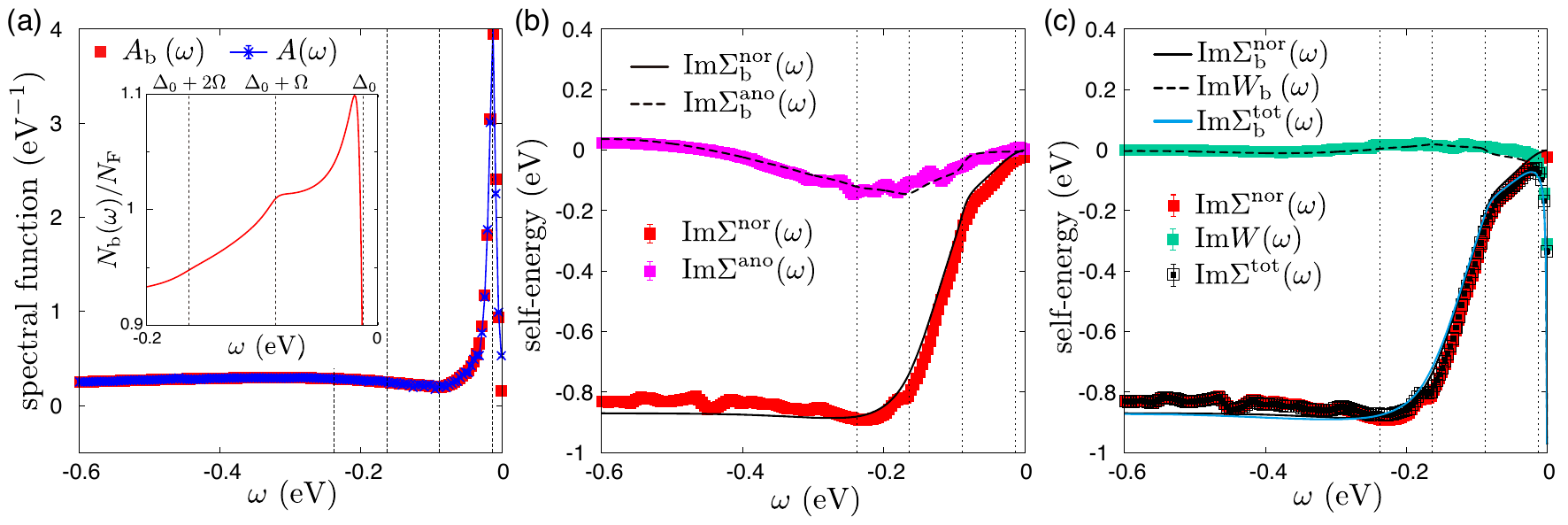}
\end{center}
\caption
{
{Spectrum and self-energies of {phonon}-mediated superconducting model.
{(a). The exact spectral function of the {phonon}-mediated superconductor, $A_{\rm b}(\omega)$, (red solid squares) is compared
with $A(\omega)$ (blue curve with crosses) obtained from the machine learning using $A_{\rm b}(\omega)$ within -1 eV $<\omega<$ 0.0 eV.
The inset shows the electronic density of states in the superconducting state normalized by the normal state density of states $N_{\rm F}$.
The vertical dotted lines show {$\omega=-\Delta_0$, $-\Delta_0-\Omega$, and $-\Delta_0-2\Omega$,} where anomalies appear reflecting
strong electron-phonon couplings and formation of the superconducting gap $\Delta_0$.
{While the spectral function at the Fermi momentum $A_{\rm b}(\omega)$ shows a dip at $\omega=-\Delta_0-\Omega$,
the density of state shows a shoulder due to the so-called kink in the renormalized dispersion that appears when a finite energy shift $\epsilon$
is introduced.}
(b) and (c): The imaginary part of the self-energies of the {phonon}-mediated superconductor,
$\Sigma^{\rm nor}_{\rm b}$, $\Sigma^{\rm ano}_{\rm b}$, {$\Sigma^{\rm tot}_{\rm b}$} and $W_{\rm b}$ (curves),
are compared with the self-energies
obtained from the machine learning (symbols). {Note that the curves for $\Sigma^{\rm tot}_{\rm b}$ and $\Sigma^{\rm nor}_{\rm b}$ are nearly overlapped in (c).}
}
}
}
\label{Fig_boson_Sigma}
\end{figure*}
{\tgtw{Here, we show that the present scheme
reproduces} the self-energies of the strong coupling BCS superconductors.
In the BCS superconductors, the amplitude of the self-energies is smaller than those in strongly correlated electron systems
such as cuprate superconductors. 
The weaker self-energy effects are suitable for perturbative treatments for forward problems such as the \tgtw{ME}
theory \tgtw{if the ME theory is justified}. 
In contrast,
the present self-energy regression from the spectral functions
works well for the strongly correlated electron systems
because the spectral functions provide more information necessary for solving the inverse problem to obtain the self-energies,
such as \tgtw{large}
superconducting gaps and \tgtw{broader}
quasiparticle peaks.
Then, the regression of the BCS superconductors
are difficult tasks to perform by using the present scheme
when the amplitude of the anomalous self-energy and its influence on the spectral function are small.
\tind{Nevertheless,} in this section, we demonstrate the \tind{successful} regression of the self-energies
for a BCS superconductor \tind{when it is} close to the strong-coupling limit, \tind{in which} we have substantial amplitude of
the anomalous self-energy.}

{As a typical model, the superconducting state described by the following Eliashberg equations
is examined.
By following the standard strong coupling theory for boson-mediated superconductors~\cite{Scalapinobook,Schriefferbook,PhysRevLett.14.108},
the superconducting gap $\Delta (\omega)$ and particle-hole symmetric component of the normal self-energy $\omega (1-Z(\omega))$
{at zero temperature}
are given by the Eliashberg equations,
\begin{widetext}
\begin{eqnarray}
\Delta (\omega) &=& \frac{N_{\rm F}}{Z(\omega)}\int_{\Delta_0}^{\omega_{\rm c}}d\omega' {\rm Re}
\left\{
\frac{\Delta (\omega')}{\left[{\omega'}^2-\Delta (\omega')^2+i\eta\right]^{1/2}}
\right\}
K_{+}(\omega,\omega'),\label{BCS_Delta}
\\
\left[1-Z(\omega)\right]\omega &=& N_{\rm F}\int_{\Delta_0}^{\infty}d\omega' {\rm Re}
\left\{
\frac{\omega'}{\left[{\omega'}^2-\Delta (\omega')^2+i\eta\right]^{1/2}}
\right\}
K_{-}(\omega,\omega')+\Sigma^{(0)}(\omega),\label{BCS_Zfac}
\end{eqnarray}
\end{widetext}
where  
{$\Delta_0={\rm Re}\Delta (\Delta_0)$},
$\omega_{\rm c}$ is the cutoff frequency, and $\eta$ is a positive broadening factor. 
Again, we assume that the noninteracting density of states is given by \tgtw{a} momentum and energy independent constant $N_F$ for simplicity and the superconducting symmetry is momentum independent $s$-wave.  In the following, we use the name ``phonon" for the boson,
\textcolor{black}{although Eqs.~(\ref{BCS_Delta}), (\ref{BCS_Zfac}), and the following Eq.~(\ref{BCS_phonon_K}) can be used to describe electrons coupled to any localized optical boson mode, irrespective of the origin of the boson mode.}
Here, we assume that the kernel functions $K_{\pm}$ originate from the Einstein phonon as 
\begin{eqnarray}
K_{\pm}(\omega,\omega')=\tind{g_{\rm el\mathchar`-ph}^2} \left[
\frac{1}{\omega'+\omega+\Omega-i\eta}
\pm
\frac{1}{\omega'-\omega+\Omega-i\eta}
\right],\nonumber\\
\label{BCS_phonon_K}
\end{eqnarray}
where $\Omega$ is the Einstein phonon frequency and \tind{$g_{\rm el\mathchar`-ph}$} is the electron-phonon coupling constant.
With the assumption that the density of states is \tgtw{a} constant around the Fermi level,
the self-energies obtained by the Eliashberg equations are independent of the electron density and dimension of the system.
Then, the normal component of the Green function is defined as 
\begin{eqnarray}
G^{\rm nor}_{\rm b}(\epsilon,\omega)=\frac{Z(\omega)\omega+\epsilon}{\{ Z(\omega)\omega \}^2-\epsilon^2-\phi(\omega)^2+i\eta},
\end{eqnarray}}
where $\phi (\omega) = Z(\omega)\Delta (\omega)$ and $\epsilon$ is the energy measured from the Fermi energy.
The spectral function is defined as $A_{\rm b}(\omega)=-(1/\pi)f_{\rm FD}(\omega){\rm Im}G^{\rm nor}_{\rm b}(\epsilon=0,\omega)$,
{ where $f_{\rm FD}(\omega)$ is the Fermi-Dirac distribution,} and
the superconducting density of states is given by
\begin{eqnarray}
N_{\rm b}(\omega) = N_{\rm F}\int d\epsilon \left|-(1/\pi){\rm Im}G^{\rm nor}_{\rm b}(\epsilon,\omega)\right|.
\end{eqnarray}
{The notation of the self-energies used in the literature on the {phonon}-mediated superconductors~\cite{Scalapinobook,Schriefferbook},
$\omega [1-Z(\omega)]$ and $\phi (\omega)$, is different from $\Sigma^{\rm nor}(\omega)$ and $\Sigma^{\rm ano}(\omega)$ in the present paper.
The normal and anomalous components of the self-energies are obtained as
\begin{eqnarray}
\Sigma^{\rm nor}_{\rm b} (\omega) &=& \omega - Z(\omega)\omega,\\ 
\Sigma^{\rm ano}_{\rm b} (\omega)
&=& \phi(\omega) = Z(\omega)\Delta (\omega). 
\end{eqnarray}
}
{ 
In Fig.~\ref{Fig_boson_Sigma}, an example of the self-energy inference for the {phonon}-mediated superconductors is shown.
Here, we set the coupling constant {$\tind{g_{\rm el\mathchar`-ph}^2} N_{\rm F} = 0.275$ eV}, the Einstein phonon frequency $\Omega = 0.075$ eV,
the cutoff frequency $\omega_{\rm c} = 4$ eV, and the broadening factor $\eta = 0.0075$ eV$^2$.   
The parameters in $\Sigma^{(0)}(\omega)$ are chosen as $a=0.008$ eV$^2$, $b=0.016$ eV$^2$, and $\alpha = 0.005$ eV$^3$.
For normalization of the spectral function, we choose {$n_0=0.3$} without tuning.
Although {its effect is negligibly small in} the inferred self-energies, the Fermi-Dirac distribution with $T=40$ K is introduced
in the spectral function used in the machine learning just by following the scheme
with finite-temperature experimental data. 
The machine learning results capture essential features of the
\tind{original}
normal and anomalous self-energies
\tind{that generate the target spectrum $A_{\rm b}$},
where the anomalous self-energy has a dip around $-(\Delta_0+\Omega)$ and $-(\Delta_0+2\Omega)$ and the normal self-energy shows a sharp drop. The dip in the anomalous self-energy, which {arises from the electron-phonon coupling and} gives the superconducting gap through the Kramers-Kronig relation, is responsible for the $s$-wave superconductivity. These features are characteristic of the strong coupling BCS (phonon-mediated) superconductors and the anomaly of $N_{\rm b}(\omega)$ at $-(\Delta_0+\Omega)$ shown in the inset of (a) is regarded as the evidence of the phonon mechanism.
\tgtw{Although the slight deviation between ${\rm Im}\Sigma^{\rm nor}_{\rm b}$ and
the inferred self-energy ${\rm Im}\Sigma^{\rm nor}$
for $\omega \lesssim -0.3$eV
\tind{due} to
the finite high-energy cutoff in ${\rm Im}\Sigma^{\rm nor}$,
in contrast to ${\rm Im}\Sigma^{\rm nor}_{\rm b}$ that stays constant even for $\omega \ll -0.3$eV,}
the machine learning well reproducing the exact results of the dip in ${\rm Im}\Sigma^{\rm ano}$ and the sharp drop in ${\rm Im}\Sigma^{\rm nor}$ indicates the reliability of the present method.

\section{Results: Cuprate Superconductors}
\label{results}
In this section, we will show the results of the regression for ARPES data of Bi2212 and Bi2201.
The prominent peak structures are found
both
\tgtw{in}
the normal and anomalous \tgtw{self-energies}, which cancel each other.

\begin{figure}[htb]
\includegraphics[width=0.475\textwidth]
{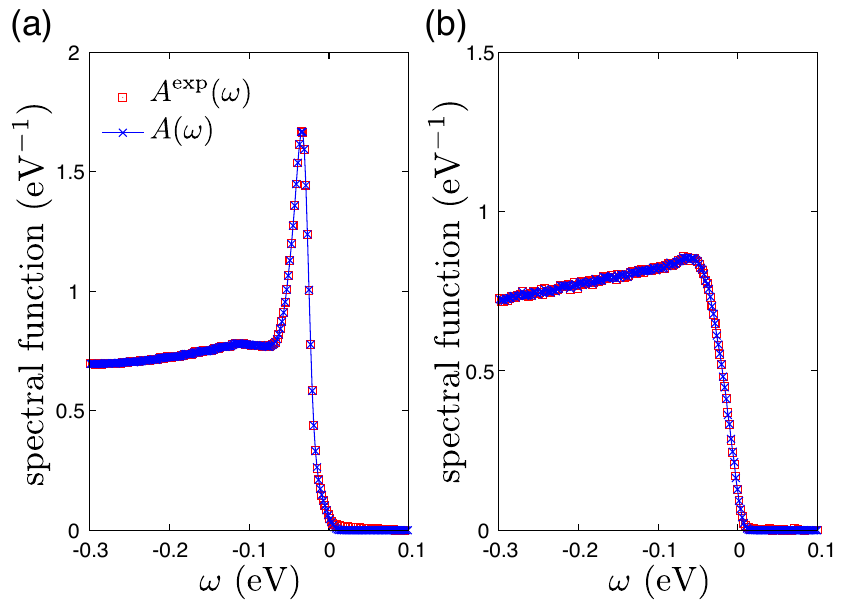}
\caption{
Comparison of experimental $A(k_{\rm AN},\omega)$ and Boltzmann-machine fitting.
Spectral function (EDC curve)  $A(k_{\rm AN},\omega)$ of (a) optimally hole-doped Bi$_2$Sr$_2$CaCu$_2$O$_{8+\delta}$ (Bi2212) at 12 K (left) (Ref.~\onlinecite{kondo2011disentangling}) and (b) underdoped Bi$_2$Sr$_2$CuO$_{6+\delta}$ (Bi2201) at 11 K (right) (Ref.~\onlinecite{kondo2009competition}) at the {antinodal} point $k=k_{\rm AF}$ (more precisely the closest point to the antinodal point, at which the momentum distribution of the quasiparticle dispersion curve is peaked). Red squares are experimental data. Blue crosses are reconstructed from the self-energies, {which are deduced from} Boltzmann machines.}
\label{FigAkw}
\end{figure}
\subsection{\textcolor{black}{Experimental data}}

We utilize
high resolution ARPES data taken for {two cuprate compounds,
\tgtw{Bi2212}
for 
\tgtw{\textcolor{black}{optimally} doped sample}
with critical temperature $T_{\rm c}\sim 90$K~\cite{kondo2011disentangling} and
\tgtw{Bi2201}
\tgtw{underdoped sample with $T_{\rm c}\sim 29$K}~\cite{kondo2009competition}}.
We analyze Bi2212 data at temperature $T=12$ K and Bi2201 at $T=11$K, which are  both well below $T_{\rm c}$.
The machine learning enables us to obtain $\Sigma^{\rm nor}$ and $\Sigma^{\rm ano}$ separately, and reveals prominent peak structures in both of them, which are
apparently hidden in the original ARPES data, because of the cancellation of these two contributions.
We elucidate 
its profound consequences for the superconducting mechanism. 



Although tremendous efforts have been devoted since the discovery of the cuprate superconductors with many fruitful clarifications,
various puzzling 
issues 
remain 
open. 
The normal-state $A(k,\omega)$ is
\tgtw{highly}
unusual including the pseudogap.
Nevertheless, the superconducting phase does not look \tgtw{so}
unusual except for the $d$-wave-type nodal gap itself and somewhat inconspicuous
``peak-dip-hump" structure (see red square symbols in Fig.~\ref{FigAkw}(a)\cite{kondo2011disentangling,kondo2009competition}):
%
%
Outside the sharp quasiparticle peak  (at $\sim -40$ meV in Fig.~\ref{FigAkw}(a)) expected at the superconducting gap edge,
$A(k,\omega)$ (energy distribution curve (EDC)) particularly at the antinodal point $k=k_{\rm AN}$ is characterized by a deeper-energy weak dip followed by a broad hump~\cite{Shen,norman1999extraction}.
{In contrast, the underdoped sample does not show
the gap-edge peak (Fig.~\ref{FigAkw}(b)),
\tgtw{although} comparable gaps 
$\sim 30$ meV open as a first look.
They are in contrast with the strong-coupling BCS superconductors, \mimt{where the solution of the Eliashberg equation using the phonon density of states predicts
prominent peaks (or saw-tooth-like) structures outside the gap in $A(k,\omega)$ \tb{(or density of states after angle integration)}, which has finger-print correspondence to the actual peak measured by the tunneling spectra,
while the peak is shown to be
\tgtw{crucial in the emergence of}
the superconductivity,
thus constitutes} the decisive testimony of the electron-phonon mechanism~\cite{PhysRevLett.14.108,PhysRevB.3.4065,Scalapino}.
}

\tgtw{There are limitation on \textcolor{black}{information} available from the ARPES measurements.
The unoccupied states invisible in the ARPES spectra may impose such \tind{a} limitation.
The bilayer nature of Bi2212 may also affect our regression scheme.
There are also issues such as photon energy dependence and
effects of matrix elements.
The influence of these
\tind{uncertainties} and issues do not change our results in the following sections,
\tind{as} examined in \tind{Appendices \ref{S2.1} and \ref{limitation_issues}}.}

\subsection{\textcolor{black}{Prominent peaks in self-energies revealed by Boltzmann machine}}
\begin{figure}[h!]
\begin{center}
\includegraphics[width=0.5\textwidth]
{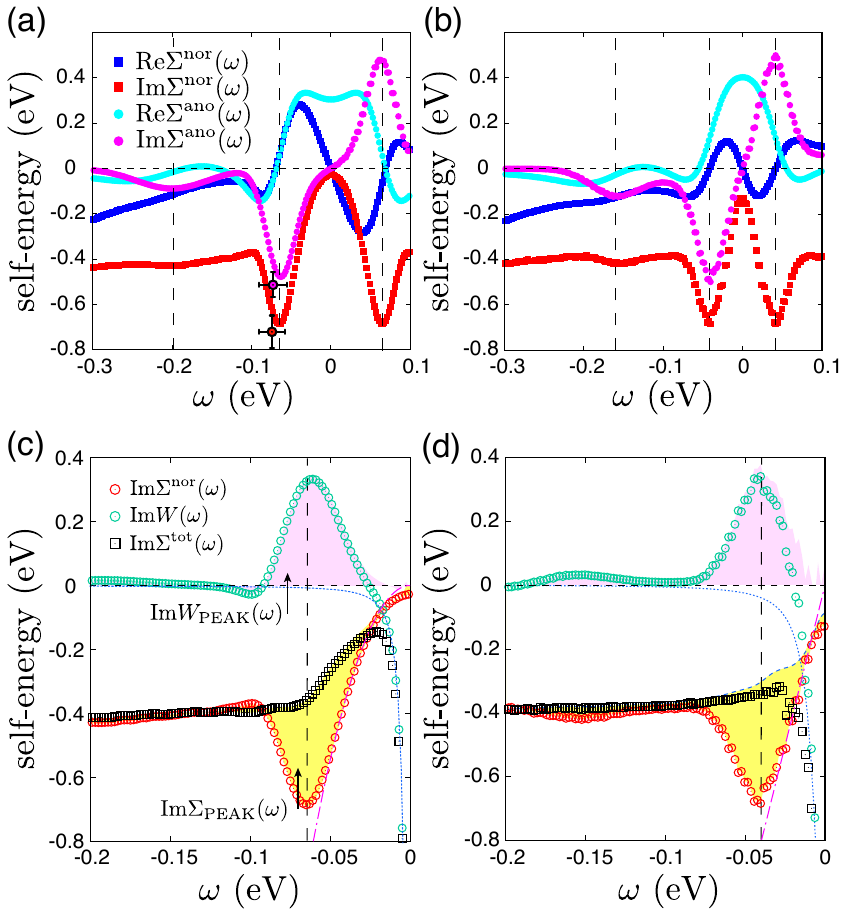}
\end{center}
\caption{
Normal and anomalous self-energies derived from machine learning and their decomposition.
Normal self-energy $\Sigma^{\rm nor}(k_{\rm AN},\omega)$ 
and anomalous self-energy $\Sigma^{\rm ano}(k_{\rm AN},\omega)$  
deduced from 
$A(k_{\rm AN},\omega)$ in Fig.~\ref{FigAkw} by the machine learning for  Bi2212 
(a) and  
Bi2201 
(b). The vertical dashed lines indicate the peak positions \tgx{$\omega_{\rm PEAK}$} in the imaginary part. \mi{The error bars for the peak energy and height are shown as horizontal and vertical error bars.}
${\rm Im} \Sigma^{\rm nor}(k_{\rm AN},\omega)$, ${\rm Im} W(k_{\rm AN},\omega)$ and ${\rm Im} \Sigma^{\rm tot}(k_{\rm AN},\omega)$ are plotted for Bi2212 (c) and Bi2201 (d). The peaks of ${\rm Im} \Sigma^{\rm nor}(k_{\rm AN},\omega)$ and  ${\rm Im} W(k_{\rm AN},\omega)$ are completely canceled in their sum ${\rm Im} \Sigma^{\rm tot}(k_{\rm AN},\omega)$. ${\rm Im} \Sigma^{\rm tot}$ is decomposed into simple BCS-type superconducting contribution $L_{\rm BCS}(k_{\rm AN},\omega)$ (a Lorentzian around $\omega=0$ ( dotted blue curve))
and the rest ${{\rm Im}\Sigma_{\rm N}(k_{\rm AN},\omega)}\equiv {\rm Im} \Sigma^{\rm tot}-L_{\rm BCS}(k_{\rm AN},\omega)$
(see Appendix~\ref{S1.5}).
The latter is fitted by a {superposition of many} Gaussian distributions (blue dashed curve).
Then the unusual structures
are identified as ${\rm Im} \Sigma_{\rm PEAK}(k_{\rm AN},\omega)\equiv {\rm Im}\Sigma^{\rm nor}(k_{\rm AN},\omega)-{\rm Im}\Sigma_{\rm N}(k_{\rm AN},\omega)$ (yellow shaded area)
and ${\rm Im} W_{\rm PEAK}(k_{\rm AN},\omega)\equiv {\rm Im} W(k_{\rm AN},\omega)-L_{\rm BCS}(k_{\rm AN},\omega)$ (pink shaded area). 
The yellow and pink areas cancel in their sum both in (c) and (d). 
The magenta dash-dotted curves show a quadratic (linear) fitting of ${\rm Im} \Sigma^{\rm nor}(k_{\rm AN},\omega)$ of Bi2212 (Bi2201) for $|\omega| < 35$ meV. \hfill
}
\label{FigSigma}
\end{figure}
By using the Boltzmann-machine learning, a dramatic consequence is revealed
for $\Sigma^{\rm nor}$ and  $\Sigma^{\rm ano}$ by reconstructing them from the mild structure of $A(k,\omega)$ given by ARPES.
{The present reconstruction is a non-linear underdetermined problem as in many of machine learning problems.
To obtain a reliable solution, we utilize physically sound constraints such as the rigorous causality encoded as the Kramers-Kronig relation.
Sparse and localized nature of ${\rm Im} \Sigma^{\rm ano}$
\tgtw{has} resulted {\it a posteriori} as the optimized solution under physical constraint as detailed in \tgtw{Sec.~\ref{prior_knowledge}}.
To represent the self-energies and incorporate the physically sound constraints,
the Boltzmann machine as universal function approximators developed in machine learning is employed.
See Sec.~\ref{detailed_method} and also 
{Figure \ref{Fig_Flow_Chart_Illustration}}
for the flow chart.}

%
%
The obtained $A(k_{\rm AN},\omega)$ (cross points in Fig.~\ref{FigAkw}) perfectly reproduces the distinct behaviors of both
the optimally doped and underdoped samples.
These EDC curves 
are constructed from  ${\rm Im} \Sigma^{\rm nor}(k_{\rm AN},\omega)$ and   ${\rm Im} \Sigma^{\rm ano}(k_{\rm AN},\omega)$ in Figs.~\ref{FigSigma}(a) and (b).  
Remarkably, prominent peaks are found in ${\rm Im}\Sigma^{\rm ano}(k_{\rm AN},\omega)$ at \mimt{{$\omega =\omega^{\rm OP}_{{\rm PEAK}} \sim \pm 70$}} meV for Bi2212 and at $\omega =\omega^{\rm UD}_{{\rm PEAK}} \sim \pm {45}$ meV for Bi2201, accompanied by weaker peaks at $\pm$ 180 meV and $\pm$ 160 meV, respectively.
We will show
\tgtw{below} that the discovered peaks are the main source of superconductivity.
{Although the peak of ${\rm Im}\Sigma^{\rm ano}(k_{\rm AN},\omega)$ had been searched for long time in analogy to the strong coupling BCS superconductors~\cite{norman1999extraction,bok2016quantitative}, its clear signature was missing in the cuprates. The machine learning now has succeeded in its identification.} 
Surprisingly, ${\rm Im}\Sigma^{\rm nor}(\omega)$ also has  distinct (positive or negative) peaks at the same energy as ${\rm Im}\Sigma^{\rm ano}(\omega)$
and\tgtw{, as we clarify below,} their contributions to the spectral function cancel \tgtw{out each other}.

The robustness of our finding 
against noise, \mi{uncertainty in experimental data} and the experimentally uncertain high-energy part is
\tgtw{demonstrated}
in {Appendix~\ref{S2.1} and \ref{S2.5}}.
In Fig.~\ref{FigSigma}(a), we also plot the error bars of the peak position and height inferred from the experimental uncertainty and the machine learning error \migr{(for detailed procedure of the error-bar estimate,
see {\textcolor{black}{Appendix~\ref{stability_against_noise}, Figure~\ref{Fignoise}, and Sec.~\ref{detailed_method}}
(``Minimization of training error" and ``Minimization of test error" as well as Appendix~\ref{S1.2}
})}.  The small error bars indicate that the existence of prominent peak is reliable.
For this analysis, we have utilized the inferred experimental noise estimated from the interpolated smooth $A(k,\omega)$ (namely, the inferred true smooth $A(k,\omega)$) obtained in the standard linear regression analysis
}.
Then the error bars of our peak estimate here is given from the optimization to hypothetical experimental data points generated with the same level of noise to the inferred true $A(k,\omega)$. See 
\textcolor{black}{Sec.~\ref{test_error}} for details.
\tm{We note that in the case of underdoped Bi2201 sample below $T_{\rm c}$, there exists subtlety in the machine learning solution. Although the superconducting solution presented here is the optimized solution that gives the smallest \tc{mean-square error} in the fitting of $A(k,\omega)$, an insulating solution is also found with larger error. This may be related to  severe competition of insulating and superconducting behaviors in the real sample.
We show the properties of \tgtw{the} superconducting solution because it gives the best optimized solution and the sample is indeed superconducting.}

Despite the peaks in ${\rm Im}\Sigma^{\rm nor}$, and ${\rm Im}\Sigma^{\rm ano}$, prominent peaks are missing in  ${\rm Im} \Sigma^{\rm tot}(\omega)$ as shown in Figs.~\ref{FigSigma}(c) and (d) (black symbols).
{We discuss below why the peaks in ${\rm Im}\Sigma^{\rm ano}$ at $\omega\ne 0$ necessarily show up and their contribution cancels with ${\rm Im} \Sigma^{\rm nor}$, when we impose physically justifiable constraints such as the Kramers-Kronig transformation.}
{
Instead of the peak in ${\rm Im}\Sigma^{\rm nor}(\omega)$ and ${\rm Im}\Sigma^{\rm ano}(\omega)$, 
a negative prominent peak~\cite{PhysRevB.57.R11093} generating the superconducting gap is found centered at $\omega\sim 0$ in ${\rm Im} \Sigma^{\rm tot}$,
which arises from the zero of the denominator in {Eq.~(\ref{eq:W})}, 
commonly to the conventional BCS superconductors.
}



\subsection{\textcolor{black}{Role of self-energy peaks to superconductivity}}
\begin{figure}[h!]
\begin{center}
\includegraphics[width=0.5\textwidth]
{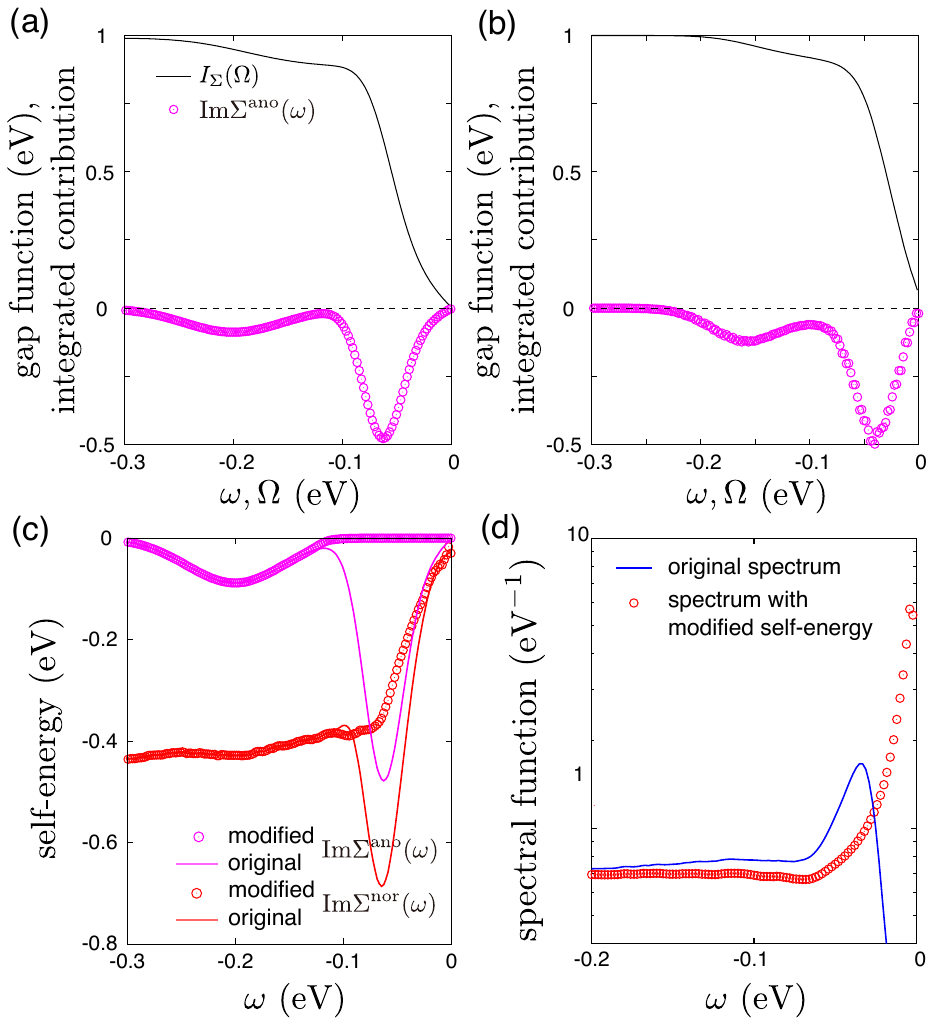}
\end{center}
\caption{
Contribution of peak of ${\rm Im} \Sigma^{\rm ano}(k_{\rm AN},\omega)$ showing dominance to superconductivity.
 $I_{\Sigma} (\Omega)$
calculated from 
${\rm Im} \Sigma^{\rm ano}(k_{\rm AN},\omega)$ 
is shown for the 
Bi2212 (a)
and for the 
Bi2201 (b)
The right negative peaks of ${\rm Im} \Sigma^{\rm ano}(k_{\rm AN},\omega)$ both contribute to more than 90\% of ${\rm Re} \Sigma^{\rm ano}(\omega=0)$.
{{(c)
{${\rm Im} \Sigma^{\rm nor}(k_{\rm AN},\omega)$ and ${\rm Im} \Sigma^{\rm ano}(k_{\rm AN},\omega)$ modified from
the originals in
Figs.~\ref{FigSigma}(a) and \ref{FigSigma^ano_I}(a) by eliminating the low-energy peaks around \mimt{$\omega = -70$} meV for Bi2212.
The peak component of ${\rm Im} \Sigma^{\rm nor}(k_{\rm AN},\omega)$ to be subtracted is ${\rm Im}\Sigma_{\rm PEAK}(k_{\rm AN},\omega)$ in Fig.~\ref{FigSigma}(c) and the subtracted ${\rm Im} \Sigma^{\rm nor}(k_{\rm AN},\omega)$ is nothing but ${\rm Im} \Sigma_{\rm N}(k_{\rm AN},\omega)$.
On the other hand, ${\rm Im} \Sigma^{\rm ano}(k_{\rm AN},\omega)$ consists of only two peaks and the right peak around \mimt{$\omega = -70$} meV can be easily subtracted
by using the sigmoid function.
Peak-subtracted ${\rm Im} \Sigma^{\rm nor}(k_{\rm AN},\omega)$ and ${\rm Im} \Sigma^{\rm ano}(k_{\rm AN},\omega)$
are 
represented by red and purple circles, respectively. (See 
Appendix~\ref{S1.5} for the prescription to decompose the self-energy and see 
Appendix~\ref{S1.6} that the peak position of the gap function $\Delta(\omega)$ is slightly shifted from $\Sigma^{\rm ano}(\omega)$.) 
}
(d) In comparison to the experimental $A(k,\omega)$ (blue thin curve), spectral function obtained from the peak-subtracted ${\rm Im} \Sigma^{\rm nor}(\omega)$ and ${\rm Im} \Sigma^{\rm ano}(\omega)$ 
is shown by red circles, where superconductivity disappears resulting in a good normal metal with a quasiparticle width comparable to the experimental resolution ($\sim 10$ meV). }
}}
\label{FigSigma^ano_I}
\end{figure}
To understand the significance of the peaks \tgx{at $\omega_{\rm PEAK}$} in 
${\rm Im} \Sigma^{\rm ano}$, we show the contribution of the peaks to {{${\rm Re} \Sigma^{\rm ano}(k,\omega=0)$}}
estimated from the normalized partial Kramers-Kronig relation~\cite{PhysRevLett.100.237001} (Cauchy relation) (see also {
Eqs.~(\ref{KK-Sigmanor}}) and (\ref{KK-Sigmaano})) \tgtw{defined by} 
\begin{equation}
{
{I_{\Sigma}(\Omega)=\frac{2\int^0_{\Omega} d\omega { {\rm Im}\Sigma^{\rm ano}(k_{\rm AN},\omega){/\omega}}}{\pi{\rm Re} \Sigma^{\rm ano}(k_{\rm AN},\omega=0)}}.
\label{KKSano}}
\end{equation}
${\rm Re} \Sigma^{\rm ano}(k,\omega=0)$ 
is a measure of the {superconducting amplitude}, because the 
gap $\Delta(k,\omega=0)$ 
is proportional to $\Sigma^{\rm ano}(k,\omega=0)$ (Eq.~(\ref{eq:Delta})).
Since $I_{\Sigma}(\Omega=-\infty)=1$, the contribution of the peak in  ${\rm Im} \Sigma^{\rm ano}$ to the superconductivity can be estimated from the increment in  $I_{\Sigma}(\Omega)$. 
{Figure~\ref{FigSigma^ano_I}(a) shows that the inner energy peak 
at 
{\mimt{$\omega_{\rm PEAK}\sim-70$} meV ($-45$ meV)} for Bi2212 (Bi2201) both contribute to more than 90\% of ${\rm Re} \Sigma^{\rm ano}(k_{\rm AN},\omega=0)$ (note that ${\rm Im} \Sigma^{\rm ano}$ is an odd function of $\omega$).
Namely, these peaks are
\tgtw{indispensable in the emergence of}
the superconductivity.  
}

To further
\tgtw{demonstrate}
the crucial role of the inner peak in ${\rm Im} \Sigma^{\rm ano}(k_{\rm AN},\omega)$,
we have  
{hypothetically} eliminated the peaks in ${\rm Im}\Sigma^{\rm nor}$ and ${\rm Im}\Sigma^{\rm ano}$ 
as shown in Fig.~\ref{FigSigma^ano_I}(c) for  Bi2212.
The resultant $A(k,\omega)$ in Fig.~\ref{FigSigma^ano_I}(d) shows the gap disappearance and switching to a normal metal with a sharp quasiparticle peak, {confirming the crucial role of the peaks to superconductivity.} 
   
%
%

%
%

Through our Boltzmann-machine analyses, $\Sigma^{\rm nor}$, $\Sigma^{\rm ano}$ and $W$ are revealed to have prominent (positive or negative) peaks, while they cancel \tgtw{each other} in the sum $\Sigma^{\rm tot}$.
It is important that this conclusion is obtained directly from experimental data without assuming any theoretical models aside from mathematical \tgtw{(causal)} requirement for the Green function.
\mir{A recent self-energy analyses of ARPES data \cite{bok2016quantitative,LiDessau} did not identify the present prominent structure. 
However, our analyses on the momentum dependence suggest that the results are not necessarily inconsistent each other for the case of Ref.~\onlinecite{bok2016quantitative}, because the momentum range in Ref.~\onlinecite{bok2016quantitative} is limited in the nodal region far from the antinodal point, where we also see that the prominent peak is missing because of the $d$-wave symmetry of the superconducting gap. It is crucially important to see the antinodal region to see the prominent peak as we discuss in 
Appendix~\ref{S2.2} and ~\ref{S2.4}.
Further, we show in
Appendix~\ref{S2.2}
that  physically inappropriate assumptions for strongly correlated electron systems posed in Refs.~\onlinecite{bok2016quantitative} and \onlinecite{LiDessau} lead to failures of identifying the existence of the peaks.}

{If a large superconducting gap is open around $\omega=0$ as in the experimental $A({k,\omega})$, it requires
\tgtw{the} corresponding famous gap structure in ${\rm Re} \Sigma^{\rm ano}$ around  $\omega=0$, where inside the two peaks at the gap energy $\omega=\pm \Delta$, ${\rm Re} \Sigma^{\rm ano}$  shows plateau
\tgtw{as shown in \tind{Fig.~\ref{FigSigma}(a)}}.}
Then consistency requires a prominent peak structure \tgx{around $\omega_{\rm PEAK}$} in ${\rm Im} \Sigma^{\rm ano}$ through the Kramers-Kronig relation. 
{The peak of ${\rm Im} \Sigma^{\rm ano}$ in turn naively anticipates prominent structures outside the gap in $A(k,\omega)$ through Eq.~(\ref{eq:Gnor}).}
{However, such structures are missing. 
This is possible only when $\Sigma^{\rm nor}$ plays a role to cancel the prominent structure in ${\rm Im} W$. {This is corroborated by the vanishing superconductivity in Fig.~\ref{FigSigma^ano_I}(d).}
{Furthermore, 
{the superconducting order accompanied by coherent quasiparticle excitations observed in experiments}
can be generated from ${\rm Im} \Sigma^{\rm ano}(\omega)$ only when electrons at $\omega$ become coherent, signaled by the reduction of 
$|{\rm Im} \Sigma^{\rm tot}|$ (or more precisely $|{\rm Im} \Sigma_{\rm N}|$) (see Figs.~\ref{FigSigma}(c) and (d), their captions, {Sec.~\ref{detailed_method}, and
Appendix~\ref{S1.5}
for the definition of $|{\rm Im} \Sigma_{\rm N}|$}) seen in the region $|\omega|<\omega^{\ast}\sim 0.07$ eV. This restricts 
\tgx{$\omega_{\rm PEAK}$} to this range. }
The present machine learning indeed reproduced this natural expectation in a physically \tgtw{transparent and} reasonable way.
} 

{We show some
analyses on
\tgtw{the}
temperature dependence
\tgtw{of the self-energies} including a case above $T_{\rm c}$ and
\tgtw{the} momentum dependence away from the antinodal point
in {Appendix~\ref{S2.3} and \ref{S2.4}},
respectively,
as supporting data of \tc{the} present analyses.}
The results confirm the validity of the present conclusion.

Although we found that the prominent peak in \tb{${\rm Im}\Sigma^{\rm ano}$} is
\tgtw{crucial for}
the high critical temperature of the curates, full understanding and mechanism of {prominent peaks} in $\Sigma^{\rm nor}(\omega)$ and $\Sigma^{\rm ano}(\omega)$, which are absent in $\Sigma^{\rm tot}(\omega)$ are \tgx{open to further analyses}. \mir{Our result is significant because the cancellation poses a severe constraint on theories of curate superconductors and calls for further consistent studies based on this finding. In Discussion, we refer to one possible explanation.} 

\section{Discussion}
\label{discussion}
\subsection{Insights into intrinsic (Planckian) dissipation } 
\mimt{In this and next subsections, we discuss further possible connection gained from the main findings of the peaks and their cancellation
to other more open issues of the cuprate superconductors in regard to the damping (incoherence) of quasiparticle and the factors that determines $T_{\rm c}$ to emphasize the significance of the findings with outlook.  }
\tr{The present machine learning is also useful to separately extract other theoretically fundamental quantities such as the \tm{momentum resolved} superconducting order parameter (the density of Cooper \tgtw{pairs} or the superfluid density) $F(k)$,
{mass} renormalization factor $z_{\rm qp}(k)$ and the 
{single-particle} relaxation time $\tau$,
which had been inferred only indirectly or only in combinations of more than one quantity in experiments in the literature,
\tb{\tc{\tbl{although} these quantities play} crucial roles independently \tgtw{of} each other below in understanding physics.}
(see \textcolor{black}{Sec.~\ref{method_Green_function}} for precise definition of the above quantities).}

\tc{How frequently the single-particle excitations are scattered is encoded in the imaginary part of the normal self-energy ${\rm Im}\Sigma^{\rm nor}$.}
\tbl{Landau's Fermi-liquid-like behavior characterized by ${\rm Im}\Sigma^{\rm nor}(\omega)\propto \omega^2$,
is satisfied} only in a  small region ($|\omega|<0.03$ eV) for Bi2212, and looks even linear ($\propto |\omega|$) in the same region for Bi2201, implying non-Fermi liquid (marginal Fermi liquid) behavior~\cite{Varma} (see Figs.~\ref{FigSigma}(c) and (d)), 
\tb{which can be fit by ${\rm Im}\Sigma^{\rm nor}(k,\omega) \sim c_0(k)+{\rm sign}(\omega)c_1(k)\hbar\omega$ \tgx{in the range 15 meV $<\omega <40$ meV}
with a \tbl{dimensionless} marginal-Fermi-liquid coefficient $c_1(k)$.}
%
\tc{
The $\omega$-linear component $c_1(k)\omega$
is disruptive to the quasiparticle picture, and manifests emergent inelastic dissipation absent in Landau's Fermi liquids.}
\tr{\tb{As \tgx{supporting} information,} tiny \tbl{quasi-particle} renormalization factor $z_{\rm qp}$ 
corroborating the non-Fermi liquid \tgx{together with its effects on pair breaking} is also shown in {
Appendix~\ref{S2.9}}
(see definition of $z_{\rm qp}$ in {Eq.~(\ref{eq:zqp})
}). 
}

\tg{
The single particle relaxation time \tb{$\tau$} \tc{is} defined by \tb{$\tau(k,\omega)^{-1}=z_{\rm qp}(k){\rm Im}\Sigma^{\rm nor}(k,\omega)/\hbar
$}.
\tc{When the carrier relaxation time is estimated from $\tau$,}
the $\omega$-linear term\tc{, $z_{\rm qp}(k) c_1(k)\omega$,} is associated with the universally observed $T$-linear resistivity in the cuprates \cite{Martin90,Takagi92} through the $\omega$-$T$ correspondence \tb{$\tau(\hbar\omega) \leftrightarrow \tau(k_{\rm B}T)$} {transformed to the self-energy of two-particle Green function for the conductivity}.
{The temperature-insensitive $z_{\rm qp}(k)c_1(k)$ shown in {Appendix~\ref{S2.9.1}}
also supports the correspondence.}
\tbl{(See Fig.~\ref{FigF} for each $z_{\rm qp}(k)$ and $c_1(k)$.)}
}

\tc{\tgx{A remarkable property of the inelastic relaxation rate $\Gamma (k)= z_{\rm qp}(k) c_1(k)$ is its high value ($\sim 1$) with only weak dependence on the doping, momentum (see Fig.~\ref{Homes_Uemura}(a)) and temperature. }
This universal behavior of $\Gamma\sim 1$-$1.5$ seems to support a local and universal mechanism of the relaxation, for instance, the Planckian dissipation mechanism of the hydrodynamic state, which claims $\tau^{-1}(T)=\Gamma k_{\rm B}T/\hbar$ or $\tau^{-1}(\omega)=\Gamma\omega$ with a universal constant $\Gamma$ of the order unity~\cite{Zaanen15,Zaanen}.
}

\tb{Although simple version of Planckian mechanism expects only an extended broad self-energy structure due to \tc{``unparticle physics,"} 
the self-energy has a broad but prominent peak structure around $\omega=\omega_{\rm PEAK}$ which is responsible for the superconductivity through the Kramers-Kronig transformation as we discussed. At the same time, 
\tgx{the actual line shape is rather broad with the width around 0.05 eV (see Figs.~\ref{FigSigma}(c) and (d)), which is comparable to $\omega_{\rm PEAK}$ itself.
\tbl{More importantly, the peak is} smoothly connected in the tail with the $\omega$-linear behavior near the zero energy, implying that the ``Planckian dissipation and hydrodynamic behavior" associated with the strange metal~\cite{Zaanen15} is
caused by the \tgtw{source} of the superconductivity.} 
The broad prominent peaks could be due to the damped pole but it could also be ascribed to \tgx{``unparticle object" generated by entangled bare electron and dark object}. 
}

\mig{The marginal fermi liquid behavior (${\rm Im}\Sigma^{\rm nor}(\omega)\propto \omega$) needs to be understood with care.  Since the present photoemission data could include extrinsic background effect, our analysis may not clarify the high-energy part of intrinsic $\omega$-linear behavior. In fact, in relation to the $T$-linear resistivity, the related $\omega$-linear behavior should show up around the gapless nodal region, while the peak of the normal self-energy vanishes at the nodal point
(see Fig.~\ref{Fig_angle_dependence}).
Therefore, the $\omega$-linear coefficient observed as the tail of the peak is not necessarily the same as the $T$-linear coefficient in the resistivity. In fact the high-energy $\omega$-linear component in Fig.~\ref{FigSigma} {(c) and (d)} (black squares) has substantially smaller slope than the present $\omega$-linear component directly associated with the prominent peak. This smaller slope at the high-energy region ($\omega<-0.1$ eV region) is consistent with the high-temperature $T$-linear resistivity~\cite{Takagi92} through the correspondence relation $\hbar \omega\leftrightarrow \pi k_{\rm B}T$ and the $\omega$-linear self-energy in the high-energy part identified in an earlier study~\cite{bok2016quantitative}.   We need further studies on the relation between these two $\omega$-linear components.}

\textcolor{black}{The $T$-linear scaling may remind the readers of quantum critical behaviors.
In the present study, we \textcolor{black}{examine} only two sets of data for different compounds with different doping concentrations.
Therefore, we could not exclude the possibility that these samples are by chance both close to quantum critical points.
In fact, the underdoped sample shows more linear behavior than the optimum doped sample,
which might imply that the underdoped sample is closer to the quantum critical point.
Alternatively, the $T$-linear scaling behavior could emerge in a distinct phase covering
a finite range of doping concentrations as discussed above as the Planckian fluid.
However, it is impossible to draw a conclusion from these two samples
only and it is left for future studies.}

\subsection{Factors that determine the superconducting critical temperature} 
\begin{figure*}[htb]
\begin{center}
\includegraphics[width=0.85\textwidth]{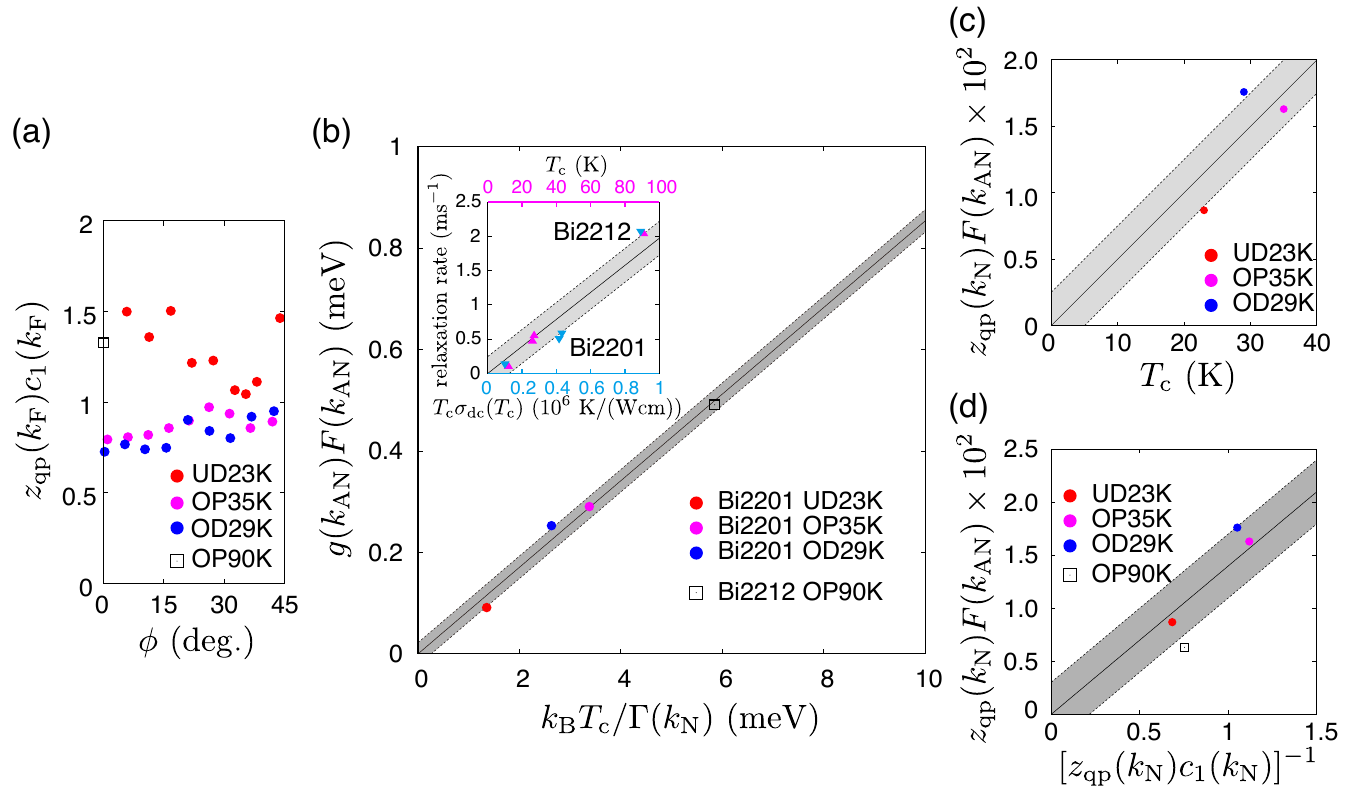}
\end{center}
\caption{
\tr{
Relation between superfluid density $F$, $T_{\rm c}$, carrier relaxation time \tg{and self-energy peak} derived from machine learning.
(a) Angle and doping dependences of $\Gamma=z_{\rm qp}c_1$ for Bi2201 and Bi2212.
(b) \tb{Proposed scaling} between  \tc{$F$}, \tgx{$T_{\rm c}$, $g$}, and \tc{$\Gamma$}.
(c) Possible \tb{scaling} between \tc{$F$} and $T_{\rm c}$ for Bi2201, which mimics the Uemura plot (see text).
(d) Possible \tb{scaling} between \tc{$F$} and $\Gamma^{-1}$ for Bi2201 and Bi2212, which mimics the Homes plot (see text).
The scaling \tc{$gF\propto T_{\rm c}/\Gamma$} in (b) gives the best 
fitting:
\tc{The standard deviation is 0.03 meV for (b),
while 0.25 for (c) (excluding the 2212 data because it is far away from the linear fitting) and 0.29 for (d).}
\tc{Even for the standard deviation of
normalized values,  
$(a_i-a_i^{\rm fit})/a_i^{\rm fit}$ with $a_i$ \tgx{($a_i^{\rm fit}$) being
the $i$th data of \tc{$\bar{Q}F$ or} $z_{\rm qp}F$ (the fitting line value)}, instead of $a_i-a_i^{\rm fit}$ employed above,
we obtain 0.023 (b), 0.19 (c) (excluding 2212), and 0.26 (d).} 
Inset of (b): Experimental plots of the muon-spin relaxation rate $R$ 
~\cite{Uemura07} vs. $T_{\rm c}$ (Uemura plot) or $T_{\rm c}\sigma_{\rm dc}(T_{\rm c})$ (Homes plot)~\cite{Ando04,Homes}
\tc{for Bi2201 and Bi2212}.
Here, the standard deviation is 
\tc{0.07 $\mu$s$^{-1}$ ( 0.25 $\mu$s$^{-1}$)}, and
\tc{the standard deviation of the normalized values is 0.22 (0.26)}
for the Uemura plot (Homes plot).}
}
\label{Homes_Uemura}
\end{figure*}
\tr{
{Fundamental quantities revealed by the machine learning provide further insight into the superconductivity through 
the scaling among experimental observables:}
The linear relation $F \propto T_{\rm c}$ between $T_{\rm c}$  and the superfluid density $F$
measured from the muon-spin relaxation rate $R$
} 
(theoretically proportional to  $F(k)z_{\rm qp}(k)$ averaged over Fermi surface momentum)
{has been examined through the Uemura plot~\cite{Uemura}}
in high-$T_{\rm c}$ superconductors
as in an example of the purple triangles in the inset of Fig.~\ref{Homes_Uemura}(b) for Bi2201~\cite{Uemura07}. 
The linearity should be satisfied for \tg{attractive interaction stronger than the effective Fermi energy scale $E_{\rm F}$, which is proportional to the carrier density in two spatial dimensions. \tbl{Here,  $E_{\rm F}$ is roughly the effective bandwidth of the dispersion $z_{\rm qp}\epsilon_k$}.
\textcolor{black}{This proposal interprets
the linearity as a signature of the Bose-Einstein condensation (BEC) regime.}
}
Homes {\it et al.}~\cite{Homes} proposed empirical but more universal fitting as plotted in an example of Bi2201 by blue upside-down triangles in the inset of Fig.~\ref{Homes_Uemura}(b)~\cite{Uemura07,Ando04},
where the dc conductivity $\sigma_{\rm dc}$ at $T_{\rm c}$ enters 
as $R\sim C T_{\rm c}\sigma_{\rm dc}(T_{\rm c})$ with a material \tbl{independent} constant $C$.
\tr{Since $\sigma_{\rm dc}$ is 
proportional to the 
\tm{momentum} relaxation time, the Homes relation proposes qualitatively different physics involving dissipation \tgx{and scattering} effects beyond the naive BEC regime. \tg{However, since $\sigma_{\rm dc}$ is \tbl{believed to be proportional to both the carrier density and the relaxation time}, it is not easy to single out the relaxation effect.}}
\tbl{Related scaling of the superfluid density $Fz_{\rm qp}$ proportional to the quasiparticle peak weight was also proposed~\cite{Feng,Ding}.} 

\mimt{
Here, we heuristically propose a better scaling for $T_{\rm c}$ by utilizing the present finding to show the power of the machine learning and the significance of the peak. The amplitude of the self-energy peak discovered here responsible for the superconductivity has to represent the scale of the effective attractive interaction for the Cooper pair in analogy to the Eliashberg formalism and should enter the $T_{\rm c}$ scaling.}  
\mimt{
Our proposal for $T_{\rm c}$ is given by 
\begin{equation}
k_{\rm B}T_{\rm c}=\Gamma(k_{\rm N})g(k_{\rm AN})F(k_{\rm AN}),
\label{eq:Tc}
\end{equation}
where $\Gamma$ is the damping introduced before and $F$ is the superfluid density. The factor $g$ is the scale of the effective attractive interaction as will be discussed later. 
It is reasonable that $T_{\rm c}$ is scaled by the mean field acting on the formation of the Cooper pair given by the product of the attractive interaction $g$ and the order parameter given by $F$ (superfluid density).  
}

\mimt{Let us first discuss how the characteristic effective attractive interaction $g$ is extracted from our self-energy analysis. We first introduce the bare attraction $\Omega_0 (k_{\rm AN})$,
which is represented by the ratio of the peak intensity $\overline{W_{\rm PEAK}}(k_{\rm AN})$ (peak intensity of $W$) in Fig.~\ref{FigSigma} to the absolute value of the peak energy $\omega_{\rm PEAK}(k_{\rm AN})$. Note that $\overline{W_{\rm PEAK}}(k_{\rm AN})/\omega_{\rm PEAK}(k_{\rm AN})$ is proportional to the gap through Kramers-Kronig relation. (Through Eq.~(\ref{eq:W}), the residues of the poles of $W$, $\Sigma^{\rm nor}$ and $\Sigma^{\rm ano}$ follow the same scaling at the self-energy peak). Here we assume that the antinodal value is the representative of this estimate, because the gap is the maximum. Now $\overline{W_{\rm PEAK}}(k_{\rm AN})=\int d\omega  {\rm Im}W_{\rm PEAK}(k,\omega)$ is the integrated intensity of the peak ${\rm Im} \Sigma_{\rm PEAK}=-{\rm Im} W_{\rm PEAK}$ plotted as the yellow and pink areas in Figs.~\ref{FigSigma}(c) and (d).}
\mimt{
We note that the quasiparticle weight constrains the coherent pairing so that the peak intensity $\overline{W_{\rm PEAK}}(k_{\rm AN})$ should involve the averaged quasiparticle weight at the peak energy. For the quasiparticle weight, instead of $z_{\rm qp}$ defined in the $\omega \rightarrow 0$ limit, we employ the renormalization factor $Q$ at $k_{\rm AN}$ (see Eq.~(\ref{eq:Q})) averaged in the self-energy peak region, namely, $\overline{Q}=\int d\omega {\rm Im}W_{\rm PEAK}(k,\omega) Q(k,\omega) /\overline{W_{\rm PEAK}}(k)$ with the integration over the interval $\omega<0$. Then 
\begin{equation}
g(k_{\rm AN})=\overline{Q}(k_{\rm AN})\Omega_0 (k_{\rm AN})
\label{eq:taud}
\end{equation}
gives the scale of the attractive interaction.}

\mimt{
The damping $\Gamma$ in Eq.~(\ref{eq:Tc}) looks counterintuitive because it tells that the strongly damped electrons would have higher $T_{\rm c}$. 
However, in this strong coupling superconductor, 
the strong damping is originated from the quantum entanglement as is discussed in the previous subsection, which may promote the quantum mechanical singlet pairing.}
In Fig.~\ref{Homes_Uemura}(b), 
the fit by Eq.~(\ref{eq:Tc}) is shown 
(the main panel of Fig.~\ref{Homes_Uemura}(b)). 

\tg{Although the Homes plot does not offer how $T_{\rm c}$ is determined because $\sigma (T_{\rm c})\propto 1/T_{\rm c}$ cancels in the relation to $F$, the present result indeed shows $T_{\rm c}$ linearly scaled by \tbl{$\Gamma(k_{\rm N})g(k_{\rm AN})F(k_{\rm AN})$}.  The linearity is \tm{crucially} different from the Uemura plot as well because 
of the dependence on the relaxation rate \tbl{$\Gamma$}.
Intuitively,  
$\tau=\hbar/(\Gamma k_{\rm B}T)$ or $\hbar/(\Gamma k_{\rm B}\omega)$ is related to the \tc{characteristic} length scale $\lambda$ \tb{for}
the extension of the quantum \tb{mechanically entangled area} through $\lambda \sim v_{\rm F}\tau$~\cite{Zaanen15},
where $v_{\rm F}$ is the \textcolor{black}{characteristic} electron velocity (``Fermi velocity").
\tb{The larger attraction generates stronger self-energy peak. It necessarily generates the steeper $\omega$-linear tail of ${\rm Im}\Sigma^{\rm nor}$ near zero energy, further enhancing more local and
\textcolor{black}{stronger pairing adiabatically continued to the BEC limit beyond the realistic cuprate regime,}
and raises $T_{\rm c}$ through Eq.~(\ref{eq:Tc}).
The strange (dissipative) metal and high $T_{\rm c}$ with the strong attraction represent the two sides of the same coin.}  
} \tgn{It is also interesting to note that Eq.~(\ref{eq:Tc}) looks compatible with the scaling
$E_{\rm c}\propto \gamma_{\rm c} T_{\rm c}^2$,
where $E_{\rm c}$ is the condensation energy and $\gamma_{\rm c}$ is the Sommerfeld constant of the specific heat~\cite{LeTacon,Stewart},
because $gF$ plays the role of the gap, which generates the energy gain. }
    
\tgx{
Note that $\Gamma (k)$ should be analyzed around $k_{\rm N}$, while $\Omega_0$ and $F$ \tbl{contribute to $T_{\rm c}$ at $k_{\rm AN}$} for better fitting. The in-plane transport \tm{and the quantum entanglement} are dominated by the contribution around the nodal region, while the pairing looks driven in the antinodal region, both of which contribute to raise $T_{\rm c}$. 
}

\tb{Equation (\ref{eq:Tc}) is the best scaling among various attempts we have made. To convince readers, we just show two examples of plot in Figs.~\ref{Homes_Uemura}(c) and (d). The first example is $z_{\rm qp}(k_{\rm AN})F(k_{\rm AN})$ vs. $T_{\rm c}$ plotted in Fig.~\ref{Homes_Uemura}(c), which, though not perfectly equivalent, apparently mimics the Uemura plot. 
The second is $z_{\rm qp}(k_{\rm N})F(k_{\rm AN})$ vs. \tgx{$1/z_{\rm qp}(k_{\rm N})c_1(k_{\rm N})$}
which mimics the Homes plot, because $T_{\rm c}\sigma(T_{\rm c})\propto 1/z_{\rm qp}(k_{\rm N})c_1(k_{\rm N})$ is expected (see Fig.~\ref{Homes_Uemura}(d)). The standard deviation is by far best for the present fit in Fig.~\ref{Homes_Uemura}(b) with Eq.~(\ref{eq:Tc}). }

The primary origin of the large difference of $T_{\rm c}$ between Bi2201 and Bi2212 is identified as the difference in $\overline{W_{\rm PEAK}}$ (namely the coupling strength of the electron with the dark object which makes the prominent self-energy peaks), 
supplemented by the difference in $\omega_{\rm PEAK}$. 
(see Table~\ref{table_gap}. \tc{$ \overline{W_{\rm PEAK}}(k_{\rm AN}) \sim 7.6\times 10^{-3}$ eV$^2$ and $\omega_{\rm PEAK}(k_{\rm AN})\sim 0.07$ eV} for the optimum Bi2201 and
\tc{$\overline{W_{\rm PEAK}}(k_{\rm AN}) \sim 1.4\times 10^{-2}$ eV$^2$ and $\omega_{\rm PEAK}(k_{\rm AN})\sim 0.045$ eV} for the optimal Bi2212 at $k=k_{\rm AN}$).
(More quantitative details, see also the list of {$\Omega_0 (k_{\rm AN})$} in {Table~\ref{table_gap}} and the angle and doping dependences of 
$\overline{W_{\rm PEAK}}(k)/\omega_{\rm PEAK}(k)$, 
and $\overline{Q}(k)$ in Fig.~\ref{FigF}.) 

\tgtw{A recent intensive study~\cite{bovzovic2016dependence} 
on clean thin films of
(La,Sr)$_2$CuO$_4$
has revealed that the Uemura plot shows
a linear scaling for $T_{\rm c} \gtrsim 10$ K and a quadratic scaling, where $T_{\rm c}$ is proportional to the square root of the relaxation rate, for $T_{\rm c} \lesssim 10$ K.
Consequently, the higher $T_{\rm c}$ linear scaling extrapolates $T_{\rm c}$ to a finite value at the zero relaxation rate limit.
The present scaling modifies the higher $T_{\rm c}$ linear scaling by taking into account $g\Gamma$.
If $g$ is approximately a constant \tind{for a given crystal structure} and, $\Gamma$ increases
by decreasing the doping,
as \tind{already}
reported in the overdoped region~\cite{legros2019universal},
\tind{the linear scaling in Fig.~\ref{Homes_Uemura}(b) is consistent with the sub-linear scaling of $T_{\rm c}$ to $F$ observed in Ref.~\onlinecite{bovzovic2016dependence}}.
We note that\tind{, if $\Gamma$ diverges as $T_{\rm c}$ decreases in the underdoped region,}
intrinsic doping dependence of $\Gamma$ may explain the square-root scaling at the \tind{underdoped limit qualitatively}.
However, \tind{the quantitative estimate of the power of the}
scaling is beyond our scope.
\tind{On the other hand,}
ARPES measurements
{on clean samples} at the low $T_{\rm c}$ limit \tind{in the overdoped region} are not available.}

\tr{The present result is significant because the whole analyses are obtained solely from the single ARPES line shape of $A(k,\omega)$ 
and contains much less ambiguity than before.
The present machine learning purely from experimental data sheds new light on understanding \tb{the superconducting mechanism, where the {energy} dissipation plays a role through the extension of quantum entanglement}.
For detailed doping concentration and momentum dependences of $T_{\rm c}$, $F(k)$, $c_1(k)$, $z_{\rm qp}(k)$, and the {superconducting} gap {$\Delta_{0}(k)$} for Bi2201 at 11K are found in {
Appendix~\ref{S2.10}}.}


\subsection{On cancellation of peaks}
\tr{\tb{On the cancellation of the two self-energy contributions which make the superconducting temperature high and the role of dissipation in determining $T_c$, it is desired to examine the present results in other cuprate compounds by measuring $A(k,\omega)$ at high accuracy and resolution.} }

The present finding also call for research to find the microscopic origin of the canceling self-energy peaks.
\mir{A two-component model was proposed~\cite{PhysRevLett.116.057003,ImadaSuzuki} for the cancellation of the self-energy poles, but even if it is the case, based on the present experimental evidence, further pursuit for the microscopic \tgtw{description of the} hidden-fermion excitation and its interaction is highly desirable. In the two-component fermion theory, electrons are fractionalized into the bare electrons and {dark} fermions {(hidden fermions)} consistently with the \tb{cluster dynamical mean-field theory (cDMFT)}~\cite{PhysRevLett.116.057003,ImadaSuzuki}. 
However, the prominent tail of the peak extending to $\omega\sim 0$ as $\omega$-linear feature suggests that the dark fermion must have strong interaction effect.
\tgtw{See also \tind{Ref.~\onlinecite{imada2021resonant}}
for the prediction of the fractionalization expected in other spectroscopic data such as the resonant inelastic X-ray scattering.}}

If the spontaneous symmetry breaking such as the stripe order coexists with the superconductivity, the cancellation may be  accounted for as an alternative interpretation~\cite{PhysRevLett.116.057003,ImadaSuzuki}.
\textcolor{black}{A phenomenological resonating valence bond theory~\cite{PhysRevB.73.174501}
also accounts for the cancellation because of the same fractionalized description as is discussed
in Ref.~\onlinecite{sakai2016hidden}.}
The present result poses severe constraints on possible theories.
Whether there exist other origins of the peaks and their cancellation rather than the above possibilities would be equally intriguing based on the present finding. 

\subsection{Kink}
A kink structure was observed in the
\tgtw{dispersion of} momentum distribution curve (MDC)
\tgtw{peak}
of Bi2212 and other compounds mainly near the nodal point~\cite{Bogdanov2000,Kaminski2001,Sato2003}.
In addition, a kink-like structure was identified in the study on the Hubbard model~\cite{Sakai2010}.  
It is an intriguing future issue to study whether this kink in the MDC
\tgtw{peak dispersion}
has any connection to the peak and accompanied sudden
change
\tind{of the slope and sign} 
in the \tind{real part of} normal self-energy found here in \tind{Figs.~\ref{FigSigma}(a) and (b)},
because the energy scale \tind{$\sim -0.07$} eV
\tind{of the sign change} 
for Bi2212 is similar
\tgtw{to the kink energy scale}.

\section{Summary and Outlook}
\label{summary_outlook}
\tgtw{We have formulated a reliable way of extracting the normal and anomalous self-energies separately from the ARPES data of superconductors by taking advantage of recently developed machine learning method. Careful benchmark tests including simple metals, conventional BCS superconductors and model systems that have
\tind{established} solutions indicate that the method offers an accurate and reliable regression of the self-energies only from the ARPES data even for challenging strongly correlated electron systems.} 

\tgtw{Then the method has been applied to cuprate superconductors, Bi2201 and Bi2212.
We have successfully extracted the normal and anomalous components of the self-energy from the ARPES spectra.}

\tgtw{In contrast to previous studies, the result shows that the imaginary part of the normal and anomalous self-energies have prominent peak structures as a function of the frequency.
However, the contributions of the normal and anomalous components
cancel in the spectral function, which accounts for the failure to identify the prominent structure for long time.
Nevertheless the peak in the anomalous self-energy has been shown to generate more than
90\% of the superconducting gap and thus turned out to be the primary \tgtw{source} of the superconductivity.
Therefore,
\textcolor{black}{the discovered these peak structures and their cancellation pose a severe constraint with insight on the mechanism of the high transition temperature of superconductivity.}}

\tgtw{The origin of the failure in identifying the peak structures in the previous studies is
elucidated in Appendix~\ref{S2.2} to be primarily the assumptions in the previous studies:
The previous studies assumed
linear energy dispersion,
\tind{and/or}
self-energies that are momentum independent along the direction perpendicular to the Fermi surface,
\tind{for example in Ref.~\onlinecite{bok2016quantitative}}. 
These assumptions are not justified in the strongly correlated electron systems.} 

\tgtw{Thus \tind{newly} obtained quantities hidden in the direct experimental measurements in the past allow us to show that the superconducting transition temperatures are well scaled by the product of the superfluid density $F$, the effective attractive interaction $g$ and the Planckian dissipation $\Gamma$.}

Present successful examples of insight obtained purely from the machine learning analysis of experimental data \tg{indicates an opening of a promising field which  allows understanding physics 
hidden in experiments, without relying on involved and specific theoretical \tgx{assumptions and
constraints that are not shown to be justifiable in strongly correlated electron systems}.
\mi{At the same time, we have shown that very accurate experimental data are required to extract hidden quantities.
\tgtw{\tind{For instance, in} Refs.~\onlinecite{he2018rapid}, \onlinecite{Chen2019}, and \onlinecite{Kondo2015},
the signal-to-noise ratio in the ARPES measurements for Bi2212 seems to be already \tind{sufficiently} small
while available momenta in the Brillouin zone are limited.
Although Ref.~\onlinecite{ai2019distinct} for Bi2212 and
Ref.~\onlinecite{he2011single} for Bi2201 reported
the ARPES spectra in a wide range of momenta, the signal-to-noise ratio does not seem to be small enough.}
It is important to improve the experimental resolution and suppress errors to increase the reliability of the machine learning inference.}}

{In addition combining with other independent measurements such as the quasiparticle interference obtained from the scanning \tgtw{tunneling} microscope in this case is important to reach better statistics.}
 By combining with other experimental data and indisputable theoretically basic constraints such as symmetry, much more powerful tool will be provided for understanding physics of complex phenomena.

{The present study will stimulates studies on the origin of the peak structures.
Indeed, there have been studies~\cite{PhysRevB.101.180510,PhysRevB.103.024525} on the origin
of the peak structures, which are \tind{inspired and motivated} by our results~\cite{yamaji2019hidden}.}

\acknowledgments
We thank Takeshi Kondo and Adam Kaminski for providing us ARPES data
published in Refs.~\onlinecite{kondo2011disentangling} and \onlinecite{kondo2009competition}.
We also thank Takeshi Kondo for discussions on the experimental results.
We are grateful to Shiro Sakai for discussions and comments on the manuscript
and Chandra Varma for clarification of the procedure of the analysis in Ref.~\onlinecite{bok2016quantitative}.
This research was supportd by MEXT as
``Priority Issue on Post-K computer" (Creation of New Functional Devices and High-Performance Materials
to Support Next-Generation Industries (CDMSI)) and ``Basic Science for Emergence and Functionality in Quantum Matter
- Innovative Strongly-Correlated Electron Science by Integration of Fugaku and Frontier Experiments -" (JPMXP1020200104) as a program for promoting researches on the supercomputer Fugaku, 
supported by RIKEN-Center for Computational Science (R-CCS) through HPCI System Research Project (Project ID: hp170263, hp180170, hp190145, hp200132 and hp210163).
Y. Y. was supported by PRESTO, JST (JPMJPR15NF). 
Y. Y. and M.I. were supported by JSPS KAKENHI
(Grant No. 16H06345). A.F. was supported by KAKENHI (Grant No. 19K03741).
{The present regression is performed by using our house code.
The numerical code
will be available upon request.}

\appendix
\section{Comparison with Previous Studies on Self-Energy}
\label{S2.2}
\begin{figure*}[htb]
\begin{center}
\includegraphics[width=0.9\textwidth]{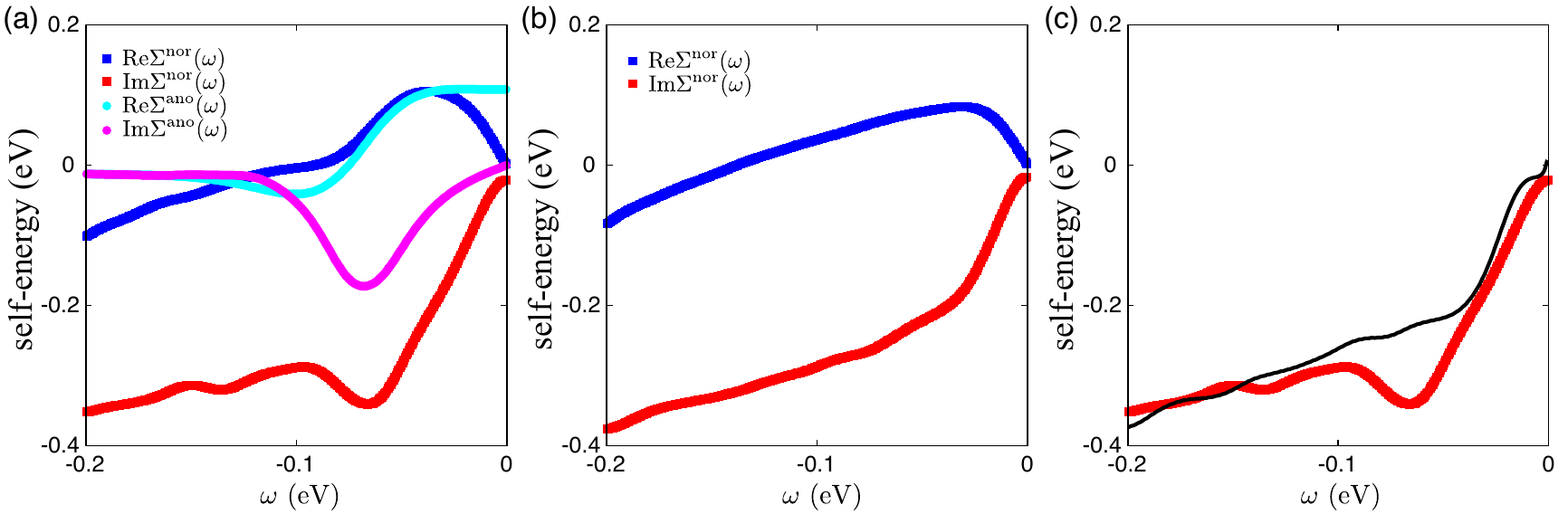}
\end{center}
\caption
{
{Imaginary part of normal self-energy estimated by the hypothetical equation $N(\omega) = \mathcal{A}_S(\omega)/\mathcal{A}_N(\omega)$ \tgtw{using the data for Bi2201 of Ref.~\onlinecite{kondo2011disentangling}}.}
{The left panel shows the self-energies in the superconducting state that are obtained around $\phi=20^{\circ}$
in Fig.~\ref{Fig_angle_dependence}(h).
The middle panel shows self-energies for the normal state, which are obtained by \tgtw{artificially} eliminating the peak structure in ${\rm Im}\Sigma^{\rm nor}$ in the left panel.
The right panels shows the estimated ${\rm Im}\Sigma^{\rm nor}$ (black solid curve) by the hypothetical relation
$N(\omega) = \mathcal{A}_S(\omega)/\mathcal{A}_N(\omega)$,
in comparison with the original ${\rm Im}\Sigma^{\rm nor}$ (red squares).
For details, see Appendix~\ref{S2.2}.
}}
\label{Fig_analysis_Bok}
\end{figure*}
\begin{figure}[htb]
\begin{center}
\includegraphics[width=0.35\textwidth]{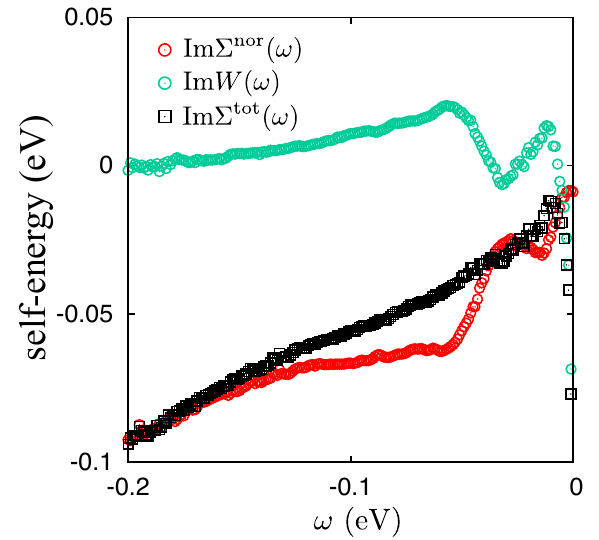}
\end{center}
\caption
{
{Imaginary parts of $\Sigma^{\rm nor}$, $W$, and $\Sigma^{\rm tot}$ obtained by Bok, {\it et al}.~\cite{bok2016quantitative}.}
{The shoulder structure in ${\rm Im}\Sigma^{\rm nor}$ (red circles) and the peak structure in ${\rm Im}W$ (green circles)
around $\omega = -0.06$ eV are canceled in ${\rm Im}\Sigma^{\rm tot}$ (black squares).
The peak structures in ${\rm Im}\Sigma^{\rm nor}$ and ${\rm Im}W$ for $\omega > -0.02$ eV,
which are interpreted as impurity effects in Ref.~\onlinecite{bok2016quantitative}, are also canceled each other in ${\rm Im}\Sigma^{\rm tot}$.
For details, see Appendix~\ref{S2.2}.}}
\label{Fig_cancell_Bok}
\end{figure}

Here, we discuss the comparison with the analysis by Bok {\it et al.}~\cite{bok2016quantitative}, \mig{which has not \migr{clearly} identified a prominent peak structure in the normal and anomalous self-energies.}
We first point out that the primary origin of the discrepancy 
may be the momentum region they studied.
They have analyzed mainly only around the nodal region and at most up to $\theta=20^{\circ}$ measured from the nodal point, which is far away from the antinodal point. This makes the identification of the prominent peak difficult.
In Appendix~\ref{S2.4}, we show the momentum dependence of the EDC curve for the optimally doped Bi2201 (see 
Fig.~\ref{Fig_angle_dependence}).
In this notation, $20^{\circ}$ from the nodal point in \cite{bok2016quantitative} nearly corresponds to the point between 
Fig.~\ref{Fig_angle_dependence}(h).
It is natural that the peak structure is not clearly visible there.
However, if we take a close look, aside from the clear difference of the featureless slope arising from the instrumental difference and presumable different background effect, the peak-like (or shoulder-like) structures at -0.06 eV in the imaginary part of the normal and anomalous self-energies in Fig. 3 of Ref.~\onlinecite{bok2016quantitative} shares a common feature with our result in
Fig.~\ref{Fig_angle_dependence}(h).  Unfortunately, this tiny signature and the cancellation in the spectral function is hardly identified conclusively because of the momentum far from the antinodal point.

{More importantly, a crucial origin of the underestimate of the peak is
the usage of the Dynes function as explained below.
It leads to the underestimation of peak-like structures in the imaginary part of the self-energies. The underestimation inevitably leads to the difference in the self-energies obtained in the present paper and Ref.~\onlinecite{bok2016quantitative}.}

{Before going into the explanation of the underestimation, we review the method used in Ref.~\onlinecite{bok2016quantitative}
to make the \tgtw{discussion} self-contained. The method assumes that the real part of the Dynes function $N(\omega)$ defined as,
\eqsa{
N(\omega) =
\left\{\frac{\omega}{\sqrt{\omega^2 - \Delta (\omega)^{2}}}\right\}\nonumber,
}
is equal to the ratio of the integrated spectral function at the normal and superconducting states as
\eqsa{
{\rm Re} N(\omega) = \mathcal{A}_S(\omega) / \mathcal{A}_N(\omega)\nonumber,
}
where the integrated spectral functions are obtained with respect to the bare dispersion $\epsilon (k_{\perp})$
along momentum perpendicular to the Fermi surface as $\mathcal{A}_S (\omega) = \int d k_{\perp} A_s (k_{\perp},\phi,\omega)$
and $\mathcal{A}_S (\omega) = \int d k_{\perp} A_n (k_{\perp},\phi,\omega)$.
Here, $A_s$ ($A_n$) is the spectral function of the superconducting (normal) state.
The relation, however, holds \tgtw{only} when the bare dispersion $\epsilon (k_{\perp})$ is
\tgtw{linear}
at the large bandwidth limit.
Practically, the method \tgtw{approximately} works around the nodal direction where the bare band is linear within a certain energy range.
However, even around $\phi = 20^{\circ}$, the approximation does not work,
where $\epsilon (k_{\perp})$ shows strong deviation from the linear dispersion, as shown below.}

{Here, we concretely demonstrate that the relationship
${\rm Re}N(\omega) = \mathcal{A}_S (\omega) / \mathcal{A}_N (\omega)$
indeed underestimates the imaginary part of the self-energies.
As the most striking example, we would like to demonstrate that, even if there are peak structures in the genuine self-energies, the self-energies estimated by using
${\rm Re}N(\omega) = \mathcal{A}_S (\omega) / \mathcal{A}_N (\omega)$
collapse to shoulder structures instead of the peak structures.}

{When there is no background $b(\omega)$ in the experimentally observed spectral function,
the scheme to extract the self-energies used by Bok {\it et al}. is summarized as follows:
\begin{description}
\item[(1)] ${\rm Re}N(\omega)$ is (initially) estimated by $\mathcal{A}_S (\omega) / \mathcal{A}_N (\omega)$. 
\item[(2)] ${\rm Im}N(\omega)$ is obtained from ${\rm Re}N(\omega)$
by using the KK transformation. Then, $\Delta (\omega)$ is obtained from
${\rm Re}N(\omega)+i{\rm Im}N(\omega)=\{\omega/\sqrt{\omega^2 - \Delta^2}\}$.
\item[(3)] By inputing $\Delta (\omega)$ into
the Green function in the superconducting phase,
$Z(\omega)=1-\Sigma^{\rm nor}(\omega)/\omega$ and $\Sigma^{\rm ano}(\omega)=\phi (\omega)
=\Delta (\omega) \Sigma^{\rm nor}(\omega)$
are obtained.
\end{description}
Below, we estimate the self-energies by following Bok {\it et al}.~\cite{bok2016quantitative}.
}

{As a model self-energies that show peak structures,
we take the self-energies at $\phi =21.24^{\circ}$ obtained in \ref{Fig_angle_dependence}(h),
which is shown in the left panel of \ref{Fig_analysis_Bok}.
We also assume a typical bare dispersion for bismuth cuprates,
$\epsilon (k)= \mu - 2t_1 ( \cos k_x+\cos k_y ) + 4t_2 \cos k_x \cos k_y - 2t_3 (\cos(2k_x)+\cos(2k_y))$,
where $\mu =405$ meV, $t_1 =360$ meV, $t_2 =108$ meV, and $t_3 =36$ meV.
Then, we integrated the spectra $\mathcal{A}_S (\omega)$ with $\Sigma^{\rm nor}$ and $\Sigma^{\rm ano}$ along the momentum cuts
(in the first quadrant of the Brillouin zone) specified by the angle $\phi$ measured from the antinode and obtain $\mathcal{A}_N(\omega)$.
To estimate $\mathcal{A}_N (\omega)$, we generate \tgtw{the} normal state self-energy $\Sigma^{\rm nor}$ by eliminating peak structures that cancel the peak structure in $\Sigma^{\rm ano}$,
which is shown in the middle panel of \tgtw{Fig.}~\ref{Fig_analysis_Bok} \tgtw{by assuming that the effect of the superconductivity appears at the peak structure only in the normal self-energy}.
Then, in the right panel of \tgtw{Fig.}~\ref{Fig_analysis_Bok},
we obtain the estimated $\Sigma^{\rm nor}$ (black curve) by assuming ${\rm Re}N(\omega) = \mathcal{A}_S (\omega)/\mathcal{A}_N (\omega)$,
in comparison with the original $\Sigma^{\rm nor}$ taken from \tgtw{Fig.}~\ref{Fig_angle_dependence}(h).
{Although we started from $\Sigma^{\rm nor}$ and $\Sigma^{\rm ano}$ that contain prominent peaks,
in the resultant $\Sigma^{\rm nor}$ derived by assuming the above Dynes function shows the disappearance of the peak.
This shows that the usage of the assumption of the Dynes function leads to the self-contradiction with the underestimate of the peak structure.}}

{As shown in the right panel of \tgtw{Fig.}~\ref{Fig_analysis_Bok},
the estimated ${\rm Im}\Sigma^{\rm nor}$ shows the shoulder structure around -50 meV,
instead of the peak around -70 meV in the original ${\rm Im}\Sigma^{\rm nor}$.
The shoulder structure of ${\rm Im}\Sigma^{\rm nor}$ found in  Fig.3D of Ref.~\onlinecite{bok2016quantitative}  around -50 meV is interpreted as
a remnant of the original peak structure in the genuine self-energies, which is consistent with our present results.
Therefore, the difference between our results and those of Ref.~\onlinecite{bok2016quantitative} is attributed to the artificial reduction in the amplitude of
${\rm Im}\Sigma^{\rm nor}$ due to the assumption
${\rm Re}N(\omega) = \mathcal{A}_S (\omega)/\mathcal{A}_N (\omega)$ \tgtw{in Ref.~\onlinecite{bok2016quantitative}}.}

{The self-energies obtained in Ref.~\onlinecite{bok2016quantitative} also show a remnant of the cancellations of the peak structures of ${\rm Im}\Sigma^{\rm nor}$ and ${\rm Im}W$ found in the present paper.
As shown in \ref{Fig_cancell_Bok},
the shoulder structure of ${\rm Im}\Sigma^{\rm nor}$ obtained in Ref.~\onlinecite{bok2016quantitative} is canceled by ${\rm Im}W$.
Then, ${\rm Im}\Sigma^{\rm tot} = {\rm Im}\Sigma^{\rm nor} + {\rm Im}W$ does not show any shoulder structure.}

{The above general underestimate (or elimination) of the self-energy peak is a universal failure of the assumption employed by the Dynes function and the assumption of the wide band limit.
The physical origin is the following: \tgtw{In} the wide band width limit, $\mathcal{A}_N(\omega)$ only contains the information of the bare band,
which does not depend on whether the system is normal or superconducting.
Then, the left-hand side of $\mathcal{A}_S(\omega)/\mathcal{A}_N(\omega)={\rm Re}\{\omega/\sqrt{\omega^2 - \Delta^2}\}$
does not contain not only the normal-state self-energy but also the bare band information, which are absent in the right-hand side of the relation.
However, even at $\phi=20^{\circ}$, the assumption of the wide band width is invalid.
Therefore, the normal-state self-energy information remains in the denominator of the ratio $\mathcal{A}_S(\omega)/\mathcal{A}_N(\omega)$,
{\begin{widetext}
\begin{eqnarray}
\mathcal{A}_S(\omega)/\mathcal{A}_N(\omega)
=
\frac{
\displaystyle
\int d k_{\perp} {\rm Im} \left[\frac{1}{Z_{\rm s}(\omega)}
\frac{\omega+i\delta+\epsilon(k_{\perp})/Z_{\rm s}(\omega)}
{
(\omega+i\delta-\epsilon(k_{\perp})/Z_{\rm s}(\omega))
(\omega+i\delta+\epsilon(k_{\perp})/Z_{\rm s}(\omega))
-\Delta (\omega)^2}\right]
}{
\displaystyle
\int d k_{\perp} {\rm Im} \left[\frac{1}{Z_{\rm n}(\omega)}
\frac{1}{\omega+i\delta-\epsilon(k_{\perp})/Z_{\rm n}(\omega)}\right]
},
\nonumber
\end{eqnarray}
\end{widetext}
where $Z_{\rm n}(\omega)$ ($Z_{\rm s}(\omega)$) is $Z(\omega)=1-\Sigma^{\rm nor}(\omega)/\omega$ in the normal (superconducting) state.}
To compensate the
\tgtw{contribution}
of $\Sigma^{\rm nor}$ in the left-hand side of the relation $\mathcal{A}_S(\omega)/\mathcal{A}_N(\omega)={\rm Re}\{\omega/\sqrt{\omega^2 - \Delta^2}\}$,
the normal-state self-energy $\Sigma^{\rm nor}$ below $T_{\rm c}$,
whose
\tgtw{contribution}
is contained in $\mathcal{A}_S(\omega)$, may resemble the self-energy $\Sigma^{\rm nor}$ above $T_{\rm c}$,
whose
\tgtw{contribution}
is contained in $\mathcal{A}_N(\omega)$.
When the normal state self-energy does not show significant structures and, thus, does not introduce pseudogap and anomalies,
such as peak-dip-hump structures, in the normal-state spectral functions, the estimated superconducting self-energies indeed do not show significant structures. 
We note that the shoulder structure consistently present in Ref.~\onlinecite{bok2016quantitative} and \ref{Fig_cancell_Bok}b in ${\rm Im} \Sigma^{\rm nor}$ without a signature of the pseudogap around 20 degrees momentum could be contributed from the background.
However, the above analysis does not change even when the background is subtracted, because it appears both in $\mathcal{A}_S(\omega)$ and $\mathcal{A}_N (\omega)$ and the effect of the subtraction cancels. 
}


\migr{In addition,
{Bok {\it et al}.~\cite{bok2016quantitative}} have assumed that the normal and anomalous self-energies are momentum independent along the direction perpendicular to the Fermi surface.
The assumption \mimg{imposes crucially restrictive condition when one infers the self-energy and superconducting gap function from the momentum distribution curve (MDC) as in Ref.~\onlinecite{bok2016quantitative} (see Eq. (S8) in the Supplementary information of their paper).}}
This assumption is adequate in the BCS superconductors in conventional weakly correlated systems,
while it is \mimg{questionable} in the present strongly correlated cases such as the cuprates, where even the normal self-energy can be singular and strongly dependent on the
{momentum}. 
{In addition to the restricted momentum dependence assumed in \cite{bok2016quantitative},
they have assumed the quasiparticle representation along the momentum perpendicular to the Fermi surface (or the Lorentzian form of the MDC) and
the spectral function that cannot be represented by the Lorentzian form are interpreted as the background,
while these have not been assumed in the present paper,  
because, though it is satisfied in the Fermi-liquid normal state, it is not clear whether these constraints are satisfied in the strong coupling cuprate superconductors.
Indeed, there exist a number of numerical evidences for the violation of the assumption: For instance, 
the singular momentum dependence of the normal self-energy \mimg{with emergence of the coexisting zeros and poles of the Green function}~\cite{PhysRevB.98.195109} and non-quasiparticle spectral function~\cite{PhysRevB.90.035111,PhysRevB.93.081107}
in doped Mott insulators.}
\migr{The machine learning is more fit in solving the present problem, because the flexible fitting of the self-energy function is required at least away from $\omega=0$ \mimg{particularly in the antinodal region, where the breakdown of the quasiparticle picture is apparent}.
Concerning the difference in the high-energy part ($\omega<-0.1 $ eV) of the original two ARPES data (namely in \cite{bok2016quantitative} and \cite{kondo2011disentangling}),
we have already analyzed the effect of the possible extrinsic high-energy part in \ref{S2.1} and has shown that it does not affect the structure of the peaks as clarified in 
\ref{Fig_dependence_on_high_energy_part} and it cannot be the origin of the difference in the peak structure.} 

\begin{figure}[htb]
\begin{center}
\includegraphics[width=0.3\textwidth]{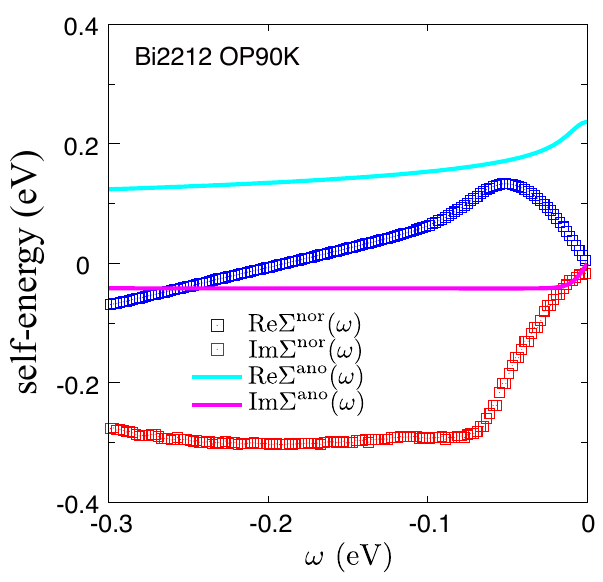}
\end{center}
\caption
{
Self-energies obtained \tind{from} a
{constrained} anomalous self-energy \tgtw{for Bi2212}.
\tgtw{The solid cyan and magenta curves represent the real and imaginary part of the anamalous self-energy,
\tind{obtained by following}
the assumption by \tind{Li {\it et al.}~\cite{LiDessau}}.
The blue and red open squares represent the real and imaginary part of the normal self-energy
given by the Boltzmann machine as in our present scheme.}
For details, see \tgtw{the text in} Appendix~\ref{S2.2}.
}
\label{Fig_LiDessau}
\end{figure}

Next we discuss the origin of discrepancy in the result by Li, {\it et al.}~\cite{LiDessau}. 
They assumed momentum independent self-energy in the MDC analysis as one sees
in Eq.~(3) of their supplementary note 2, whose basis is unclear.
More crucially, they assumed the imaginary part of self-energy in the forms
Eq.~(5) or (6) in Supplementary Information of Ref.~\onlinecite{LiDessau} 
for the normal part and
\eqsa{
{\rm Im} \Sigma^{\rm ano}(\omega)=1/(e^{(\omega-E_3)/W_3}+1)+1/(e^{(\omega-E_4)/W_4}+1),\nonumber\\
\label{LiDessauEq7}
}
with constant fitting parameters $W_3,W_4,E_3$ and $E_4$,
for the anomalous part,
which is unjustified.
Particularly,
{the assumed form Eq.~(\ref{LiDessauEq7})}
does not allow the formation of peak or dip and it does not allow the cancellation with the structure in the normal contribution in the spectral function as we discovered.
It is crucially important to allow the flexibility of the self-energy form and the machine learning is one of the best way to incorporate it while the attempt by {Li, {\it et al}.} failed in implementing the flexibility.
We have attempted to fit the self-energy with the constraint of Eq.~(\ref{LiDessauEq7})
for the anomalous part and found the resultant optimized $\chi^2$ is 
$\textcolor{black}{\chi_{\rm MLF}^2}=6.1\times 10^{-6}$,
which is much higher than the present result $\textcolor{black}{\chi_{\rm ML}^{2}}=2.1 \times 10^{-6}$.
Because the experimental resolution is $\chi_{\rm exp }^2=1.4\times 10^{-6}$ 
as mentioned above, the intrinsic $\chi^2$ 
defined by $\delta\chi_{\rm MLF}^2=\chi_{\rm MLF}^2-\chi_{\rm exp}^ 2$ is $4.7\times 10^{-6}$.  Then the  intrinsic machine learning error in the unit of the experimental standard deviation is
$\textcolor{black}{\delta\chi_{\rm MLF}/\chi_{\rm exp}=1.8}$, which is \textcolor{black}{close to} the twice of the standard deviation.
Namely, the probability that this
{constrained} choice is true is \textcolor{black}{less than 7\%}. 
The resultant self-energy does not show any appreciable peak as it should be in contrast to the present result as one sees in 
Fig.~\ref{Fig_LiDessau}.
Note that, despite the
{constrained} anomalous part, due to the unbiased choice of the normal self-energy here, the self-energies in 
Fig.~\ref{Fig_LiDessau} should be much better fit than the more
{constrained} ones (including the normal part) in {Li, {\it et al}~\cite{LiDessau}}.
The spectral function obtained from the additional
{constraint} in the normal part, Eq.~(5) or (6) in Supplementary Information of Ref.~\onlinecite{LiDessau} must give even higher $\chi^2$ than $\chi_{\rm MLF}^2$.

\tgtw{In addition, the form of the imaginary part of the anomalous self-energy is
unphysical.
The form assumes the attractive interaction ranging to the infinite frequency scale (namely, instantaneous attractive interaction),
which can not exist in the real experiments.}

{In summary, constraints or assumptions unjustified {\it a priori} such as momentum independent self-energy, strictly constrained but unjustifiable self-energy form, and Eliashberg equation in the previous works
\tgtw{do not allow prominent peak structures and at the same time}
give higher errors in the regression of the experimental data than our result indicating the superiority of our analysis. It supports that the cancellation of the normal and anomalous contribution in the total self-energy found in our analysis and not in the former studies should be considered seriously.}

\section{Pair Breaking Effect
\label{S2.9}}
\begin{figure}[htb]
\begin{center}
\includegraphics[width=0.5\textwidth]
{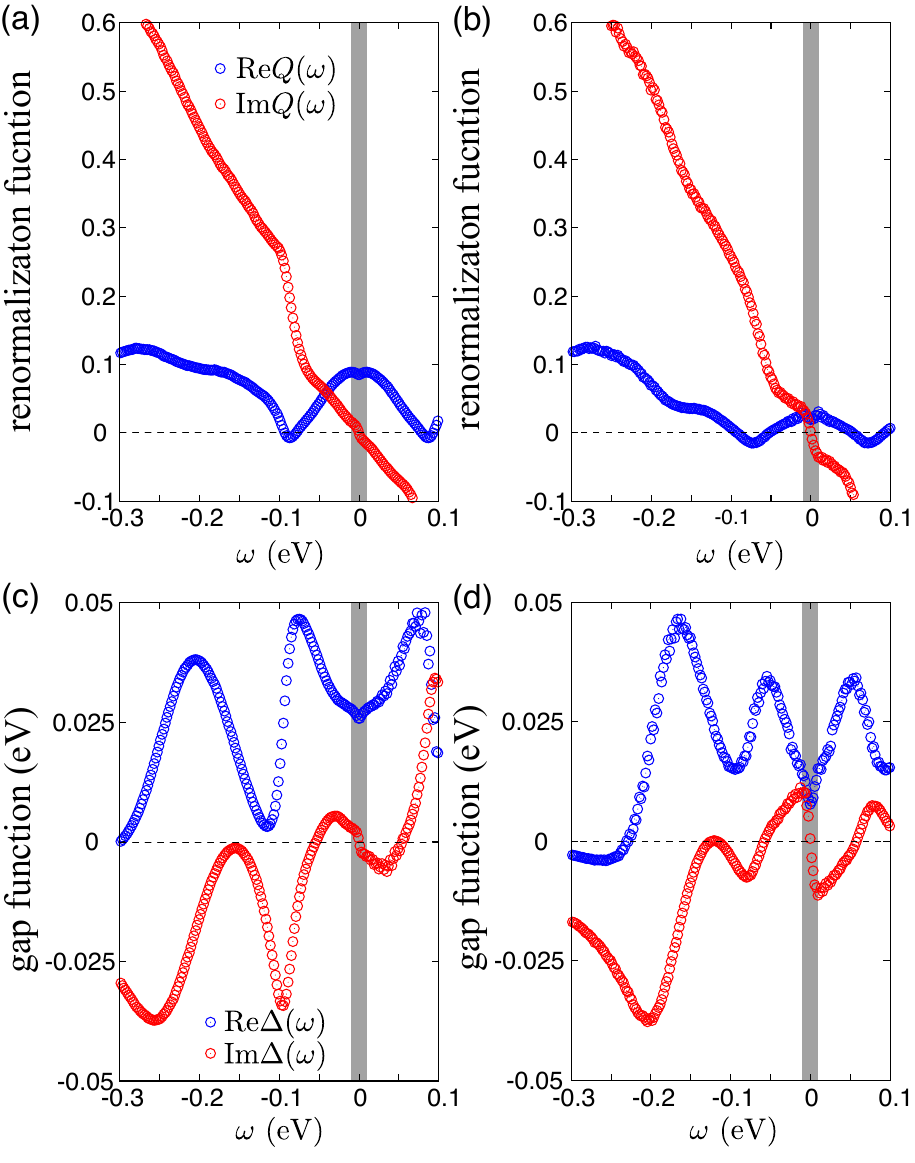}
\end{center}
\caption
{
Renormalization function $Q(k_{\rm AN},\omega)$ and Gap function $\Delta(k_{\rm AN},\omega)$ obtained by machine learning.
$Q(k_{\rm AN},\omega)$ ( (a) and (b)) and $\Delta(k_{\rm AN},\omega)$ ( (c) and (d)) are plotted 
for optimally doped 
Bi2212 ((a) and (c)) and 
 underdoped 
Bi2201 ((b) and (d))
 for the experimental data shown in 
Fig.~1. 
{The {width of} shaded area shows the experimental resolution (see Sec.~\ref{training_error} and \ref{test_error}).}
{The positive part in  ${\rm Im} \Delta(k_{\rm AN},\omega)$ near $\omega=0$ indicates the pair breaking contributing to destroy the superconductivity.}
}
\label{FigQ}
\end{figure}
\tr{The renormalization factor (quasiparticle residue) \tgx{estimated from the expression 
\begin{equation}{
z(k) =Q(k,\omega)|_{\omega\rightarrow 0}
\label{eq:z}
}
\end{equation}
theoretically equivalent to $z_{\rm qp}$ defined
in {Eq.~(\ref{eq:zqp})} 
is the weight of the quasiparticle, which can be substantially reduced from the noninteracting value $z(k)=1$ due to the interaction effects.
The renormalization factor estimated from the fitting of Eq.~(\ref{eq:z})
is $z_{\rm qp}\sim 0.1$ for Bi2212 and $z_{\rm qp}\sim 0.03$  for Bi2201} (see 
Fig.~\ref{FigQ}) supporting the
non-Fermi liquid behavior especially in the underdoped case. }

\tr{As shown in 
Figs.~
\ref{FigQ}(c) and (d),
the non-Fermi-liquid-like 
${\rm Im}\Sigma^{\rm nor}(\omega)$
{affects} the gap function $\Delta (\omega)$ (in 
Eq.~(\ref{eq:Delta})
through $Q(\omega)$
{(see Eq.~\ref{eq:Q})}.
In general, negative ${\rm Im} \Delta (\omega)$ for $\omega<0$ enhances  ${\rm Re} \Delta(\omega=0)$
through the Kramers-Kronig relation and is indeed negative in most of $\omega$ in 
Fig.~\ref{FigQ}(c) and (d).
However, ${\rm Im} \Delta (\omega)$ is positive at {$|\omega|<0.04$} {eV} ($|\omega|<0.06$ eV) for Bi2212 (Bi2201).
Because ${\rm Im} \Sigma^{\rm ano}$ is found to be always negative for $\omega<0$, {it is ascribed to the pair breaking effect of $Q$, arising  from poles of  $\Sigma^{\rm nor}$
inside the superconducting gap 
as already pointed out~\cite{PhysRevLett.116.057003}.}
The pair breaking is much more prominent for underdoped sample, Bi2201.}

\tr{
Although a similar conclusion for the underdoped Bi2201 suggests a universal nature, the prominent non-Fermi liquid behavior and the pair breaking could be accounted for by an alternative at $k=k_{\rm AN}$, namely the pole of $\Sigma^{\rm nor}$ shifts to the energy $\omega\sim 0$ and destroys $\Sigma^{\rm ano}$ accompanied by an insulating gap. Although such a solution gives worse $\chi^2$ in our analysis, a momentum selective insulating behavior at the antinodal point deserves to be explored further together with the full momentum and temperature dependences.}

\section{Gap Functions}
\label{S1.6}
\subsection{Resolution of gap functions}
{The gap function $\Delta (\omega)$ defined in 
Eq.~(\ref{eq:Delta}) can show significant $\delta$ dependence
near the small $\delta$ limit around $\omega \sim 0$.
The $\delta$ dependence originates from the finite imaginary part of the normal self-energy ${\rm Im}\Sigma^{\rm nor}(k,\omega=0)$ {inevitable in the experimental data}.
When we modify $Q$ as
\begin{eqnarray}
Q(k,\omega)=\frac{1}{\displaystyle 1-\frac{\Sigma^{\rm nor}(k,\omega+i\delta)-\Sigma^{\rm nor}(k,-\omega-i\delta)^{\ast}}{2(\omega+i\delta')}},\nonumber\\
\end{eqnarray}
we obtain stable behaviors of $\Delta (\omega)$ for $|\omega|> 10$ meV by keeping $\delta =10$ meV and {restricting to} $\delta' < \delta$.
In 
Fig.~\ref{FigQ}, we use $\delta' = 2.5$ meV.
}

\subsection{Peak shift in gap function}
\label{S1.7}
Fig.~\ref{FigQ}(c) and (d) show the gap function $\Delta$ defined in 
{Eqs.~(\ref{eq:Delta}) and (\ref{eq:Q})}.
It reveals that the peak positions in $\Delta(k_{\rm AN},\omega)$ are different from those in ${\rm Im}\Sigma^{\rm ano}$, which is consistent with the hidden fermion theory~\cite{PhysRevLett.116.057003}. In fact, the peak positions of ${\rm Im} \Delta(k_{\rm AN},\omega)$ ({$\sim \pm 80$ and $\pm$220 meV} for Bi2212 and {$\sim \pm 80$ and $\pm$210 meV} for Bi2201) are nearly the same as the peak positions of ${\rm Re}\Sigma^{\rm ano}$, while the peak positions of ${\rm Re} \Delta(k_{\rm AN},\omega)$ {($\sim \pm 50$ and 
{$\pm 180$} meV for Bi2212 and $\sim \pm 50$ and 
{$\pm 160$} meV} for Bi2201) are nearly the same as the peak positions of ${\rm Im}\Sigma^{\rm ano}$.  This is because the imaginary part of $Q$ is dominant in the relevant frequency region ($\sim 100-200$ meV) as shown in 
Fig.~\ref{FigQ}. 
The shift of the peak positions indicates the strong renormalization effect in the normal quasiparticle contained in $Q$.  
In any case, in the contribution to the real order parameter of the superconductivity $\Delta_{\rm AN}= {\rm Re} \Delta(k_{\rm AN},\omega=0)$
{is expected to be} contributed mostly from the two peaks in ${\rm Im} \Delta(k_{\rm AN},\omega)$ through the Kramers-Kronig relation.
The $d$-wave gap amplitude $\Delta_{\rm AN}$ is 30 meV for the optimally doped  Bi2212 while it is around 10 meV. However, at small energy ($\sim 60$ meV), the gap amplitude is both around 40 meV, which is comparable to the peak energy of ${\rm Im} \Sigma^{\rm ano}$ and ${\rm Im} \Sigma^{\rm nor}$ suggesting the similar pseudogap energy for optimum and underdoped samples.

\section{
Accuracy, Stability and Robustness Tested by Solvable Benchmarks
\label{S2.5}}
%

\begin{figure*}[htb]
\begin{center}
\includegraphics[width=1.0\textwidth]{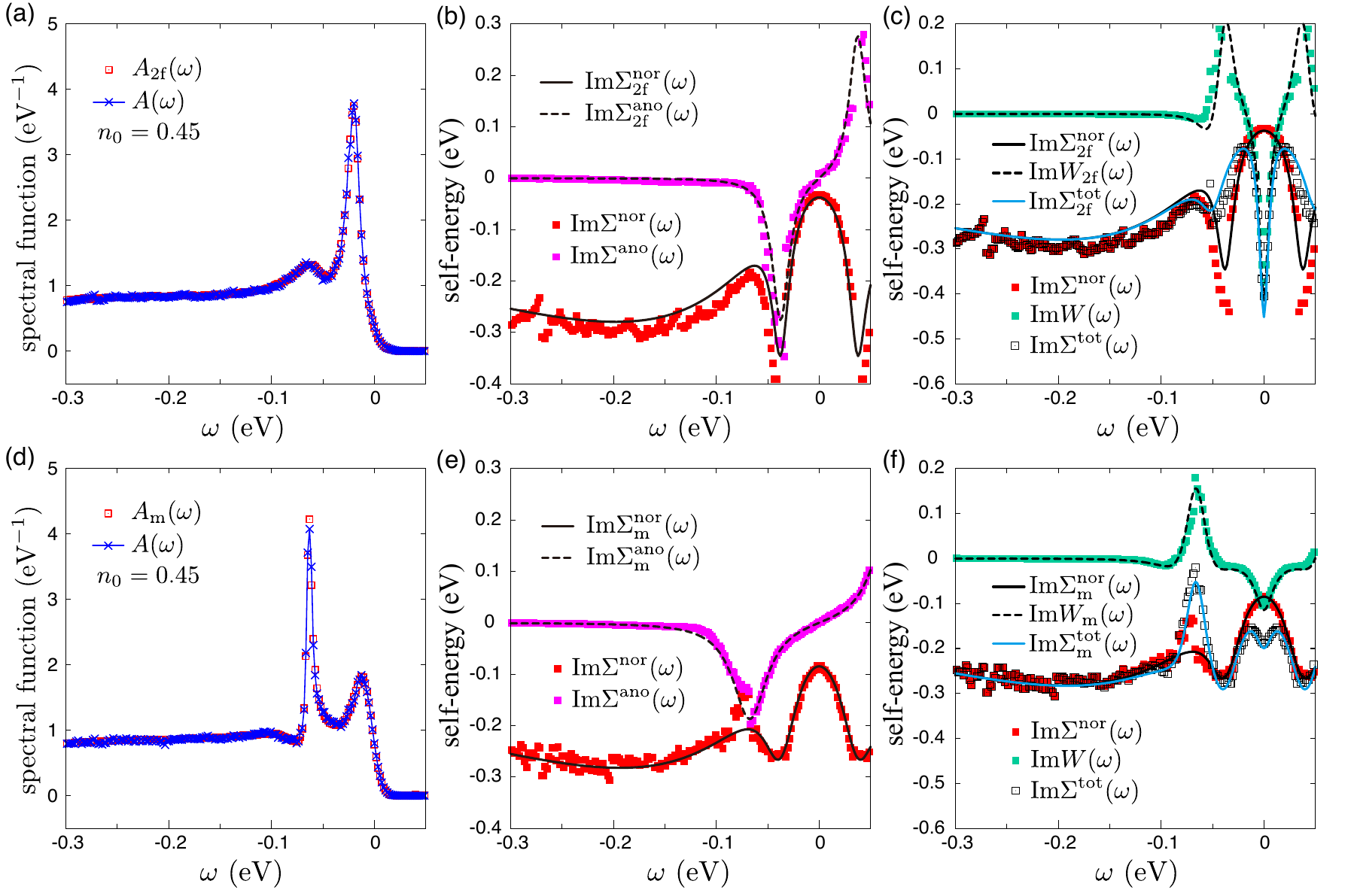}
\end{center}
\caption
{
{Spectrum and self-energies of two-component fermion model and its modified version.
{(a). The exact spectral function 
of the two-component fermion model, 
$A_{2f}(\omega)$, (red open squares) is compared
with $A(\omega)$ obtained from the machine learning using $A_{2f}(\omega)$ (blue curve crosses) within -0.55 eV $<\omega<$ 0.05 eV.
(b) and (c): The imaginary part of the self-energies of the two-component fermion model,
$\Sigma^{\rm nor}_{2f}$, $\Sigma^{\rm ano}_{2f}$, and $W_{2f}$ (curves),
are compared with the self-energies
obtained from the machine learning (symbols).
(d). The exact spectral function of the modified two-component fermion model, $A_{\rm m}(\omega)$, (open squares) is compared
with $A(\omega)$ obtained from the machine learning using $A_{\rm m}(\omega)$ within -0.55 eV $<\omega<$ 0.05 eV.
(e) and (f): The imaginary part of the self-energies of the modified two-component fermion model,
$\Sigma^{\rm nor}_{\rm m}$, $\Sigma^{\rm ano}_{\rm m}$, and $W_{\rm m}$ (curves),
are compared with the self-energies
obtained from the machine learning (symbols).
}
}
}
\label{Fig_two_fermion_Sigma}
\end{figure*}

{In this section, we employ
exactly solvable models as benchmarks.}
\tgtw{We} examine whether our machine learning correctly reproduces the exact self-energies (with prominent peak structures), if the exact solution indeed shows the cancellation of the normal and anomalous self-energy contributions in the total self-energy and the spectral function $A(\omega)$ shows only a weak peak-dip-hump structure.
To our knowledge, exact solution, which shows such a cancellation is not found except for the case of the two-component fermion model. Then, as a benchmark, we inferred the self-energy of a superconducting two-component fermion model 
defined by the following Lagrangian,
\begin{eqnarray}
L(\omega)&=&\sum_{k,\sigma}[(\omega +i\delta - \epsilon_c(k) - \Sigma^{(0)} (\omega ) )c_{k,\sigma}^{\dagger}c_{k,\sigma} -\epsilon_d d_{k,\sigma}^{\dagger}d_{k,\sigma}  \nonumber 
\\
&-& V_1 (c_{k,\sigma}^{\dagger}d_{k,\sigma} +{\rm H.c.})
-D_1 ( d_{k,\sigma}^{\dagger}d_{-k,-\sigma}^{\dagger} + {\rm H.c})
],
\label{TCfermionAnomalous} 
\end{eqnarray}
which is
essentially the same form as that introduced in 
Ref.~\onlinecite{PhysRevLett.116.057003} and discussed in Ref.~\onlinecite{ImadaSuzuki}.
{In the following discussion, we assume that the noninteracting density of states  determined from $\epsilon_c(k)$ is a constant $N_F$ and focus on a specific momentum $k$ at the Fermi momentum just for simplicity. Because of the momentum independence, this consideration at a specified momentum does not cause loss of generality.}
Here, we add $\Sigma^{(0)}(\omega)$ {at the above momentum} defined by
\begin{eqnarray}
\Sigma^{(0)}(\omega)&=&\frac{\alpha}{b-a}\left\{\frac{\sqrt{b}}{\omega+i\sqrt{b}} - \frac{\sqrt{a}}{\omega+i\sqrt{a}}\right\}
\label{S(0)}
\end{eqnarray}
in addition to $\epsilon_c(k)$ to mimic the additional normal Fermi-liquid-like component seen in the experimental result arising from interaction effect for the part not represented by the coupling to the $d$ fermion,
where $a$, $b$ and $\alpha$ are constants. 
The self-energy in the exact solution of this two-fermion model is given as,
\begin{eqnarray}
\Sigma^{\rm nor}_{2f}(\omega) &=& \frac{V_1^2 (\omega+i\delta+\epsilon_{d})}{(\omega+i\delta)^2-(\epsilon_{d}^2+D_1^2)}
+ \epsilon_{c}+\Sigma^{(0)}(\omega),\nonumber\\
&&  \label{Snor}
\\
\Sigma^{\rm ano}_{2f}(\omega) &=&-\frac{V_1^2 D_1}{(\omega+i\delta)^2-(\epsilon_{d}^2+D_1^2)}, \label{Sano}
\end{eqnarray}
For simplicity, we have dropped the momentum dependence in the solutions (\ref{Snor}) and (\ref{Sano}).
In our calculation, we set $a=0.008$ eV$^2$, $b=0.2$ eV$^2$, $\alpha=0.08$  eV$^3$, $V_1=0.075$ eV, $D_1=0.0375$ eV, and $\epsilon_{d}=\epsilon_{c}=0$.
{The present choice of the parameters is enough to generate the spectral function observed at the Fermi momentum {we focus, and the dependence on the doping and dimension of the system {\it etc.} are implicitly contained in $N_F$.}
By using the exact solution for the spectral function $A_{2f}(\omega)$, we add small but finite noise, where $\sigma^2$ of the noise is set to be $6\times 10^{-4}$ to simulate the role of noise in experiments and perform the machine learning using
this noisy $A_{2f}(\omega)$.
Although it is irrelevant to the inferred self-energies, the Fermi-Dirac distribution with $T=40$ K is introduced
in the spectral function used in the machine learning just by following the scheme
{with finite-temperature experimental data}.
In 
Figs.~\ref{Fig_two_fermion_Sigma}(a)-(c), the spectral function of the two-component fermion model $A_{2f}(\omega)$ and the self-energies are shown for the exact solutions (solid and dashed curves) and the machine learning results (symbols).
In the self-energy inference, we choose $n_0=0.45$ without fine tuning and use the spectrum within 
$-0.55\ {\rm eV}\ <\omega<0.05\ {\rm eV}$.
The peak position in ${\rm Im}\Sigma^{\rm nor}$ and ${\rm Im}\Sigma^{\rm ano}$ and the Fermi liquid-like normal contribution in the exact solution 
{(shown with the index $2f$ such as $A_{2f}(\omega)$ illustrated by solid and dashed curves)} are well reproduced by our machine learning results (symbols). The peak cancellation in 
{$\Sigma_{2f}^{\rm tot}$} in the exact solution is also well reproduced.}

{In 
Figs.~\ref{Fig_two_fermion_Sigma}(d)-(f),
we show an artificial case, where the pole of $\Sigma^{\rm ano}$ is shifted 0.03 eV from the solution (\ref{Sano}), where the pole cancellation in the total self-energy
does not occur any more
and the spectral function shows weird two peaks. 
The concrete representation of the self-energies of the modified two-component model is given by the following form,  
\begin{eqnarray}
\Sigma^{\rm nor}_{\rm m}(\omega) &=& \frac{V_1^2 (\omega+i r \delta+\epsilon_{d})}{(\omega+ir \delta)^2-(\epsilon_{d}^2+D_1^2)}
+ \epsilon_{c}+\Sigma^{(0)}(\omega),\nonumber\\
&& \label{Smnor}
\\
\Sigma^{\rm ano}_{\rm m}(\omega) &=&-\frac{V_1^2 (D_1+\Delta D_1)}{(\omega+i r \delta)^2-(\epsilon_{d}^2+(D_1+\Delta D_1)^2)}, \label{Smano}
\end{eqnarray}
where the pole of $\Sigma^{\rm ano}$ is shifted from that of $\Sigma^{\rm ano}$ because of $\Delta D_1 =0.03$ eV and a factor $r=2$ is introduced to avoid singular spectrum.
The spectrum and self-energies of the modified two-component model are denoted with the index m as $A_{\rm m}(\omega)$.
Even in this case, the machine learning results well reproduce all the line shapes.
This indicates that our machine learning flexibly and accurately reproduces the exact solution irrespective of the presence or absence of the peak cancellation.}

\section{Numerical Optimization Procedure }
\label{S1.4}
\subsection{Details in minimization of training error}
\label{S1.4.1}
\noindent
The parameters in the Boltzmann machine, $\bvec{\alpha}^{\rm nor}=(b,\{W_{\ell m}\})$ and $\bvec{\alpha}^{\rm ano}=(w_{\lambda},\{b_{\ell}^{\lambda}\},\{V^{\lambda}_{\ell m}\})$, 
are optimized by using the standard gradient method.
The parameters at the $k+1$th step, $\bvec{\alpha}^{\rm nor}_{k+1}$ and $\bvec{\alpha}^{\rm ano}_{k+1}$, are updated from the $k$-th values as 
\begin{eqnarray}
\bvec{\alpha}^{\rm nor}_{k+1}
&=&\bvec{\alpha}^{\rm nor}_k
-\epsilon
\left(
\|
S^{-1}\bvec{g}^{\rm nor}_k
\|_1\right)^{-1/2}
S^{-1}
\bvec{g}^{\rm nor}_k,
\\
\bvec{\alpha}^{\rm ano}_{k+1}
&=&\bvec{\alpha}^{\rm ano}_k
-\epsilon'
\left(\|
\bvec{g}^{\rm ano}_k
\|_1\right)^{-1/2}
\bvec{g}^{\rm ano}_k,
\end{eqnarray}
where
\begin{eqnarray}
S_{\mu\nu}&=&
\frac{1}{N_{\rm d}}\sum_{j}
\frac{\partial {\rm Im}\Sigma^{\rm nor} (\omega_j)}{\partial \alpha^{\rm nor}_{\mu}}
\frac{\partial {\rm Im}\Sigma^{\rm nor} (\omega_j)}{\partial \alpha^{\rm nor}_{\nu}}
,
\\
\bvec{g}^{\rm nor}_k
&=& 
\frac{
\partial \chi^2
}{\partial \bvec{\alpha}^{\rm nor}},
\\
\bvec{g}^{\rm ano}_k
&=& 
\frac{
\partial \chi^2
}{\partial \bvec{\alpha}^{\rm ano}},
\end{eqnarray}
and {$\|\cdots \|_1$} represents $L_1$ norm.
The factors $\left(\|
S^{-1}\bvec{g}^{\rm nor}_k
\|_1\right)^{-1/2}$
and $\left(\|
\bvec{g}^{\rm ano}_k
\|_1\right)^{-1/2}$
are introduced to accelerate the optimization.
Here, we use the natural gradient method to optimize the variational parameters in ${\rm Im}\Sigma^{\rm nor} (\omega_j)$ because of its efficiency,~\cite{Amari1998,Sorella1998,Sorella2001}
while the simple steepest descent method is employed to optimize the part of ${\rm Im}\Sigma^{\rm ano} (\omega_j)$ because the natural gradient method assumes {that}
the optimized distribution is positive or negative definite, \mimg{while ${\rm Im}\Sigma^{\rm ano} (\omega_j)$ does not satisfy this condition.}
{During the optimization of the Boltzmann machine, we may introduce a regularization term by $L_1$ norm of
the mixture of the Boltzmann machines as $\lambda_w \sum_{\lambda}|w_{\lambda}|$.
While $\lambda_w=10^{-3}$ will accelerate the optimization, the results of the optimization is confirmed to be insensitive
if $\lambda_w \leq 10^{-3}$. In the actual fitting, we employed $\lambda_w=10^{-3}$.}

\noindent
\subsection{Parameters in optimization}
\label{S1.4.2}\noindent
In the present paper, first, we optimize the Boltzmann machine with {$L=8$} visible nodes and {$2L=16$} hidden nodes for the part ${\rm Im} \Sigma^{\rm nor}$
and, then, we enhance the resolution
with {$L=9$} visible nodes and {18} hidden nodes to obtain better resolution with reasonable numerical cost.
In the optimization with {$L=9$}, we \mimg{skip the outer loop (the update of the center of mass by the Bayesian process) to reduce the computational cost} and perform longer minimization steps up to $2\times 10^4$. 
We employ the broadening factor {$\delta = 10$ meV} throughout this paper. 
{We show in 
Appendix~\ref{S2.8} that the result does not sensitively depend on the choice of $\delta$.}

\subsection{ Effects of resolution $\delta$
\label{S2.8}}
\begin{figure}[htb]
\begin{center}
\includegraphics[width=0.3\textwidth]
{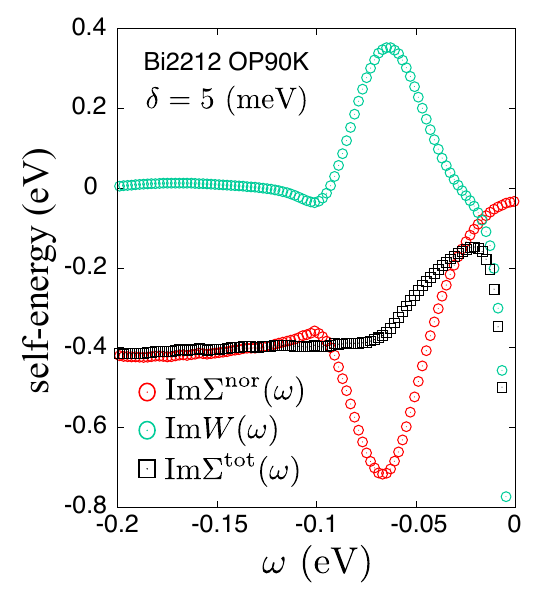}
\end{center}
\caption{
Resolution ($\delta$) dependence of self-energies.
Self-energies are obtained from the machine learning using the ARPES EDC curves with $\delta$=5 meV.
The ARPES EDC is taken from the experimental data of Bi2212 at optimum doping at 11K, which are supplied by Kondo {\it et al.}\cite{kondo2009competition}.
The peak position and their cancellation between $W$ and $\Sigma^{\rm nor}$ 
{remain essentially the same} even for $\delta$ smaller than the experimental resolution.} 
\label{Fig_W_delta}
\end{figure}
{In the present study, the small imaginary part $i\delta$ utilized in the Green functions is
chosen to be equal to the experimental resolution.
When 
{substantially} larger resolution $\delta$ is taken, the detailed spectra are trivially not reproducible.
On the other hand, when smaller resolution $\delta$ is taken, the spectra may be easily fitted.
Here, we examine how the smaller $\delta$ affects the inferred self-energy.
As a typical example, we take $\delta$=5 meV, which is a half of $\delta$ used in the main article,
and confirmed that the smaller $\delta$ does not change the qualitative structure of the self-energy.
As shown in 
Fig.~\ref{Fig_W_delta}, the peak structures in ${\rm Im}\Sigma^{\rm nor}$ and ${\rm Im}W$,
and the cancellation between them are reproduced.}

\subsection{
Gaussian distribution represented by Boltzmann machine
\label{S2.6}}

When we choose the parameters as
\begin{eqnarray}
W_{\ell m}^{\lambda} &=& -\frac{1}{2s_{\lambda}^2}\left(\frac{\Lambda}{2^L}\right)^2 2^{\ell+m},\\
b_{\ell}^{\lambda} &=& \frac{1}{s_{\lambda}^2} (\Lambda/2+x_{\lambda})\frac{\Lambda}{2^L}2^{\ell},\\
w_{\lambda} &=& \frac{w_{0\lambda}}{\sqrt{2\pi s_{\lambda}^2}}e^{-\frac{1}{2s_{\lambda}^2}(x_{\lambda}+\Lambda/2)^2},
\end{eqnarray}
in Eqs~(26) and (27), the Boltzmann machine easily represents the Gaussian distribution with the center $x_{\lambda}$, variance $s_{\lambda}^2$, and weight $w_{0\lambda}$, which is a localized sparse distribution. Superposition of the Gaussian distribution can easily be expressed by Eq.~(\ref{eq:mixture}) by {taking $M$ larger than 1 (typically we take $M$ several)}.

\section{Robustness of Machine Learning}
\label{S2.1}
The present use of machine learning is categorized to a general class of regression analysis as addressed
\textcolor{black}{in the first paragraph of Sec.~\ref{method_regression}}.
In the standard simple case of the regression task, training data set is simply given by the observed $A$ at
\tr{discrete number of} $x$ and we infer the functional form of $A(x)$. In the present case, it is more involved and the training data is the experimentally measured \tr{discrete and limited number of} $A$ and $\omega$, and the regression task is to determine $\Sigma$ as a continuous function of $\omega$. 
In terms of the optimization with the machine learning, our task is to minimize the difference between the measured data $A$ and that obtained from the inferred $\Sigma(\omega)$, which is a continuous function of $\omega$. Therefore, our work is categorized to the machine learning application to a regression task, one of the most widely applied machine learning fields. 
Our regression scheme is illustrated in 
Figure 5.

{
In the regression analysis, it is helpful to examine the reliability of the machine learning by using solvable cases as the benchmark, as in other type of the regression task found in the problem of solving quantum many-body problems and classical statistical physics problems\cite{Carleo}
 It is also important to test the stability of the procedure by adding noises.  In this section we show the robustness against the noise and in \ref{S2.5}, we show several benchmark tests for solvable models.} 

\subsection{Stability against noise}
\label{stability_against_noise}
\begin{figure}[htb]
\begin{center}
\includegraphics[width=0.5\textwidth]
{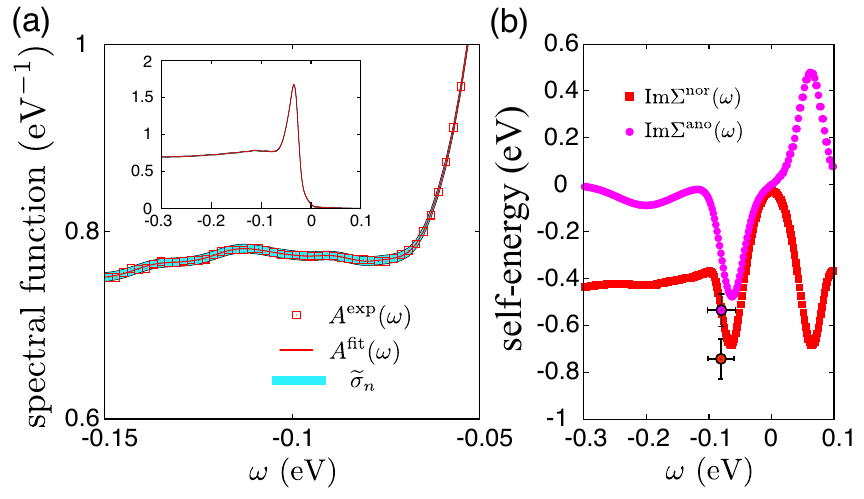}
\end{center}
\caption
{
Robustness of the machine-learning procedure
\noindent 
\migr{(a). Examples of 
{synthetic} spectral function with
{amplified noise $\widetilde{\sigma}_n=4\sigma_n$}
shown for Bi2212
at 12 K (a) obtained 
at the antinodal point in the following way:
The fitting of the experimental data (open red squares) by linear regression is shown as red curves and the standard error (or noise) of the experimental data are estimated as the blue belt. The inset shows the zoom out to see the overall feature.
Then 
{synthetic} random noise with this standard deviation 
{$\widetilde{\sigma}_n$} is added to the red curve to generate 
{synthetic} experimental samples and the machine learning solution of the self-energy for this 
{synthetic} data provides us with the error bar for the self-energy in (b).
For the method of imposing noise, see Sec.~\ref{detailed_method} and \ref{S2.1}. 
(b). Imaginary part of normal self-energy ${\rm Im} \Sigma^{\rm nor}(k_{\rm AN},\omega)$, ${\rm Im} W(k_{\rm AN},\omega)$ and ${\rm Im} \Sigma^{\rm tot}(k_{\rm AN},\omega)$ deduced
by the present machine learning from $A(k,\omega)$.
The error bars are those for the dip energy (horizontal bar) and the dip depth (vertical bar) derived in the procedure mentioned above.
}}
\label{Fignoise}
\end{figure}
We examine stability of the present machine-learning scheme.
By using $A^{\rm fit}$ and $\sigma_n^2$ introduced in \textcolor{black}{Sec.~\ref{test_error}}, 
we can generate 
{synthetic} experimental data with \mi{the same or} larger amplitude of noise than the original data.
Here, we use the 
{synthetic} data to examine the input data dependence of the present scheme.

\mi{{Here, $\chi^2$ of the optimized $A(k,\omega)$ by the machine learning
from the synthetic ARPES spectrum $A^{\rm syn}$ generated by the maximally-likelihood inference of the ARPES spectrum
is given by $\chi_{\rm ML}^2\equiv \overline{\chi^2}=\sum_s^{N_{\rm d}}\sum_{r}^{N_r} (A^{\rm ML}(\omega_s^{(r)})-A^{\rm syn}_r (\omega_s^{(r)}))^2/N_{\rm d}N_r
=2.1\times 10^{-6}$
{(defined in Eq.~(\ref{overline_chi2}))},
which is the same level as the experimental $\chi^2$, namely
$\chi_{\rm exp}^2 \equiv \sigma_{n}^2 = \sum_i^{N_{\rm d}} (A^{\rm exp}(\omega_i)-A^{\rm fit}(\omega_i))^2/N_{\rm d}=1.4\times 10^{-6}$
\mig{obtained in Eq.~(\ref{eq:experror})}.
The same level of $\chi^2$ value indicates that the machine learning optimization to fit the experimental $A(k,\omega)$ is successfully achieved within the limit of the level of the experimental noise.}
The standard deviation of the experimental uncertainty is around $\chi_{\rm exp}=1.2\times 10^{-3}$.
To estimate the likelihood (degree of certainty) of the present solution as the experimental interpretation,
we used a standard index (for noise, variance and bias decomposition, see Ref.~\onlinecite{bishop2006pattern}) expressed as
{$\delta\chi_{\rm ML}=\sqrt{\chi_{\rm ML}^2-\chi_{\rm exp}^2}=0.8\times 10^{-3}$}.
This is nothing but the pure generalization error/test error derived after subtracting the experimental noise. 
Here $\delta\chi_{\rm ML}/\chi_{\rm exp}=0.7$ is the intrinsic machine learning error in the unit of the experimental standard deviation. This is well within the experimental error bar. 
We show in \ref{S2.2} that other example of optimization without peak structure shows much larger standard error.}
\mi{If we assume that the inferred \mig{$A(k,\omega )$} follows the probability distribution 
$P=\exp[-\chi_{\rm ML}^2 /2\chi_{\rm exp}^2]$ given from the maximum likelihood inference (see Ref.~\onlinecite{bishop2006pattern}), one can estimate the \mig{corresponding} variance of the inference for the self-energy \mig{by sampling the variation of the peak structure}. The variance is plotted in 
Fig.~\ref{Fignoise}(b) for the peak part of ${\rm Im} W$. This indicates that the variance for the peak position and the weight is small and the existence of the peak is robust. 
}

\mi{To further examine the reliability of the emergence of the peak,} 
\migr{in 
Fig.~\ref{Fignoise}(a), 
we first show the estimated $A^{\rm fit}$ and 
{amplified noise $\tilde{\sigma}_{n}=4\sigma_n$} for the optimum doped Bi2212. 
With $\tilde{\sigma}_n$ we generate many 
{synthetic} experimental samples.
\mimg{The reason why we take $\tilde{\sigma}_{n}$ instead of $\sigma_n$ is to secure the stability of the peak structure in the presence of the experimental noise with the safety factor 4.}
Then we perform the machine learning and extract the self-energy from the
{synthetic} $A(k,\omega)$, which provides us with the error bars of the estimated self-energies in our machine learning.
As shown in 
Fig.~2(c), the variance of { ${\rm Im}\Sigma^{\rm nor}$, and ${\rm Im}\Sigma^{\rm ano}$} 
thus obtained from the 
{synthetic} data is reasonably small with the peak structure in the imaginary part of the self-energy, which indicates that our solution of the inverse problem is numerically stable. \mimg{Note that the error bars are somewhat overestimated here (namely, larger error bars than those of Fig.2(c)) because of the factor 4 above, but still the peak structure is reasonably retained.}
\mi{However, further increase of the noise to several times of $\tilde{\sigma}_n$ smears out the peak structure, implying that very accurate experimental data in the present ARPES quality are required to reveal the peak structure.} }

The stability in the present inverse problem shown here clearly indicates the difference from notorious ill-conditioned problems such as the analytic continuation from the imaginary time (Matsubara frequency) variable to the real frequency typically studied by the maximum entropy method.
In contrast to the non-sparse and involved nature of the analytic continuation from the Matsubara frequency, the present stability numerically shown here is consequences of the sparse structure of the transformation between the spectral function and the self-energy. The mapping between $A$ and $\Sigma$ is, though strongly nonlinear, diagonal in the variable $\omega$ and transformation is sparse confined in a limited frequency range. Imposed physical requirement further ensures the stability.  The machine learning including the Boltzmann machine in known to be powerful to strongly nonlinear transformation\cite{Theodoridis,Roux,GaoDuan2017,MontufarAy2011,ackley1985learning} such as Eqs.~(1) and (2). 

Of course if the noise is too high, the peak structure is smeared out. Our synthetic data analysis tells that 64 times higher noise level than the estimated experimental noise washes out the peak structure (not shown) and the factor 10 to the present experimental level would be the limit for the meaningful quantitative analysis.


\subsection{\textcolor{black}{Dependence on initial guess}}
\begin{figure*}[htb]
\begin{center}
\includegraphics[width=1.0\textwidth]
{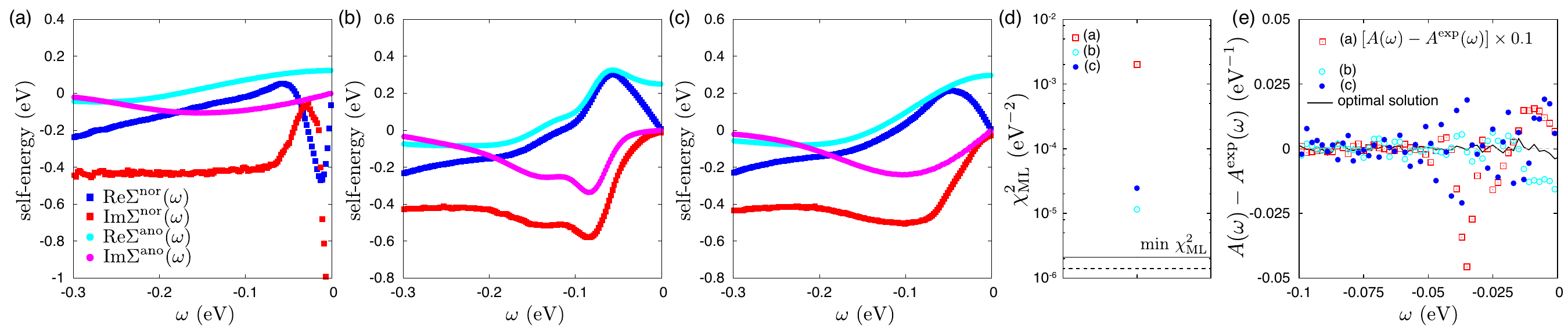}
\end{center}
\caption
{
\textcolor{black}{Typical examples of self-energies obtained by random initial guesses for Bi2212.
From random initial guesses for the imaginary part of the anomalous self-energy,
we often obtain solutions with cost functions, $\chi_{\rm ML}^2$, higher than the current best value, $2.1\times 10^{-6}$ eV$^{-2}$. 
(a)-(c) Three examples of different self-energies obtained from the different random initial guesses for ${\rm Im}\Sigma^{\rm ano}$ are shown.
The self-energies in (a) seems to correspond to a pseudogap or insulating phase, which is characterized by a peak of ${\rm Im}\Sigma^{\rm nor}$
around $\omega = 0$.
On the contrary,
(b) and (c) are interpreted as variants of superconducting solutions.
In (b), small signature of peaks are found in the imaginary parts of the self-energies.  
(d) The cost functions are given for the solutions with the self-energies summarized in (a)-(b).
The horizontal solid line shows the minimum value $\chi_{\rm ML}^2=2.1\times 10^{-6}$ eV$^{-2}$ while the horizontal broken line
shows the experimental noise level $\chi_{\rm exp}^2 = 1.6\times 10^{-6}$ eV$^{-2}$.
(e) The differences between the regression models $A(\omega)$ and the experimental data $A^{\rm exp}(\omega)$ are shown.
For the regression model with the self-energies (a), the difference, $A(\omega) - A^{\rm exp}(\omega)$, multiplied by a factor $0.1$
is shown because the difference is larger than those with the self-energies shown in (b) and (c).
The black solid curve shows the difference, $A(\omega) - A^{\rm exp}(\omega)$, for the optimal solution with $\chi_{\rm ML}^2=2.1\times 10^{-6}$ eV$^{-2}$ (shown in Fig.~\ref{FigSigma}(a)).
}
}
\label{Fig_rand_int}
\end{figure*}
\textcolor{black}{The stability of the optimal solution is examined in the previous subsection.
While a single local minimum of the cost function is analyzed above,
in the present regression scheme,
a multi-valley structure of the cost function in
the parameter space may appear due to
the non-linear nature of the regression.
To explore the nature of the possible multi-minima,
we perform the outer loop optimization,
which is illustrated in Figs.~\ref{Fig_Flow_Chart_Illustration_simplified}
and \ref{Fig_Flow_Chart_Illustration}.
In the outer loop optimization, we update the initial condition of the imaginary
part of the anomalous self-energy to prepare for the next iteration of the inner loop optimization.
During the practical optimization, the initial guess for the (R)BM parameters at the first stage of
the optimization shown in Fig.~\ref{Fig_Flow_Chart_Illustration} will affect the optimized
self-energies.}

\textcolor{black}{To illustrate the initial guess dependence, here,
we examine typical solutions for Bi2212 at 12K obtained from
the randomly chosen initial guesses for ${\rm Im}\Sigma^{\rm ano}$.
As explained in Sec.~\ref{wavelet},
we initialize ${\rm Im}\Sigma^{\rm ano}$ as a linear combination of the Gaussian distributions.
We randomly chose the center of mass, height, and width of these Gaussian distributions.
From the physical constraint, we choose the center of mass within $|\omega|<$ 0.3 eV.
In Fig.~\ref{Fig_rand_int}, the typical examples of the solutions are shown.}

\textcolor{black}{There are two kinds of the solutions: Superconducting and pseudogap solutions
are found.
When an initial guess for ${\rm Im}\Sigma^{\rm ano}$ generates a large enough superconducting gap,
superconducting solutions are obtained.
The superconducting solutions show a minimum of the amplitude of ${\rm Im}\Sigma^{\rm nor}$ around $\omega=0$,
while the amplitude of ${\rm Re}\Sigma^{\rm ano}$ around $\omega=0$ is substantial enough to generate a
quasiparticle gap in the spectral function.
In contrast, if an initial guess for ${\rm Im}\Sigma^{\rm ano}$ cannot generate a large enough superconducting gap,
the amplitude of the normal component ${\rm Im}\Sigma^{\rm nor}$ shows a (negative) peak around $\omega=0$ to generate a gap in the spectral function.
We call such a solution the pseudogap solution.
A typical example of the pseudogap solution is shown in Fig.~\ref{Fig_rand_int}(a),
which shows a three order of magnitude larger cost function as illustrated in Fig.~\ref{Fig_rand_int}(d).
Two superconducting solutions with cost functions larger than the minimum value $\chi^2_{\rm ML}=1.6\times 10^{-6}$
are shown in Figs.~\ref{Fig_rand_int}(b) and (c).
Similarly to the optimal solution shown in Fig.~\ref{FigSigma}(a),
the superconducting solution shown in Fig.~\ref{Fig_rand_int}(b)
exhibits peak structures of ${\rm Im}\Sigma^{\rm ano}$.
However, due to the shift in the peak position, the solution gives an one order of magnitude larger cost function.
A featureless ${\rm Im}\Sigma^{\rm ano}$ also generates a superconducting solution with a large cost function
as shown in  Figs.~\ref{Fig_rand_int}(c) and (d).}

\textcolor{black}{Here we note that the larger cost functions originate from systematic deviation of the regression model
$A(\omega)$ from the experimental data $A^{\rm exp}(\omega)$.
The difference between them, $A(\omega) - A^{\rm exp}(\omega)$, is shown for the three solutions
in Fig.~\ref{Fig_rand_int}(e).
The pseudogap solution (Fig.~\ref{Fig_rand_int}(a)) and the superconducting solution with the featureless ${\rm Im}\Sigma^{\rm ano}$
(Fig.~\ref{Fig_rand_int}(c)) overestimate the spectral function within the quasiparticle gap:
$A(\omega)$ is larger than $A^{\rm exp}(\omega)$ for $\omega\sim 0$ eV.
In contrast, the solution with a peak of ${\rm Im}\Sigma^{\rm ano}$ at a higher energy scale
shows stronger superconducting gap, which results in $A(\omega) < A^{\rm exp}(\omega)$ around $\omega= 0$ eV.}

\textcolor{black}{As examplified by the solutions from randomly chosen initial guesses in Fig.~\ref{Fig_rand_int},
the optimal self-energies shown in Fig.~\ref{FigSigma} indeed give the spectral function closer to the experimental data.
Within our many attempts, we found the unique solution that has a cost function value comparable to the estimated experimental error,
as shown in Fig.~\ref{FigSigma}.}

\subsection{Stability against energy cut-off and background}
\label{stability_background}
\begin{figure*}[htb]
\begin{center}
\includegraphics[width=0.72\textwidth]
{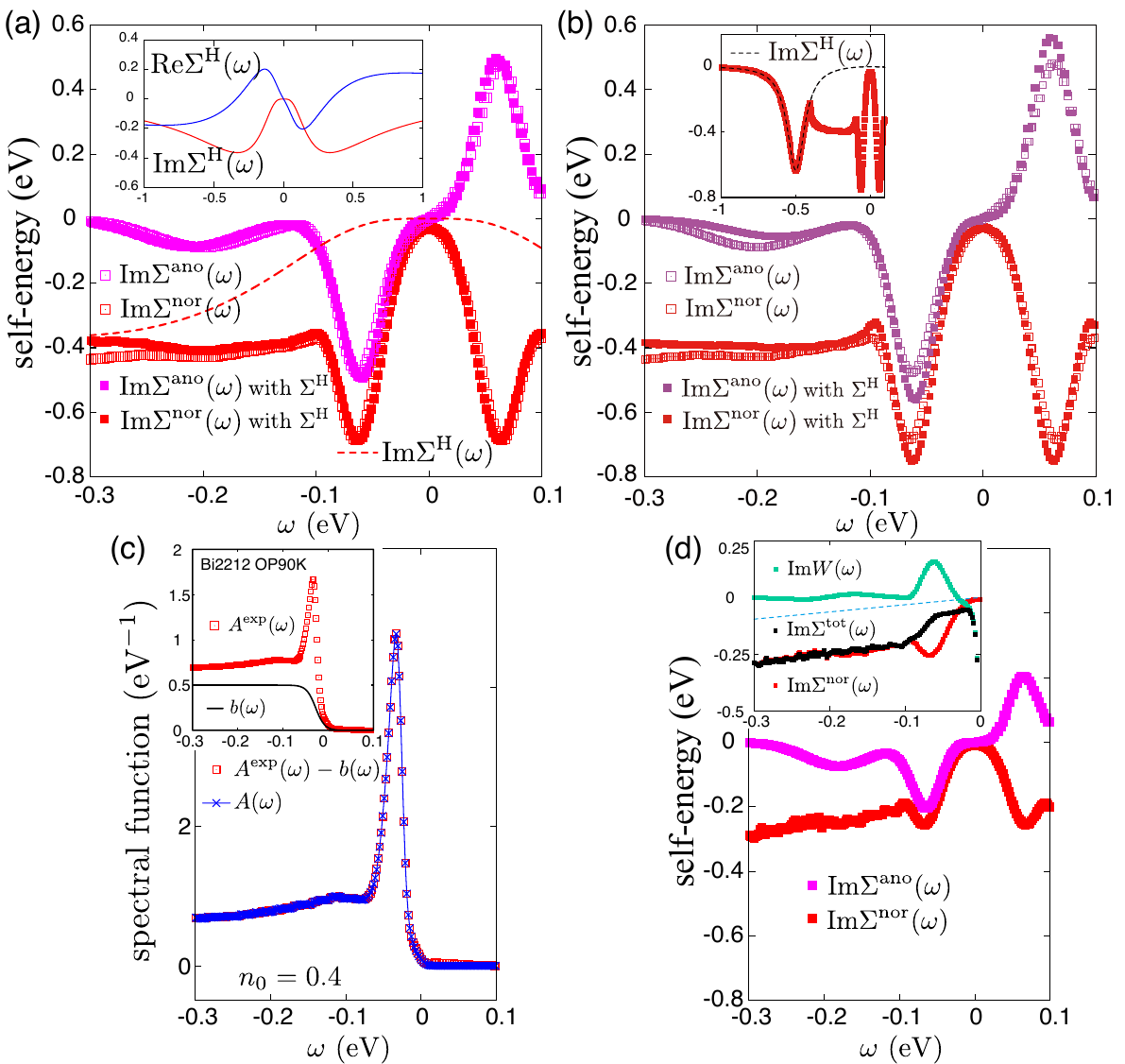}
\end{center}
\caption
{
Effect of high-energy contributions and background
on the self-energy structure \tgtw{for Bi2212}.
\migr{(a). Artificial normal self-energy $\Sigma^{\rm H}$ added by hand in the high-energy region shown in the inset does not have appreciable effect for the self-energies near the Fermi level and the deep dips (peaks) as shown in the main panel. This artificially added normal self-energy is chosen to satisfy the Kramers-Kronig relation in the form of Eq.~(\ref{high-energy_Sigma}).
\mimg{Note that the self energies with (filled symbols)  and without (open symbols)
the high energy 
{contribution} are mostly closely overlapped except for the region near -0.3 eV.  See the text for details of the procedure.}
{(b). Similarly to (a), artificial Lorentzian peak added by hand in the high-energy region as $\Sigma^{\rm H}$
shown in the inset by broken curve does not have appreciable effect 
on
the deep dips (peaks) as shown in the main panel.
}
(c). The spectral function obtained by subtracting the possible extrinsic origin $b(\omega)$ (thin black curve in the inset) from the experimental data $A^{\rm exp}(k,\omega)$ (open red squares in the inset) is given as the open red squares in the main panel. 
{Here, the spectral function is rescaled with $n_0=0.4$.}
The machine learning result to fit $A^{\rm exp}(k,\omega)-b(\omega)$ is plotted as blue crosses and blue fitting curves. $b(\omega)$ can be regarded as a hypothetical background contribution similar to the form in Ref.~\onlinecite{Kaminski2004}.
(d). The self-energies ${\rm Im}\Sigma^{\rm nor}(k,\omega)$ and  ${\rm Im}\Sigma^{\rm ano}(k,\omega)$ obtained by the machine learning of the procedure in (c). Inset: ${\rm Im}\Sigma^{\rm nor}(k,\omega)$ and  ${\rm Im} W(k,\omega)$ together with ${\rm Im}\Sigma^{\rm tot}(k,\omega)$, showing the robust cancellation of ${\rm Im}\Sigma^{\rm nor}(k,\omega)$ and  ${\rm Im} W(k,\omega)$ in the peak (dip). The blue dashed line $\omega/\pi$ has a similar slope with ${\rm Im}\Sigma^{\rm tot}(k,\omega)$, implying a
universal origin of this marginal Fermi-liquid behavior. }
}
\label{Fig_dependence_on_high_energy_part}
\end{figure*}

{The present machine-learning scheme is based on the imaginary parts of the self-energy within a finite frequency range $-\Lambda < \omega < \Lambda$,
where $\Lambda\simeq 0.4$ eV,
because the experimental data observed within $-0.4 \lesssim \omega \lesssim 0.2$.
Therefore, in the genuine self-energy, there is a possible unknown contribution from the outside of the cutoff energy $\Lambda$.
However, as explained below, such a contribution is a monotonic and bounded function of $\omega$, and, thus, possible errors due to the lack of information can be estimated.}

{Due to the Kramers-Kronig relation, the real part of the self-energy can be affected by the cutoff energy $\Lambda$.
Because the imaginary part of the normal self-energy is expected to extend over the cutoff energy, the real part of the normal self-energy has a monotonic and bounded contribution from the outside of the cutoff energy.
On the other hand, because the anomalous self-energy is finite only within the cutoff energy scale, the real part of the anomalous self-energy can be affected by the cutoff only through
the normal self-energy.}

{In the main text, we ignored the contribution of the high-energy part of normal self-energy.} To critically examine the possible contribution from the outside of the cutoff energy,
here, {we assume a possible distribution of the imaginary part of the normal self-energy outside the cutoff:}
The imaginary part of the normal self-energy outside the cutoff is assumed to be confined within $\Omega' -W'/2 \lesssim \omega \lesssim \Omega' + W'/2$ centered at $\Omega'$, where $|\Omega'| > \Lambda \gtrsim W'$  and the amplitude of the imaginary part is approximately constant within this energy range.
Then, contribution to the real part is given by
\begin{eqnarray}
\Delta {\rm Re}\Sigma^{\rm nor}(\omega) \simeq \frac{{\rm Im}\Sigma^{\rm nor}(\Omega')}{\pi} \int_{\Omega'-W'/2}^{\Omega'+W'/2}\frac{d\omega'}{\omega-\omega'},
\end{eqnarray}
which is monotonic for $-\Lambda < \omega < \Lambda$.
When 
{$|\Omega'| >  W'$} is assumed,
{the contribution} $\Delta {\rm Re}\Sigma^{\rm nor}(\omega=0)$
and
{its derivative} 
$\left. \partial \Delta {\rm Re}\Sigma^{\rm nor}(\omega)/\partial \omega \right|_{\omega=0}$
are approximately estimated as
{$\left[-{\rm Im}\Sigma^{\rm nor}(\Omega')/(\pi \Omega')\right] \left[W'+\mathcal{O}({W'}^2/{\Omega'})\right]$}
and 
{$\left[{\rm Im}\Sigma^{\rm nor}(\Omega')/(\pi {\Omega'}^2)\right] \left[W'+\mathcal{O}({W'}^2/{\Omega'})\right]$, respectively.}
If we consider formation of a lower Hubbard band, for instance, we may assume that $|\Omega'| \sim |{\rm Im}\Sigma^{\rm nor} (\Omega')| \sim \mathcal{O}(1)$ eV and
{$W' \lesssim \Lambda$}.
Then, the contribution from the outside of the cutoff is bounded as 
{$|-{\rm Im}\Sigma^{\rm nor}(\Omega') W'/(\pi \Omega') | \lesssim \Lambda/\pi$}
and ${\rm Im}\Sigma^{\rm nor}(\Omega') W'/(\pi {\Omega'}^2) < 1/(2\pi)$.

{As we show in 
Fig.~\ref{Fig_dependence_on_high_energy_part}(a), the qualitative feature of the peak structure at $\omega<-0.4$ eV does not change even when we add an artificial high-energy part for the normal self-energy,
where we compare the self-energy inferred from the experimental ARPES data with
the self-energy inferred with an additional high-energy part $\Sigma^{\rm H}$. 
Here, we added the following fixed high-energy part $\Sigma^{\rm H}$ in the optimization process,
\begin{eqnarray}
\Sigma^{\rm H}(\omega) &=&
\frac{sb}{(c-a)(b-c)}\left\{\frac{\sqrt{b}}{\omega + i\sqrt{b}}-\frac{\sqrt{c}}{\omega + i\sqrt{c}}\right\}
\nonumber\\
&+&
\frac{sa}{(a-b)(c-a)}\left\{\frac{\sqrt{c}}{\omega + i\sqrt{c}}-\frac{\sqrt{a}}{\omega + i\sqrt{a}}\right\},
\nonumber\\
\label{high-energy_Sigma}
\end{eqnarray}
where $s=0.15$eV$^3$, $a=0.3$eV$^2$, $b=0.025$eV$^2$, and $c=0.01$eV$^2$, whose imaginary part becomes substantial for $\omega>-0.4$ eV as shown in the inset of 
Fig.~\ref{Fig_dependence_on_high_energy_part}(a).
Then we fit the experimental ARPES data within the experimentally measured energy range using this self-energy form with the added high-energy tail.
Indeed, in the solution, the artificial high-energy part does not affect the prominent peak structures at all in ${\rm Im}\Sigma^{\rm nor}$ and ${\rm Im}\Sigma^{\rm ano}$.}
When the energy range covered by the measured $A(k, \omega)$ becomes wider \mi{after excluding the extrinsic background}, the uncertainty becomes of course further reduced.

{To examine effects of high-energy structure of ${\rm Im}\Sigma^{\rm nor}$ further,
especially effects of a possible peak structure responsible for the waterfall structure around -0.5eV~\cite{Inosov2007,Kordyuk2005,Graf2007,Xie2007,Valla2007,Meevasana2007},
we conducted the following analysis by adding $\Sigma^{\rm H}$ that has a peak around -0.5eV to the Boltzmann-machine representation of ${\rm Im}\Sigma^{\rm nor}$.
Here, we assume a Lorentzian peak around $\omega =$ -0.5eV as $\Sigma^{\rm H}$ instead of monotonic $\omega$ dependence
at high energies and find optimized $\Sigma + \Sigma^{\rm H}$ to fit the experimentally observed ARPES data of the optimally doped Bi2212.
As shown in 
Fig.~\ref{Fig_dependence_on_high_energy_part}(b),
the presence of the peak structure around -0.5eV does not essentially change
the structures of the low-energy self-energy in the region $\omega >$ -0.2eV.
Here, we note that larger amplitude and/or wider width of the Lorentzian cannot fit the experimental spectral function,
and we also note that the anomalous cusp in ${\rm Im}\Sigma^{\rm nor}$ near the peak structure around -0.5eV is
inevitable to minimize the cost function if one keeps the Lorentzian peak structure.
Aside from details, thus, the potential peak structure does not alter the peak structure of our interest.
As examined in theoretical and numerical studies
(for example, Ref.~\onlinecite{Macridin2007}),
we also note that the realistic waterfall structure can be derived from broader structures in ${\rm Im}\Sigma^{\rm nor}$ than $\Sigma^{\rm H}$ in 
Fig.~\ref{Fig_dependence_on_high_energy_part}(b). In this case the effect of that broad structure becomes smaller than the present critical test.
}

\mig{ \mimg{As an alternative way, one can also examine whether the peak structure is insensitive to the possible slowly varying extrinsic energy dependence or not, by studying the effect of the possible background onto the spectral function.
The origin of the background in the experimental spectral function may be the electronic incoherent part or the extrinsic experimental setup extended in the low- and high-energy regions. 
The background is expected to have a broad (slowly $\omega$ dependent) structure. }
Such a background has been studied before~\cite{Kaminski2004} and we mimic such background structure as a possible effect to see the sensitivity to the peak (dip) structure around $\omega \sim -0.07$ eV. 
In 
Fig.~\ref{Fig_dependence_on_high_energy_part}{(c)},
we show the optimized solution when a model background $b(\omega)$ shown in the inset is subtracted from the spectral function by hand.
The high energy offset in the experimental $A(k,\omega)$ does not appreciably depend on  temperature and momentum, implying such an extrinsic origin. 
The way of optimization to minimize $\chi^2$ in {Eq.~(\ref{chi_square})} is
the same as before except
that we fit $A^{\rm exp}(k,\omega)-b(\omega)$ with a $\omega$ independent rescaling of the amplitude to satisfy the optimal $n_0=0.4$, instead of $A^{\rm exp}(k,\omega)$.
The result shows again that such a broad structure does not affect the prominent peak structure and the peak cancellation between the normal and anomalous self-energies is retained.
{While the cancellation
\tgtw{continues to be retained}
even after the subtraction of the background, the amplitude of the peak structures becomes smaller after the subtraction of the background, as shown in \ref{Fig_dependence_on_high_energy_part}(d). 
\tgtw{Although the} reduction of the peak intensity
\tind{is obtained as a consequence of the regression, it has a simple physical interpretation}
\tgtw{originating} from the renormalization \tgtw{effect:}
\tind{The} \tgtw{inferred} renormalization factor is affected by the amplitude of the imaginary part of the normal self-energy for $\omega <$ -0.1 eV.
After the background subtraction, the amplitude of the imaginary part becomes smaller and, thus, the renormalization factor becomes larger.
While the superconducting gap amplitude estimated in the quasiparticle peak does not depend on the background, the normalization factor depends.
To reproduce the gap amplitude, the smaller amplitude of ${\rm Im}\Sigma^{\rm ano}$ is required when the renormalization becomes larger.
Therefore, the peak structures of ${\rm Im}\Sigma^{\rm nor}$, which \tgtw{have to} cancel the peaks in ${\rm Im}\Sigma^{\rm ano}$
\tgtw{to reproduce the spectral function}
become smaller after the subtraction of the background.}
By the subtraction of the hypothesized background which is essentially constant for $\omega <-0.1$ eV, we see that the high-energy imaginary part of the normal self-energy has a marginal fermi liquid feature ${\rm Im} \Sigma^{\rm nor}(k,\omega)\propto \omega$ consistent with the coefficient of the $T$-linear resistivity in experiments.}

{Here we \tgtw{remark} that the realistic background estimated from the ARPES data
has a similar form to $b(\omega)$, but with much smaller amplitude.
As a standard method to estimate the background, we utilize the EDC curve at the momentum far outside the Fermi surface
\tgtw{\tind{provided} by Kondo~\cite{Kondoprivate}}.
The estimated background at the antinodal point can be well fitted by a modified sigmoid function, $w(1-a\omega)/\{1+\exp[(\omega-b)/c]\}$, where $a=2.24$ eV$^{-1}$, $b=-0.02$ eV, $c = 0.014$ eV, and the weight of the background $w$ is determined to constitute 20\% of the original spectrum for -0.4 eV $<\omega<$ 0.1 eV, which is nearly a quarter of $b(\omega)$ in amplitude employed in
Fig.~\ref{Fig_dependence_on_high_energy_part}(c).
Therefore, the analysis in
Fig.~\ref{Fig_dependence_on_high_energy_part}(c) is regarded as the effect of unrealistically exaggerated background as a extreme case. In the realistic background, the modification of the self-energy peak by the background must be much smaller.}

\begin{figure}[htb]
\begin{center}
\includegraphics[width=0.5\textwidth]
{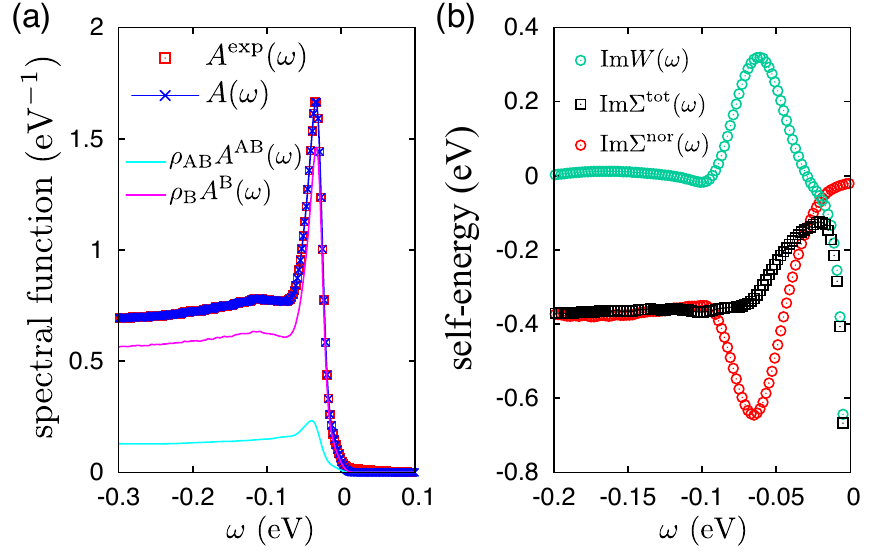}
\end{center}
\caption
{
{Self-energy learning based \tgtw{on two-band Green function for Bi2212}.}
{The total spectral function of the bilayer model $A(\omega)$ (blue crosses) is shown in (a),
which reproduces the experimental data (red squares).
Here, the spectral function is rescaled to fulfill $n_0=0.3$, where $n_0$ is the integrated spectral function for -0.4 eV $< \omega <$ 0.1 eV.
See \tgtw{the text at the end of} Appendix~\ref{S2.1}
for the definition of the integrated spectral function.
We choose \tgtw{that} the bilayer splitting $\epsilon^{\rm AB}-\epsilon^{\rm B}$ at the antinodal region is equal to 0.19 eV by following the tight-binding Hamiltonian for an underdoped Bi2212 with $T_{\rm c} = 78$ K reported by Drozdov {\it et al.}~\cite{drozdov2018phase}.
The matrix element of the antibonding band $\rho_{\rm AB}$ is equal to 0.25 by following Kordyuk {\it et al.}~\cite{kordyuk2002origin}, where $\rho_{\rm B}=1-\rho_{\rm AB}$.
The optimized Boltzmann machine self-energies shown in (b) reproduce the spectral function $A(\omega)$ (blue crosses)
\tgtw{in (a)}.
The cancellation between ${\rm Im}\Sigma^{\rm nor}$ and ${\rm Im}W$ is illustrated in (b).
The contribution from the bonding band $A^{\rm B}(\omega)$ (magenta curve)
and antibonding band $A^{\rm AB}(\omega)$ (cyan curve) is shown in (a).}}
\label{Fig_bilayer_nature}
\end{figure}

{ARPES data are basically not available in the unoccupied part and therefore the inferred behavior in the positive energy side of the spectral function has relatively larger uncertainty. In fact, electron-hole asymmetry was suggested in the scanning \tgtw{tunneling} microscope data~\cite{Fischer2007}. If any kind of data in the positive energy side can be analyzed together, the asymmetry can be analyzed more quantitatively. However, this issue is beyond the scope of the present work. Nevertheless, the asymmetry  expected in the optimally doped Bi2212 (our main target of the analysis) is modest or weak  if it exists as one sees from the STM data and the level of quantitative uncertainty in our analysis does not essentially alter the peak structure in the negative energy side. Technically this is analogous to the case of the uncertainty at the tail part of negative energy side beyond the accessibility by ARPES or the effect of the background discussed above. In fact, consideration of the background assumed only in the negative energy side employed here necessarily introduces electron-hole asymmetry in the measurement and the insensitivity to the background is interpreted for the asymmetry as well.} 

\subsection{Bilayer nature}
\label{appendix_bilayer}
\tgtw{In the bilayer cuprate, the bare band structure consists of the bonding and antibonding band.
However, we have analyzed the ARPES data by using the single-band description of the Green function, which describes the bonding band, because it crosses the Fermi energy.}
We \tgtw{here} examine the effects from the antibonding band, which is located above the bonding band in the $\omega$ axis,
and show that \tgtw{the contribution from the} antibonding band does not change the qualitative results.
Below, we explicitly take into account the \tgtw{two-band} nature of Bi2212
and demonstrate that the results from the multiband treatment are consistent with those from the \tgtw{single-band} treatment.

{Due to the symmetry of the Bi2212 crystal structure, the bare band structure is diagonalized by
using the bonding and antibonding Wannier orbitals that are given by the bonding and antibonding combination of the $d_{x^2-y^2}$ orbitals in two adjacent CuO$_2$ planes.
Even when the self-energies are taken into account, the Green functions for the bilayer cuprates are
diagonal if the interlayer components of the normal and anomalous self-energies are negligible.}

{When we neglect the interlayer components
\tgtw{of the normal and anomalous self-energies,}
the self-energies in the two adjacent CuO$_2$ planes are identical.
Therefore, we only introduce the self-energies, $\Sigma^{\rm nor}(k,\omega)$ and $\Sigma^{\rm ano}(k,\omega)$,
for both of the adjacent CuO$_2$ planes.
The Green functions for the bonding and antibonding band are obtained as follows, respectively: 
\begin{widetext}
\begin{eqnarray}
G^{\rm B} (k,\omega) &=\displaystyle \frac{\omega+i\delta+\epsilon^{\rm B}_k+\Sigma^{\rm nor}(k,-\omega)^{\ast}}{
\left[\omega+i\delta-\epsilon^{\rm B}_k-\Sigma^{\rm nor}(k,\omega)\right]
\left[
\omega+i\delta+\epsilon^{\rm B}_k+\Sigma^{\rm nor}(k,-\omega)^{\ast}
\right]
-\Sigma^{\rm ano}(k,\omega)^2},\nonumber\\
G^{\rm AB} (k,\omega) &=\displaystyle \frac{\omega+i\delta+\epsilon^{\rm AB}_k+\Sigma^{\rm nor}(k,-\omega)^{\ast}}{
\left[\omega+i\delta-\epsilon^{\rm AB}_k-\Sigma^{\rm nor}(k,\omega)\right]
\left[
\omega+i\delta+\epsilon^{\rm AB}_k+\Sigma^{\rm nor}(k,-\omega)^{\ast}
\right]
-\Sigma^{\rm ano}(k,\omega)^2}.\nonumber
\end{eqnarray}
Here, we note that the self-energies are identical even after diagonalization
to obtain the bonding and antibonding band,
\tgtw{irrespective of the amplitude of the interlayer hopping constituting the bare band}.
\end{widetext}
Then, the spectral function, $A(\omega)$, is decomposed into the contribution from the bonding orbitals, $A^{\rm B}(\omega)$,
and antibonding orbitals, $A^{\rm AB}(\omega)$, as follows:}
\begin{eqnarray}
A(\omega) =
\rho_{\rm B}A^{\rm B}(\omega)
+
\rho_{\rm AB}A^{\rm AB}(\omega).
\nonumber
\end{eqnarray}
Here, the orbital dependent matrix elements $\rho_{\rm B}$($=1-\rho_{\rm AB}$) and $\rho_{\rm AB}$ are taken into account.

{Then, we extract the self-energies by using the two band decomposition of $A(\omega) =
\rho_{\rm B}A^{\rm B}(\omega)
+
\rho_{\rm AB}A^{\rm AB}(\omega)$.
As same as in the single band picture, we train the self-energies by minimizing the training error between the theoretical spectral
function $A(\omega)$ and experimental spectral function $A^{\rm exp}(\omega)$.
By following Kordyuk {\it et al}.~\cite{kordyuk2002origin},
we assume that the integrated weight of the contribution from the antibonding band is 25\% as $\rho_{\rm AB}=0.25$.
In \tgtw{Fig.}~\ref{Fig_bilayer_nature}, the optimized spectral function $A(\omega)$ and self-energies are shown.}


{Even when we take the multiband nature into account,
we obtain the self-energies 
that are essentially same as those obtained in the single band analysis.
The cancellation between ${\rm Im}\Sigma^{\rm nor}$ and ${\rm Im}W$ is also observed in \tgtw{Fig.}~\ref{Fig_bilayer_nature}b.}


\section{Details of Wavelet Analysis and Improved 
${\bf Im} \Sigma$ to Reduce Test Error}
\label{S1.2}

{Here,} we discuss why we employ this wavelet formalism.
The idea of using the binary representation {Eq.~(\ref{eq:sigma})}
is to represent a complex function of $\omega$ by successive coarse graining. If a function has $\omega$ dependence with various frequency scales, this hierarchical structure can be efficiently picked up by wavelet with different scales and each wavelet is represented by each digit of the binary number
{$\sigma_j$ $(j=0,1,\dots,L-1)$}.
For example, the last digit $\sigma_{L-1}$ represents the slowest nonzero modulation of frequency dependence (or in other words, short real-time value), namely with the period of the half of our frequency range.  
$\sigma_0$ picks up the most rapid modulation in frequency (or in other words, long real-time value) alternating in the period of the frequency grid mesh $\Delta \omega =\omega/2^L$ {\it etc.}. 
One can have an analogy to Fourier series analysis, where the first digit of the binary number $\sigma_0$ corresponds to the largest time component and the last digit $\sigma_{L-1}$ corresponds to the shortest nonzero time component in the form of $e^{i\omega t}$. 
It was shown that the wavelet can represent the $\omega$ dependence in the orders of magnitude different scales simultaneously and has a flexible representability in the regression problem with small number of parameters because of the logarithmic description (in the present case, $C$ and $D$ with only $L$ arguments, where each of the $L$ components represent logarithmically different scales of frequency dependences)
 for any discrete data-point set~\cite{mallat1989theory,mallat2008wavelet,akansu2001multiresolution}. 

\mi{The grid mesh $\Delta \omega$ is chosen to be smaller than or comparable to the experimental energy resolution ($\sim 10 $meV~\cite{kondo2009competition,kondo2011disentangling}) to fully reproduce the experimental $A(k,\omega)$ within the resolution of the grid size $\Delta \omega $. }

%
%

If the experimental noise level is comparable to the energy resolution, the step-wise representation is sufficient. However, when the experimental spectral data contains only small noise,
the step-wise representation for the imaginary part of the self-energies
may introduce a \mimg{systematic} increase in the test errors.
\mimg{To reduce the possible increased error, we introduce a piecewise-linear representation
instead of the step-wise representation ${\rm Im}\Sigma^{\rm nor/ano}$.
Namely, we interpolate the self-energies between $\omega=\omega_I$ and $\omega=\omega_{I+1}$ linearly as
\eqsa{
{\rm Im}\Sigma_{f}^{\rm nor/ano}(\omega)&=&
\frac{\omega-\omega_I}{\omega_{I+1}-\omega_I}
{\rm Im}\Sigma^{\rm nor/ano}(\omega_{I+1})
\nonumber\\
&+&
\frac{\omega_{I+1}-\omega}{\omega_{I+1}-\omega_I}
{\rm Im}\Sigma^{\rm nor/ano}(\omega_{I}).
\label{piecewise_linear}
}
for $\omega_{I}\leq \omega < \omega_{I+1}$, where $\omega_I$ is the midpoint of the $I(\bvec{\sigma})$th interval defined by $\omega_I \equiv \Lambda (I+1/2)/2^L - \Lambda/2$.
Since the estimated noise is very small for the experimental ARPES spectra of the optimally doped Bi2212 ($T_{\rm c}=90$K) at 11K
(analyzed in Figures 6(a), 7(a), 7(c), 8(a), 8(c), 8(d) and 9 of the main text;
Figs. 10(a), 10(c), 12, 13, 14, 15, 16, 19, and 23 in Appendix),} 
the piecewise linear representation is helpful to achieve the comparable size of test errors with the noise in the experimental data
(see Appendix~\ref{S2.1} for the quantitative discussion).

\section{Details in Real Part of Self-Energy}
\label{S1.3}
\begin{widetext}
The real part of the retarded self-energy is obtained through the Kramers-Kronig relation as Eqs.~(28) and (29).
For example, the real part of normal self-energy in the stepwise representation 
derived from {Eq.~(\ref{eq:Snor})} through the discretized Kramers-Kronig relation is given as
\begin{equation}{
{\rm Re}\Sigma^{\rm nor}(\omega)
=-\sum_{\bvec{{S}}}\frac{{\mathcal{C}}(\bvec{{S}})}{2\pi}\ln \frac
{
\{\Lambda (1+I(\bvec{\sigma}))/2^L -\Lambda/2 -\omega \}^2 + \delta^2
}
{
\{\Lambda I(\bvec{\sigma})/2^{L}-\Lambda/2 -\omega \}^2 + \delta^2
}, 
}
\end{equation}
where we introduce a broadening factor $\delta$
to represent a principal value,
\eqsa{
\mathcal{P}\int d\omega' \frac{f(\omega')}{\omega'-\omega}\nonumber}
by
\eqsa{{\rm Re}\int d\omega' \frac{f(\omega')}{\omega'+i\delta-\omega}.\nonumber}

When the piecewise linear representation in Eq.~(\ref{piecewise_linear}) is \mimg{employed},
the real part of the self-energy is \mimg{corrected by $f_I$} as
\eqsa{
{\rm Re}\Sigma^{\rm nor}_f(\omega) = {\rm Re}\Sigma^{\rm nor}(\omega) + \sum_I f_I (\omega), 
}
where the \mimg{correction term $f_I$ from the stepwise representations}
in each interval $[\omega_I,\omega_{I+1})$ is calculated as 
\eqsa{
&&
f_I(\omega)
=\frac{\Delta {\rm Im}\Sigma (\omega_I)}{\pi}
\left\{
1
+\frac{\omega-\omega_I}{2\Delta\omega}\ln\frac{(\omega-\omega_I-\Delta\omega/2)^2+\delta^2}{(\omega-\omega_I+\Delta\omega/2)^2+\delta^2}
\right.
\nonumber\\
&&
\left.
-\frac{\delta}{\Delta\omega}\left[
\tan^{-1} \left(\frac{\omega-\omega_I+\Delta\omega/2}{\delta}\right)
-
\tan^{-1} \left(\frac{\omega-\omega_I-\Delta\omega/2}{\delta}\right)
\right]
\right\},
}
\end{widetext}
where the width of the interval is given by {$\Delta\omega = \Lambda/2^{L}$}
and the \mimg{increment} of the imaginary part is  $\Delta {\rm Im}\Sigma (\omega_I)={\rm Im}\Sigma^{\rm nor} (\omega_{I+1})-{\rm Im}\Sigma^{\rm nor} (\omega_I)$.
The amplitude of the contribution $f_I$ has the extremum significantly smaller than the \mimg{increment} $\Delta {\rm Im}\Sigma (\omega_I)$, because 
\eqsa{
\left| f_I (\omega)\right|\leq \frac{\Delta {\rm Im}\Sigma (\omega_I)}{\pi}
\left(1-\frac{2\delta}{\Delta\omega}\tan^{-1}\frac{\Delta\omega}{2\delta}\right),
}
\mimg{
for $\delta>\Delta \omega$ and thus
\eqsa{
\left| f_I (\omega)\right|
\sim 10^{-2} \times \frac{\Delta {\rm Im}\Sigma (\omega_I)}{\pi},
}
 is satisfied} when $\delta = 10$meV and $\Delta\omega=\Lambda/2^{L}\sim 3.2$meV.
Thus, the piecewise linear representation introduces a negligible correction to the real part.

\section{ Optimization of $n_0$ 
\label{S2.7}}
\begin{figure*}[htb]
\begin{center}
\includegraphics[width=1.0\textwidth]
{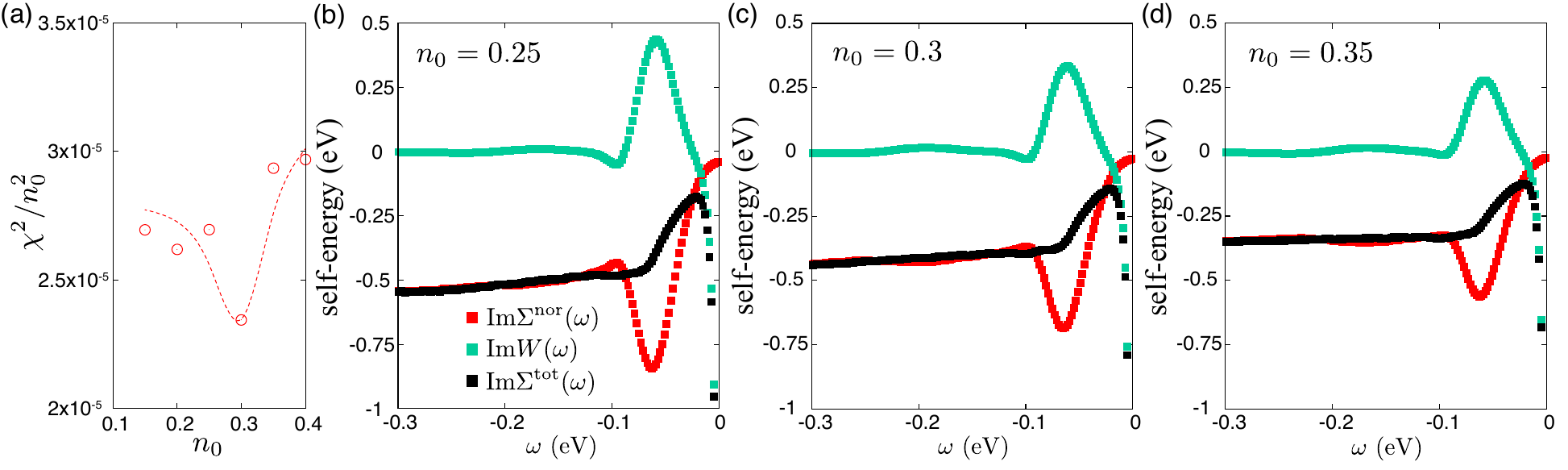}
\end{center}
\caption
{
Determination of $n_0$ \tgtw{for Bi2212}.
In the main text, we have employed $n_0=0.3$.
To assess the validity of this choice, we show $n_0$ dependence of $\overline{\chi^2}$ defined in
Eq.~(\ref{eq:experror}).}
\label{Fig_n0_dependence}
\end{figure*}
We employed $n_{0}=0.3$ in the present analyses. This choice is justified by the least square fit.
We have examined the optimum choice by the least square fit of $\overline{\chi^2}$ by taking several choices of $n_0$.
{Here, we note that $\overline{\chi^2}$ is trivially scaled by the square of the amplitude of $A^{\rm exp}(\omega)$,
and thus is scaled by the square of $n_0$.
Therefore, we need to optimize $\overline{\chi^2}/n_0^2$.}
Fig.~\ref{Fig_n0_dependence}a shows that $\overline{\chi^2}$ normalized by 
{$n_0^2$}
for optimally doped Bi2212 at the antinodal momentum has indeed minimum at $n_0=0.3$,
which indicates that the machine learning suggests that this choice is the optimized value of $n_0$.
The obtained self-energies do not sensitively depend on the choice of $n_0$ 
{as one sees in 
Figs.~\ref{Fig_n0_dependence}(b)-(d)} and we see no qualitative change in the feature of the pronounced peaks in ${\rm Im} \Sigma^{\rm ano}$ and ${\rm Im} \Sigma^{\rm nor}$ at the same energy together with their cancellation in ${\rm Im} \Sigma^{\rm tot}$.

\section{Unoccupied states, bilayer nature, and matrix element}
\label{limitation_issues}
\subsection{Unoccupied states}
{One might be concerned \tgtw{with} the effect of
the unoccupied states, especially the effect of the particle-hole asymmetry in the scanning tunneling spectroscopy 
reported in the literature~\cite{mcelroy2003relating,randeria2005particle}.
\tgtw{Our present scheme does not assume the particle-hole symmetry and is capable of the asymmetry if it exists, although the asymmetry at the energies far from the Fermi level may have some uncertainty because of the lack of the data.}
Aside from the possible origin of the asymmetry~\cite{randeria2005particle,PhysRevLett.102.037001},
the cancellation of the peak structures in the self-energies \tgtw{turns out to} be robust.}

{First of all,
in the superconducting phase, the low-energy spectra around the Fermi level are plausibly particle-hole symmetric.
Therefore, due to the lack of the information \tgtw{about} the unoccupied states,
the optimized self-energies are almost symmetric.}
{Then, the asymmetric behaviors \tgtw{could originate} from the self-energies away from the Fermi level.
However, these high energy asymmetric behaviors hardly affect the cancellation.
We have already demonstrated the robustness of the cancellation under the high-energy perturbation in Appendix~\ref{S2.1}.
Thus, the essence of the present results is not affected by the asymmetry, which may originate from the high-energy
or unoccupied \tgtw{parts of the} spectrum.
\tind{See also Appendix \ref{S2.1}2.}}

\subsection{Matrix elements}
\tgtw{One might also be concerned with the effects of the matrix elements.
Even when the energy dependent matrix elements exist,
the self-energies obtained by using our method is robust as far as the matrix elements are smooth functions of $\omega$.
This is because, when the energy dependent matrix elements are treated as a smooth function of $\omega$,
the effects of the matrix element can be simply taken into account as a smooth background.
As already examined in Appendix \ref{stability_background},
a smooth background does not change the qualitative and essential conclusion of the present paper.
Only if the high resolution ARPES measurements on the entire Brillouin zone, especially at the antinodal region, become available from different experimental setups, quantitative examination of the effects of the matrix elements and background will be conducted and these are highly desirable. However, the data are not available yet and clearly beyond the scope of the present paper.}

\section{Decomposition of Self-Energy}
\label{S1.5}
{While the normal and superconducting components of the total self-energy, ${\rm Im}\Sigma^{\rm nor}(k,\omega)$ and ${\rm Im}W(k,\omega)$,
show the prominent peak structures, which are absent in the Bardeen-Cooper-Schrieffer (BCS) mean-field theory~\cite{bardeen1957theory},
there are an extended background and a BCS-like superconducting contribution, in addition to the peaks.
To highlight \mi{the peak part}, we decompose ${\rm Im}\Sigma^{\rm nor}(k,\omega)$ and ${\rm Im}W(k,\omega)$
into the peaks and other components.
As proposed in Ref.~\onlinecite{PhysRevB.57.R11093}, $\Sigma^{\rm tot}(k,\omega)$ may consist of a single pole that generates a superconducting gap
and a smooth normal state component.
Then, we decompose ${\rm Im}\Sigma^{\rm tot}(k,\omega)$ as
\begin{eqnarray}{
{\rm Im}\Sigma^{\rm tot}(k,\omega)=
{\rm Im}\Sigma_{\rm N}(k,\omega) + L_{\rm BCS}(k,\omega).
}
\label{SigmaDecomp}
\end{eqnarray}
Here, while the BCS-like superconducting contribution is represented by a Lorentzian,
\begin{eqnarray}{
L_{\rm BCS}(k,\omega) = -\frac{1}{\pi}\frac{\Delta_{0}^2{\Gamma}}{(\omega+\epsilon_k)^2 + \Gamma^2},
}
\end{eqnarray}
where $\Delta_0$ and $\Gamma$ are phenomenological parameters that correspond to a BCS-like superconducting gap and
life time of quasiparticles, respectively.
The background ${\rm Im}\Sigma_{\rm N}(k,\omega)$ \mi{can be} represented by a linear combination of many Gaussian distributions, {where its large amplitude signals \mi{either} the incoherence of electrons at that energy \mi{or some extrinsic origin arising from the experimental setup or background}.}
Then, the peak contribution canceled in ${\rm Im}\Sigma^{\rm tot}(k,\omega)$ is, if it exists, obtained from ${\rm Im}\Sigma^{\rm nor}(k,\omega)$ as
\begin{eqnarray}{
{\rm Im}\Sigma_{\rm PEAK}(k,\omega)
=
{\rm Im}\Sigma^{\rm nor}(k,\omega) - {\rm Im}\Sigma_{\rm N}(k,\omega),
}
\end{eqnarray}
and from ${\rm Im}W (k,\omega)$ as
\begin{eqnarray}{
{\rm Im}W_{\rm PEAK}(k,\omega)=
{\rm Im}W (k,\omega)
 - L_{\rm BCS}(k,\omega),
}
\end{eqnarray}
where ${\rm Im}\Sigma_{\rm PEAK}(k,\omega) = - {\rm Im}W_{\rm PEAK}(k,\omega)$ holds.
}

\section{ Momentum Dependence}
\label{S2.4}
\begin{figure*}[htb]
\begin{center}
\includegraphics[width=1.0\textwidth]{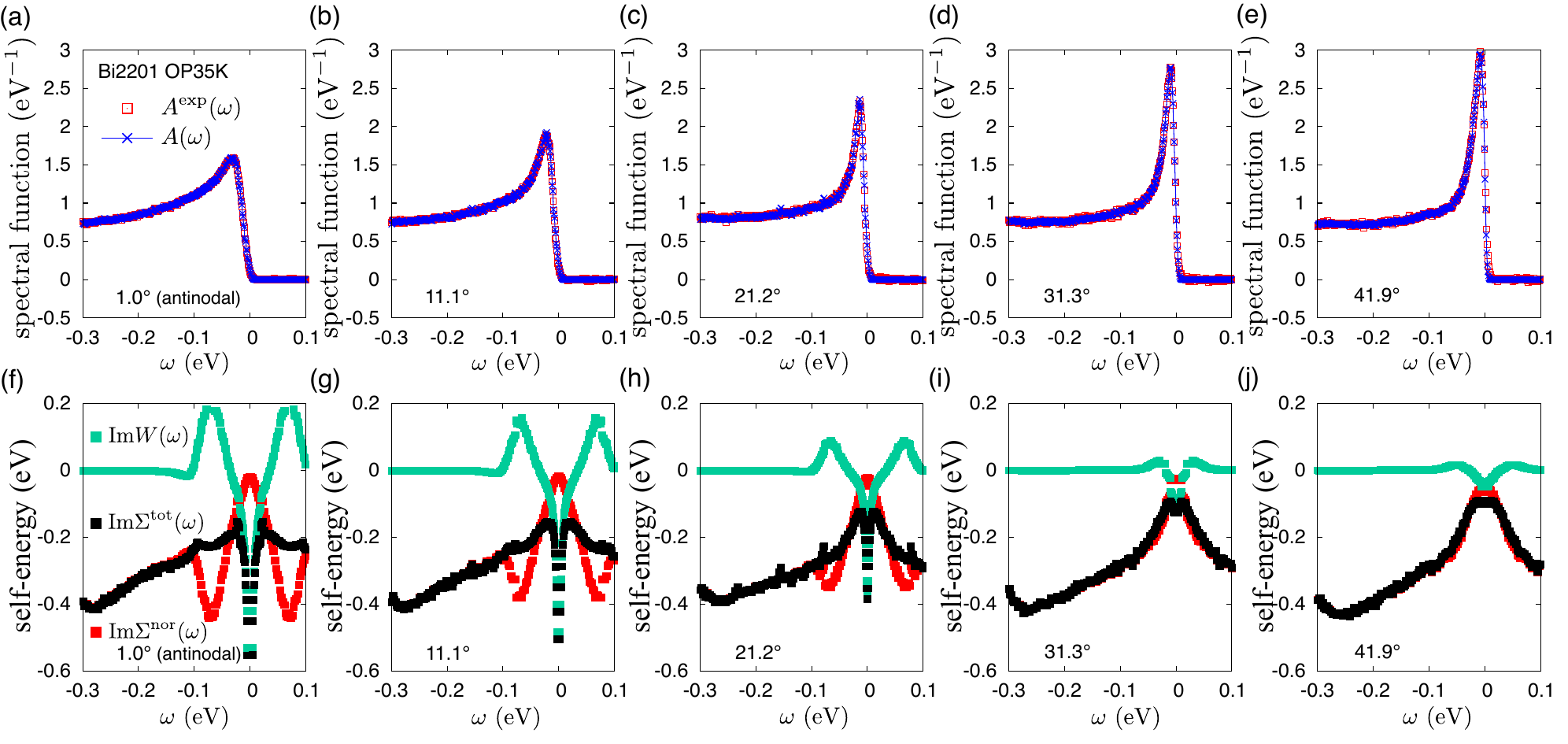}
\end{center}
\caption
{
Momentum (angle) dependence of self-energies.
 Self-energies are obtained from the machine learning using the ARPES EDC curves plotted in the upper panel and taken from the experimental data of Bi2201 at optimum doping at 
\tc{11.3 K at the angle 1.0$^{\circ}$ ((a) and (f)), 11.1$^{\circ}$ ((b) and (g)),
21.2$^{\circ}$ ((c) and (h)), 31.3$^{\circ}$ ((d) and (i)), and 41.9$^{\circ}$ ((e) and (j))}
supplied by Kondo {\it et al.}~\cite{kondo2011disentangling}.
Although the quasiparticle peak becomes sharper when the nodal point is approached, prominent peaks are found at all angles in imaginary parts of the normal and anomalous self-energies around \tc{$\pm 0.07$ eV}, which are missing in $W$ at all the angles,
though the peaks become less pronounced and are \tc{almost missing at 41.9$^{\circ}$} (nearly nodal point).}
\label{Fig_angle_dependence}
\end{figure*}
We have shown the machine learning results in the main text at the antinodal point, because the remarkable structure of the pronounced anomalous self-energy peak, coexisting with the normal self-energy peak is most clearly identified with its dominant contribution to the superconductivity.   However, the momentum dependence of the peak structure provides us with useful insight. We here show the momentum dependence of self-energy structure. 
Fig.~\ref{Fig_angle_dependence}  shows the imaginary part of the normal and anomalous self-energies together with $W$ for the ARPES measurement angle \tc{1.0$^{\circ}$, 11.1$^{\circ}$, 21.2$^{\circ}$, 31.3$^{\circ}$, and 41.9$^{\circ}$} obtained by the machine learning result of Bi2201 at optimum doping at {11.3 K}~\cite{kondo2011disentangling}. 
Note that 0$^{\circ}$ is the antinodal and 45$^{\circ}$ is the nodal points. Although the peaks in the normal and anomalous self-energies become less significant with approaching the nodal point as is expected, the cancellation of ${\rm Im} \Sigma~{\rm nor}$ and $W$ always holds and the prominent peak is missing in ${\rm Im} \Sigma^{\rm tot}$ similarly to the case at the antinodal point. This result further corroborates the universal mechanism of the peak cancellation and the dominant contribution to the superconductivity. 

\section{ Machine Learning Results above $T_{\rm c}$}
\label{S2.3}
\begin{figure*}[htb]
\begin{center}
\includegraphics[width=1.0\textwidth]{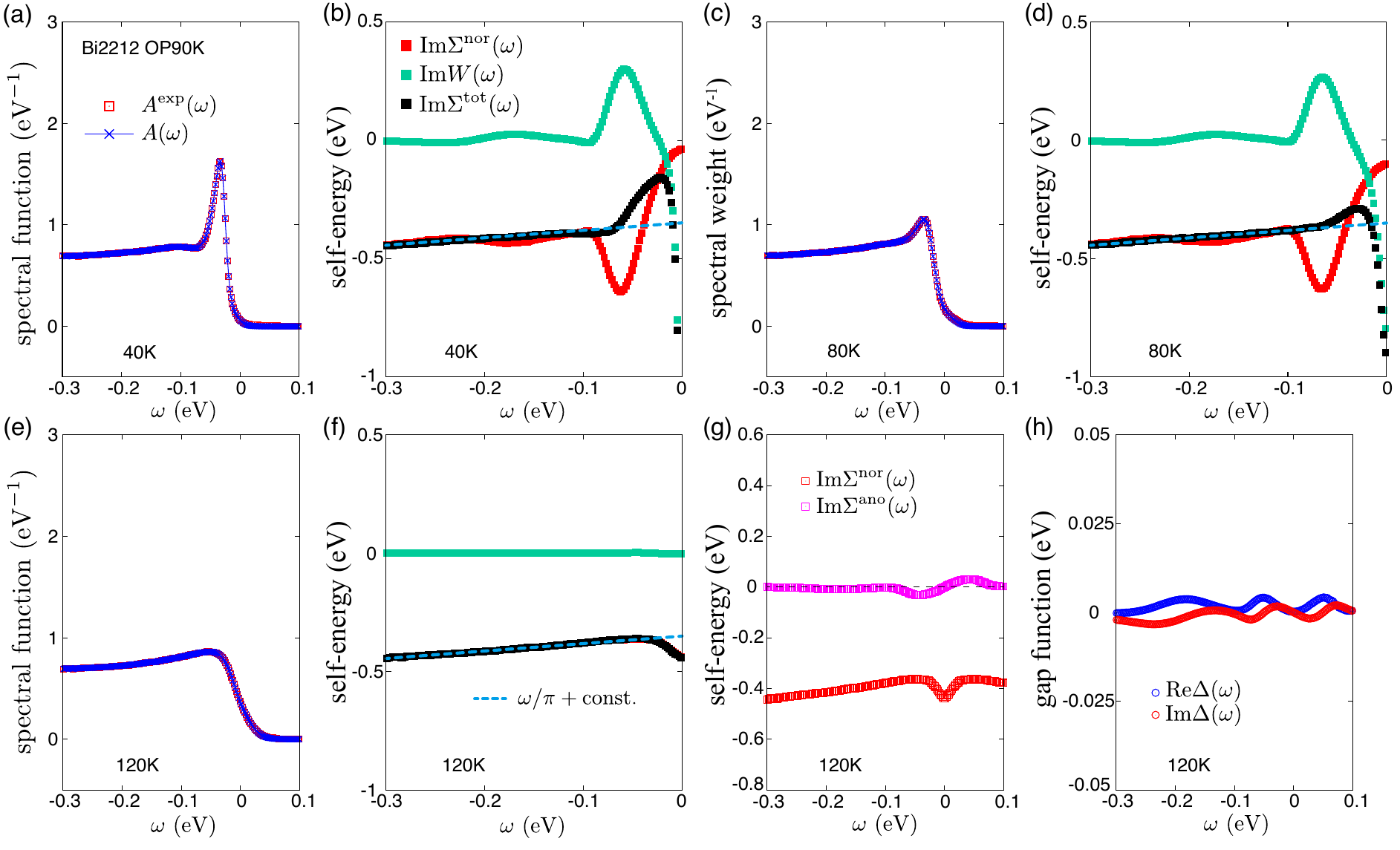}
\end{center}
\caption
{
Temperature dependence of spectral function at {the antinode}
and corresponding self-energies
{((a) and (b) at 40K, (c) and (d) at 80K, (e) and (f) at 120K)} obtained by the machine learning.
{(a), (c), and (e):} Blue crosses and curves are the machine learning results to fit the experimental data shown as open red squares (data of \tr{Bi2212} at optimum doping supplied by Kondo {\it et al.}~\cite{kondo2011disentangling}).
(b), (d), and (f): Self-energies obtained from the machine learning yielding the spectral functions in the left panels. Although the peak (dip) cancellation between $W$ and $\Sigma^{\rm nor}$ around -0.07eV still exists at 40K and 80K, the peak and dip of ${\rm Im} \Sigma^{\rm ano}$ and ${\rm Im} W$ \tgx{essentially} vanish above $T_{\rm c}$, while the dip of ${\rm Im} \Sigma^{\rm nor}$ at $\omega<0$ below $T_{\rm c}$, shifts to the energy around $\omega=0$, indicating the formation of the pseudogap. Furthermore, the total self-energy (black symbols) below the peak energy ($<0.07$ eV) shows a constant slope approximately given by $\omega/\pi$ as drawn as blue dashed lines, supporting marginal Fermi liquid behavior except for the constant value of possible background. 
(g). Comparison of imaginary parts of normal and anomalous self-energies at 120K. While the peak (dip) structure of ${\rm Im} \Sigma^{\rm nor}$ at $\omega =0$ (as shown in {\bf f}) introduces the pseudogap even at 120K in the spectrum ({\bf e}). 
(h). Gap function at 120K. Even though $\Delta (\omega)$ is very small but finite at $\omega\neq 0$,
$\Delta (\omega)$ is more strictly vanishing at $\omega =0$.
}
\label{Fig_temperature_dependence_1}
\end{figure*}
We have shown the machine learning results in the main text for the superconducting phase well below $T_{\rm c}$ to show the remarkable structure of the pronounced anomalous self-energy peak with its dominant contribution to the superconductivity. 
However, the question how the cancellation of the normal and anomalous self-energy evolves with raising temperatures provides us with further insight on its role in the superconductivity. 
Fig.~\ref{Fig_temperature_dependence_1} shows machine learning results of 
the normal and anomalous self-energies together with $W$ for the ARPES measurement obtained by the machine learning of Bi2212 at optimum doping at 40, 80 and 120K~\cite{kondo2011disentangling}. 
The anomalous self-energy peak vanishes above $T_{\rm c}$ as it should be,
which further confirms the validity of the present machine learning scheme.
${\rm Im}\Sigma^{\rm nor}(\omega)$ at 120 K (red curve overlapped with the black curve, ${\rm Im}\Sigma^{\rm tot}(\omega)$)
shows small signature of pseudogap (small dip around $\omega=0$).
{The pseudogap, though not a standard behavior as observed in the underdoped region, is clearly seen in the spectrum
(\ref{Fig_temperature_dependence_1}(c)) if one would perform the electron-hole symmetrization and was analyzed in detail in the original experiments 
(Figs.~\ref{Fignoise}c and 1d in Ref.~\onlinecite{kondo2011disentangling}). In Ref.~\onlinecite{kondo2011disentangling}, it was even argued that the (incoherent) electron pairing is formed below 150K.} 
The energy of the dip of ${\rm Im}\Sigma^{\rm nor}(\omega)$ shifts with raising temperatures  and crosses the quasiparticle peak energy when $T$ crosses $T_{\rm c}$ consistently with the results in Ref.~\onlinecite{PhysRevLett.116.057003}.
\mimt{Then the pseudogap is interpreted as generated by the peak of the normal self-energy above $T_{\rm c}$, which is continued from the peak below $T_{\rm c}$, while it is hidden in the spectral function in the superconducting phase below  $T_{\rm c}$ because of the cancellation with the anomalous self-energy. It turns out that the $d$-wave superconducting gap has an entirely different origin, which emerges as the sharp drop around $\omega=0$ in ${\rm Im} W$, namely as arising from a pole of $W$, distinct from the pole of the self-energies $\Sigma^{\rm ano}$ and $\Sigma^{\rm nor}$. Although the superconducting gap has a different origin, the main contribution to the superconducting order is attributed to the peak (ideally pole) of ${\rm Im} \Sigma^{\rm ano}$ around $\omega=-0.07 (0.04)$ eV for Bi 2212 (Bi2201) (see Figs. 2 and 3). This indicates a tight relation of the pseudogap (manifested by the normal self-energy peak) to the superconductivity (contributed from the peak in ${\rm Im} \Sigma^{\rm ano}$) through their cancellation.}
%

\section{ 
Local and Temperature Insensitive Scattering Rate $z_{\rm qp}(k)c_1(k)$
\label{S2.9.1}}
\begin{figure}[htb]
\begin{center}
\includegraphics[width=0.25\textwidth]{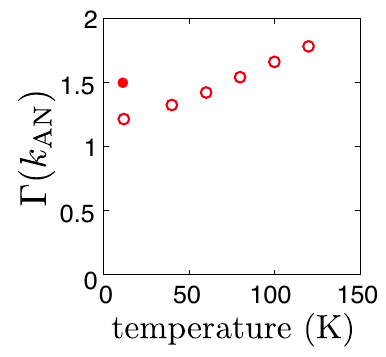}
\end{center}
\caption 
{
Single-particle relaxation time derived from machine learning.
Temperature dependence of $z_{\rm qp}c_1$ of the underdoped Bi2201 with $T_{\rm c}=23$ K at 
\tc{$k_{\rm AN}$}.
At the lowest temperature, the results of two different samples are shown:
The closed circle represents $z_{\rm qp}c_1$ inferred from the underdoped Bi2201 reported in Ref.~\onlinecite{kondo2009competition}
and the open circles denote $z_{\rm qp}c_1$ inferred from the underdoped Bi2201 reported in Ref.~\onlinecite{kondo2011disentangling}.
}
\label{Homes_Uemura0}
\end{figure}

Angle (momentum) dependence of 
$z_{\rm qp}(k_{\rm F})c_1(k_{\rm F})$ plotted for Bi2201 in Fig.4{\bf a} \tb{and resultant $\tau(k)$}
shows that it is only weakly dependent 
around the unity on the angle and doping concentration.
(See also 
Fig.~\ref{FigF} for the plots for each $z_{\rm qp}(k_{\rm F})$ and $c_1(k_{\rm F})$.)
Even for the optimal Bi2212 at the antinodal point,
despite the large difference in $T_{\rm c}$, the value of $z_{\rm qp}(k) c_1(k)$ is similar ($\sim 1.4$).
(Note that the value \tb{$z_{\rm qp}(k_{\rm F})c_1(k_{\rm F})$} is somewhat large ($\sim 1.5$)
\tb{at the nodal point} for the underdoped Bi2201 sample, consistently with the increasing slope of the $T$-linear resistivity in the underdoped region~\cite{Ando04}.
This could be related to the effect of competing insulating behavior.)

With raising temperatures, experimentally observed $T$-linear resistivity has to lead to the temperature insensitive $z_{\rm qp}(k)c_1(k)$ at least near the node,
because the transport is governed by the nodal region.  
Temperature dependence of $z_{\rm qp}(k)c_1(k)$ in the normal state at $k_{\rm AN}$,
for instance for the underdoped Bi2201 is also weak with a large constant offset
shown in 
Fig.~\ref{Homes_Uemura0},
which \tgx{implies} that the $T$-linear dependence ($\propto z_{\rm qp}(k)c_1(k)T$) is preserved irrespective of the momentum. It supports the local nature of dissipation saturated against temperature below and above $T_{\rm c}$ and intrinsically quantum mechanical.




\section{
Momentum and Doping Dependences of $F(k), z_{\rm qp}(k), c_1(k)$ and $\Delta_{\rm qp}(k)$
\label{S2.10}}

\begin{figure*}[htb]
\begin{center}
\includegraphics[width=1.0\textwidth]{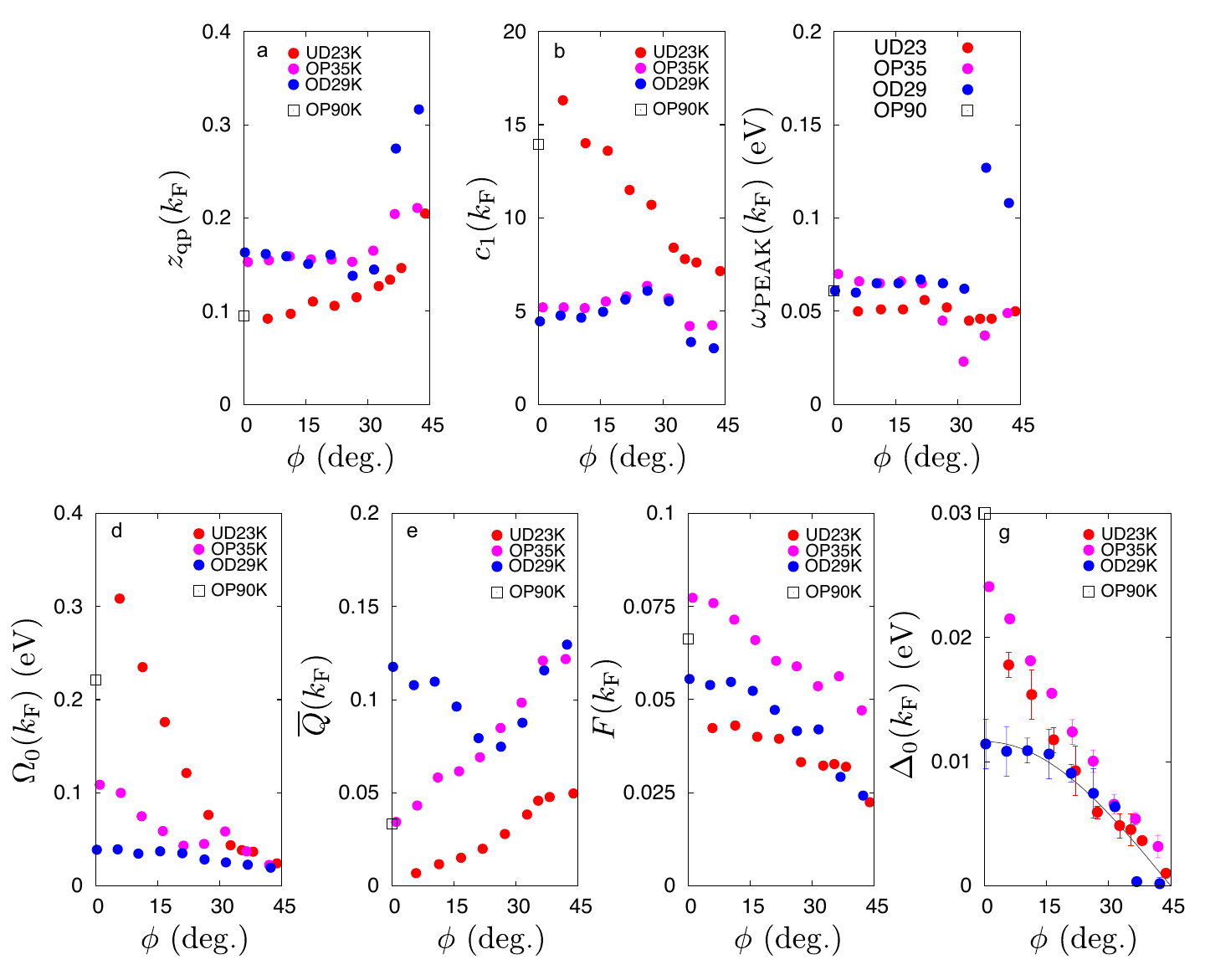}
\end{center}
\caption{
Doping and momentum dependences of \tc{$z_{\rm qp}(k)$, $c_1(k)$, $\omega_{\rm PEAK}(k)$,
$\overline{W_{\rm PEAK}}(k)/\omega_{\rm PEAK}(k)$, $\bar{Q}(k)$, $F(k)$, and $\Delta_{\rm 0}(k)$ for Bi2201 and Bi2212} at $T=11$K. 
{Angle $\phi$ (taken at Fermi momentum $k_{\rm F}$) dependences of 
\tc{(a). $z_{\rm qp}(k)$ {(Eq.~(\ref{eq:zqp}))}, (b). $c_1(k)$ (Eq.~(\ref{eq:c1k})), (c). $\omega_{\rm PEAK}$, 
(d). $\Omega_0 (k)=\overline{W_{\rm PEAK}}(k)/\omega_{\rm PEAK}$, $\overline{W_{\rm PEAK}}(k)\equiv \int d\omega {\rm Im} W_{\rm PEAK}(k,\omega)$,
(e). $\bar{Q}(k)=\int d \omega {\rm Im}W(k,\omega) Q(k,\omega)/\overline{W_{\rm PEAK}}$,
(f). $F(k)$ ({Eq.~(12)}), and
(g). $\Delta_{\rm 0}(k)$ {(Eq.~(\ref{eq:Dqp}))},} 
for three choices of Bi2201 samples with $T_{\rm c}\sim$ 23K (UD), 35K (OP) and 29K(OD) are plotted
\tgtw{by filled symbols}.
Data for Bi2212 with $T_{\rm c}\sim$ 90K (OP) at the antinode ($\phi\sim 0$) are also added \tgtw{by open symbols}.
The solid curve in the most right panel is a cosine curve fitted to $\Delta_0(k)$ for 29K (OD).}}
\label{FigF}
\end{figure*}

\begin{table*}[htb]
\begin{center}
\caption{
{\bf 
Doping dependence of superconducting order parameters and quasiparticle gap.}
Physical quantities are calculated by the self-energy inferred from ARPES data of
three Bi2201 samples and Bi2212 at the optimum doping~\cite{kondo2011disentangling}.
The doping dependence of the superconducting order parameter $F(k)$,
\tb{gap amplitude determined 
from the peak position in EDC, the superconducting gap amplitude $\Delta_0(k)$ obtained by the
present Boltzmann machine learning (BML),}
\tg{the quasiparticle renormalization factor $z_{\rm qp}$, $c_1(k)$, $\overline{Q}$,  $\overline{W_{\rm PEAK}}$, $\omega_{\rm PEAK}$ and $\Omega_0$} are shown mainly at $k=k_{\rm AN}$, where the doping $p$ is estimated by doping dependence of $T_{\rm c}$ in Ref.~\onlinecite{YAndo}.
The average over the Fermi surface of $F$, $\overline{F(k_{\rm F})}$ is also estimated.}
\begin{tabular}{lcclcl}
Sample & & Doping &
Order Parameter & Gap from EDC & Gap from BML 
\\  
& $T_{\rm c}$ (K) 
& $p$ 
& $2F(k_{\rm AN})\left( 2\overline{F(k_{\rm F})} \right)$  
& 
(meV) 
& $\Delta_0(k_{\rm AN})$ (meV) 
\\
\hline
Bi2201 UD & 23 & 0.12 & 0.085 (0.075) & 60  & 18$\pm 1$
\\
Bi2201 OP & 35 & 0.15 & 0.152 (0.126) & 30  & 24.1 $\pm$ 0.2
\\
Bi2201 OD & 29 & 0.18 & 0.108 (0.089) & 12  & 11 $\pm 2$
\\
Bi2212 OP & 90 &  -   & 0.13 (-) & 33  & 30
\\
\hline
\hline
\end{tabular}
\begin{tabular}{lcllll}
Sample & &
$Z$-factor & Coefficient & $Z$-factor 
& Coefficient
\\  
& $T_{\rm c}$ (K) 
& $z_{\rm qp}(k_{\rm N})$
& 
$c_1(k_{\rm N})$
& $z_{\rm qp}(k_{\rm AN})$ 
& $c_1(k_{\rm AN})$ 
\\
\hline
Bi2201 UD & 23 & 0.205 & 7.14 & 0.092  & 16.3  
\\
Bi2201 OP & 35 & 0.211 & 4.24 & 0.155  & 5.20
\\
Bi2201 OD & 29 & 0.316 & 3.00 & 0.16   & 4.45
\\
Bi2212 OP & 90 & - & - & 0.095   & 13.3
\\
\hline
\hline
\end{tabular}
\begin{tabular}{lcllll}
Sample & & Renormalization & Weight 
& Peak Energy&  Energy Scale 
\\  
& $T_{\rm c}$ (K) 
& 
$\overline{Q}(k_{\rm AN})$ 
& $\overline{W_{\rm PEAK}}(k_{\rm AN})$ (eV$^2$) 
& $\omega_{\rm PEAK} (k_{\rm AN})$ (eV)  
& $\Omega_{0}(k_{\rm AN})$ (eV)  
\\
\hline
Bi2201 UD & 23 & 0.007  & 0.015
& 0.05 & 0.31
\\
Bi2201 OP & 35 & 0.035  & 0.0076
& 0.07 & 0.11 
\\
Bi2201 OD & 29 & 0.118  & 0.0024
& 0.061 & 0.039 
\\
Bi2212 OP & 90 & 0.033  & 0.014
& 0.061 & 0.22
\\
\hline
\end{tabular}
\label{table_gap}
\end{center}
\end{table*}

\tr{
In 
Fig.~\ref{FigF}, doping concentration dependences of the superconducting carrier density $F(k)$, the \tb{mass} renormalization factor $z_{\rm qp}(k)$, $c_1(k)$ defined as the $\omega$-linear component of ${\rm Im} \Sigma^{\rm nor}(k,\omega)$,  and $\Delta_{0}(k)$ at the Fermi momentum $k_{\rm F}$ are plotted. Here, $c_1(k)$ is defined by 
\begin{equation}
\tb{c_1(k)=\frac{\partial {\rm Im} \Sigma^{\rm nor}(k,\omega)}{\partial \omega}\mid_{\omega\sim 0}}
\label{eq:c1k}
\end{equation}
 obtained from the linear fitting of  ${\rm Im} \Sigma^{\rm nor}(k,\omega)$ in the range of \tb{15meV$<\omega< 40$meV}. 
}

\tr{
Discrepancy between the doping dependence of $T_{\rm c}$ and
quasiparticle gap amplitude, established in the literature~\cite{Ding,LeTacon}, is \tb{further examined} by
the present self-energy learning.
{In Table~\ref{table_gap}},
the doping dependences of
the density of Cooper pairs
$F(k)$,
the gap amplitude 
\tb{estimated from} the peak position in EDC,
\tb{the superconducting gap $\Delta_0$ obtained by the Boltzmann machine learning}, \tg{ the quasiparticle renormalization factor $z_{\rm qp}$,   $c_1(k)$, $\bar{Q}$,  $P$, $\omega_{\rm PEAK}$ and $\Omega_0$}}
obtained from ARPES data of underdoped (UD), optimally doped (OP), and overdoped (OD) Bi2201 samples~\cite{kondo2011disentangling} are summarized.
While the order parameter \tb{$F$ and superconducting gap $\Delta_0$} show dome-like doping dependence as $T_{\rm c}$ does,
the gap amplitude \tb{estimated from the peak position in EDC} monotonically decreases upon increasing doping~\cite{Ding,LeTacon}.
\tg{On the other hand, $z_{\rm qp}(k_{\rm AN})$ monotonically increases.} They are all consistent with the observed trend in the cuprates.

\bibliography{MachineLearning_updated0825}

\end{document}